\documentstyle[12pt,epsf,rotate]{article} 
\if@twoside \oddsidemargin 21pt \evensidemargin 59pt \marginparwidth 85pt
\else \oddsidemargin 0pt \evensidemargin 0pt
\fi
\renewcommand{\vec}[1]{\mbox{\boldmath $#1$}}
\newcommand{\weg}[1]{#1}
\newcounter{saveeqn}                  
\newcommand{\alpheqn}[1]{\refstepcounter{equation}\label{#1}%
\setcounter{saveeqn}{\value{equation}}%
\setcounter{equation}{0}%
\renewcommand{\theequation}
{\mbox{\arabic{saveeqn}\alph{equation}}}}
\newcommand{\reseteqn}{\setcounter{equation}{\value{saveeqn}}%
\renewcommand{\theequation}{\arabic{equation}}}
\newcounter{savefig}

\renewcommand{\bibitem}[1]{\\[0.2cm]\vphantom{#1}}
\begin{document}
\title{A Mathematical Model for the Behavior of
Individuals in a Social Field}

\author{\mbox{ }\\
Dirk Helbing} 
\maketitle
\clearpage
\thispagestyle{empty}
\vspace*{2cm}
\begin{center}                              
{\LARGE A Mathematical Model for the Behavior of\\[3mm] 
Individuals in a Social Field}
\end{center}
\vfill
\begin{abstract}
\mbox{ }\\[-0.5cm]
Related to an idea of {\sc Lewin}, a mathematical model for behavioral
changes under the influence of a social field is developed. The social field
reflects public opinion, social norms and trends. It is not only given by
external factors (the environment)
but also by the interactions of individuals. Two
important kinds of interaction processes are distinguished: Imitative and
avoidance processes. Variations of individual behavior are taken into 
account by ``diffusion coefficients''.\\[7mm]
{\sc Key words:} Behavioral model, field theory, diffusion model, decision
theory, pair interactions, imitative and avoidance processes
\end{abstract}
\vfill
\clearpage
\setcounter{page}{1}
\begin{center}
{\LARGE\bf A Mathematical Model for the Behavior of\\[3mm] 
Individuals in a Social Field}
\end{center}
\section{Introduction} \label{s1}

Many models have been developed for behavioral changes, but only a few are
formulated in terms of mathematical relations. For example, 
\begin{itemize}
\item {\em game theory} 
({\sc von Neumann} and {\sc Morgenstern}, 1944), based on the concept of 
success of meeting strategies, is used for the
description of cooperation and competition processes among individuals,
\item {\em decision theories} ({\sc Domencich} and {\sc McFadden}, 1975,
{\sc Ort\'{u}zar} and {\sc Willumsen}, 1990), 
assuming the maximization of utility,
successfully model the choice behavior among several alternatives,
\item {\em diffusion models} ({\sc Coleman}, 1964, 
{\sc Bartholomew}, 1967, {\sc Granovetter}, 1983, {\sc Kennedy}, 1983)
mathematically describe the spread of
behaviors or opinions, rumors, innovations, etc.
\end{itemize}
All these models are related to a 
more general behavioral model discussed
in the following. This model is based on {\sc Boltzmann}-{\em like equations}
and includes {\em spontaneous} (or {\em externally induced}) 
behavioral changes and behavioral changes by
{\em pair interactions} of individuals (sect. \ref{s2}). 
These changes are described 
by {\em transition rates}. They reflect the results of mental
and psychical processes, which could be simulated with the help
of {\sc Osgood} and {\sc Tannenbaum}'s (1955) 
{\em congruity principle}, {\sc Heider}'s (1946) 
{\em balance theory} or {\sc Festinger}'s (1957) {\em dissonance
theory}.
However, it is sufficient for our model to determine the transition rates
empirically (sect. \ref{s5}). The {\em ansatz} 
used for the transition rates distinguishes
{\em imitative} and {\em avoidance processes}, 
and assumes {\em utility maximization} by a variant of the 
{\em multinomial logit model} ({\sc Domencich} and {\sc McFadden}, 1975,
{\sc Ort\'{u}zar} and {\sc Willumsen}, 1990) (sect. \ref{s2.1}).
\par
In section \ref{s3} a consequent mathematical 
formulation related to an idea of {\sc Lewin} (1951) 
is developed, according to which the behavior of individuals is
guided by a {\em social field}. This formulation is achieved by a
second order {\sc Taylor} {\em approximation} 
of the {\sc Boltzmann}-like equations
leading to {\em diffusion equations}. 
Because of their relation with the {\sc Boltzmann} equation
({\sc Boltzmann}, 1964) and the
{\sc Fokker-Planck} equation ({\sc Fokker}, 1914, {\sc Planck}, 1917)
they will be called the {\sc Boltzmann-Fokker-Planck} {\em equations}. 
According to these new equations the most
probable behavioral change is given by a 
vectorial quantity that can be interpreted
as {\em social force} (sect. \ref{s3.1}). 
The social force results from external
influences (the environment) as well as from individual 
interactions. In special cases the social 
force is the derivative (gradient) of
a potential. This potential reflects public opinion, 
social norms and trends, and will be called the {\em social field}. 
By {\em diffusion coefficients} individual variation of the behavior
(the ``freedom of will'') is taken into account. In section \ref{s4} 
representative cases are illustrated by computer simulations. 
\par
The {\sc Boltzmann-Fokker-Planck} model for the behavior of individuals under
the influence of a social field shows some analogies with the physical model
for the behavior of electrons in an electric field (e.g. of an atomic
nucleus) ({\sc Helbing}, 1992a,c). 
In particular, individuals and electrons influence the concrete
form of the {\em effective} social, respectively, electric field.
However, the behavior of electrons is governed by a {\em different}
equation: the {\sc Schr\"odinger} {\em equation} 
({\sc Schr\"odinger}, 1926, {\sc Davydov}, 1976).

\section{The {\sc Boltzmann}-like behavioral model} \label{s2}

Let us consider a population consisting
of a great number $N\gg 1$ of individuals.
Concerning a special topic of interest, these individuals show a 
behavior $\vec{x}$ out of several possible behaviors in the set $\Omega$.
\par
Due to ``freedom of the will'' one cannot expect a deterministic theory
for the temporal change $d\vec{x}/dt$ of the individual behavior
$\vec{x}(t)$ to be realistic. However, one can construct a model for
the change of the {\em probability distribution} $P(\vec{x},t)$ of
behaviors $\vec{x}(t)$ within the given population ($P(\vec{x},t) \ge 0$,
$\displaystyle \sum_{\weg{x}\in \Omega} P(\vec{x},t) = 1$). 
A theory of this kind is, of course,
stochastic. In order to take into account several 
{\em types} $a$ of behavior, we
may distinguish $A$ {\em subpopulations} 
$a$ consisting of $N_a \gg 1$ individuals
($\displaystyle \sum_{a=1}^A N_a = N$). Then, the following relation holds:
\begin{equation}
 P(\vec{x},t) = \sum_{a=1}^A \frac{N_a}{N} P_a(\vec{x},t) \, .
\end{equation}
Our goal is now to find a suitable equation for the probability distribution
$P_a(\vec{x},t)$ of behaviors within subpopulation $a$
($P_a(\vec{x},t) \ge 0$, $\displaystyle \sum_{\weg{x}\in \Omega} 
P_a(\vec{x},t) = 1$). If we neglect
memory effects (cf. sect. \ref{s6.1}), the desired equation is
of the form
\begin{equation}
 \frac{d}{dt} P_a(\vec{x},t) = \mbox{inflow into $\vec{x}$} - 
 \mbox{outflow from $\vec{x}$} \, .
\end{equation}
Whereas the {\em inflow} into $\vec{x}$ is given as the sum over all
absolute transition rates describing 
changes from an {\em arbitrary} behavior $\vec{x}'$
to $\vec{x}$, the {\em outflow} 
from $\vec{x}$ is given as the sum over all
absolute transition rates describing 
changes from $\vec{x}$ to {\em another} behavior
$\vec{x}'$. Since the absolute transition rate of changes from $\vec{x}$
to $\vec{x}'$ is the product $w^a(\vec{x}'|\vec{x};t)P_a(\vec{x},t)$
of the {\em relative transition rate} $w^a(\vec{x}'|\vec{x};t)$ for
a change to behavior $\vec{x}'$ {\em given} $\vec{x}$, 
and the probability $P_a(\vec{x},t)$ of behavior $\vec{x}$,
we arrive at the explicit equation
\begin{equation}
 \frac{d}{dt} P_a(\vec{x},t) = \sum_{\weg{x}'\in \Omega \atop
 (\weg{x}' \ne \weg{x})} \Big[
 w^a(\vec{x}|\vec{x}';t) P_a(\vec{x}',t) 
 - w^a(\vec{x}'|\vec{x};t) P_a(\vec{x},t) \Big] \, .
\label{Boltz}
\end{equation}
$w^a(\vec{x}'|\vec{x};t)$ has the meaning of a
transition probablility from $\vec{x}$
to $\vec{x}'$ per unit time and takes into account the behavioral variations
between the individuals (occurring even within the same type $a$ of behavior!). 
\par
In the following we have to specify the 
relative transition rates $w^a(\vec{x}'|\vec{x};t)$, which will turn out
to be {\em effective} transition rates. If we restrict the model to 
spontaneous (or externally induced)
behavioral changes and behavioral changes due to pair
interactions, we have ({\sc Helbing}, 1992a,c):
\begin{equation}
 w^a(\vec{x}'|\vec{x};t) := w_a(\vec{x}'|\vec{x};t)
+ \sum_{b=1}^A \sum_{\weg{y}\in \Omega} \sum_{\weg{y}'\in \Omega} 
N_b \, \widetilde{w}_{ab}
(\vec{x}',\vec{y}'|\vec{x},\vec{y};t) P_b(\vec{y},t) \, .
\label{effrate}
\end{equation}
$w_a(\vec{x}'|\vec{x};t)$ describes the rate of spontaneous 
(resp. externally induced) transitions from
$\vec{x}$ to $\vec{x}'$ for individuals of subpopulation $a$.
$\widetilde{w}_{ab}(\vec{x}',\vec{y}'|\vec{x},\vec{y};t)$
is the transition rate for two individuals of types $a$ and $b$
to change their behaviors from $\vec{x}$ and $\vec{y}$ to
$\vec{x}'$ and $\vec{y}'$, respectively, due to pair interactions. 
The total frequency of these
interactions is proportional to the probability $P_b(\vec{y},t)$
of behavior $\vec{y}$ within subpopulation $b$ and the number $N_b$ of
individuals of type $b$. We have to sum up over $b$, 
$\vec{y}$, and $\vec{y}'$ since all specifications of these
variables are effectively connected with transitions from $\vec{x}$ to
$\vec{x}'$ of individuals of subpopulation~$a$.
\par
Inserting (\ref{effrate}) into (\ref{Boltz}), we now obtain the socalled
{\sc Boltzmann}-{\em like equations} 
({\sc Helbing}, 1992a,c)\alpheqn{Boltzlike}
\begin{eqnarray}
 \frac{d}{dt} P_a(\vec{x},t) &=& \sum_{\weg{x}'\in \Omega} \Big[
 w_a(\vec{x}|\vec{x}';t) P_a(\vec{x}',t) 
 - w_a(\vec{x}'|\vec{x};t) P_a(\vec{x},t) \Big] \\
 &+& \sum_{b=1}^A \sum_{\weg{x}'\in \Omega} \sum_{\weg{y}\in \Omega} 
 \sum_{\weg{y}'\in \Omega} 
 w_{ab}(\vec{x},\vec{y}'|\vec{x}',\vec{y};t) P_b(\vec{y},t)P_a(\vec{x}',t)
 \nonumber \\
 &-& \sum_{b=1}^A \sum_{\weg{x}'\in \Omega} \sum_{\weg{y}\in \Omega} 
 \sum_{\weg{y}'\in \Omega} 
 w_{ab}(\vec{x}',\vec{y}'|\vec{x},\vec{y};t) P_b(\vec{y},t)P_a(\vec{x},t)
\end{eqnarray}\reseteqn
with
\begin{equation}
 w_{ab}(\vec{x}',\vec{y}'|\vec{x},\vec{y};t)
 := N_b \, \widetilde{w}_{ab}(\vec{x}',\vec{y}'|\vec{x},\vec{y};t) \, .
\end{equation}
Obviously, (\ref{Boltzlike}b) depends nonlinearly (quadratically) on the
probability distributions $P_a(\vec{x},t)$ (resp. $P_b(\vec{y},t)$) which
is due to the pair interactions. 
\par
The {\sc Boltzmann}-like equations originally had been developed for the
description of the kinetics of gases 
({\sc Boltzmann}, 1964). However, they have also
been applied to attitude formation 
({\sc Helbing}, 1992b,c) and the avoidance behavior
of pedestrians ({\sc Helbing}, 1992c,d).
\par
It is possible to generalize the model to simultaneous
interactions of an arbitrary number of individuals (i.e., higher order
interactions) ({\sc Helbing}, 1992a,c). 
However, in most cases behavioral changes are dominated
by pair interactions ({\em dyadic} 
interactions). Many of the phenomena occurring
in social interaction processes can already be understood in terms 
of pair interactions.

\subsection{The form of the transition rates} \label{s2.1}

For models of behavioral changes the following special form of the
effective transition rates (\ref{effrate}) has been found to be suitable
({\sc Helbing}, 1992b,c,e):
\begin{equation}
 w^a(\vec{x}'|\vec{x};t) := \nu_a(t) R_a(\vec{x}'|\vec{x};t)
 + \sum_{b=1}^A \nu_{ab}(t) 
 \Big[ f_{ab}^1(t) P_b(\vec{x}',t) + f_{ab}^2(t) P_b(\vec{x},t) \Big] 
 R^a(\vec{x}'|\vec{x};t) \, .
 \label{concrates}
\end{equation}
Here,
\begin{itemize}
\item $\nu_a(t)$ is a measure of the rate of spontaneous 
(or externally induced) behavioral changes
within subpopulation $a$.
\item $R_a(\vec{x}'|\vec{x};t)$ [resp. $R^a(\vec{x}'|\vec{x};t)$] is the
{\em readiness} of an individual of subpopulation $a$ to change behavior
from $\vec{x}$ to $\vec{x}'$ spontaneously [resp. in pair interactions].
\item $\nu_{ab}(t)\equiv N_b \, \widetilde{\nu}_{ab}(t)$ 
is the {\em interaction
rate} of an individual of subpopulation $a$ with individuals of subpopulation
$b$.
\item $f_{ab}^1(t)$ is a measure for the frequency of 
{\em imitative processes}
\begin{equation}
 \vec{x}',\vec{x}' \longleftarrow \vec{x},\vec{x}' \qquad (\vec{x} \ne
 \vec{x}')\, ,
\end{equation}
where an individual of subpopulation $a$ takes over the behavior $\vec{x}'$
of an individual of subpopulation $b$. The total frequency of imitative
processes is proportional to the probablility $P_b(\vec{x}',t)$ of behavior
$\vec{x}'$ within subpopulation $b$.
\item $f_{ab}^2(t)$ is a measure for the frequency of {\em avoidance 
processes}
\begin{equation}
 \vec{x}', \vec{x} \longleftarrow \vec{x}, \vec{x} \qquad
 (\vec{x} \ne \vec{x}') \, ,
\end{equation}
where an individual of subpopulation $a$ changes the behavior $\vec{x}$ to
another behavior $\vec{x}'$ if meeting an individual of subpopulation $b$
with the same behavior (defiant behavior, snob effect). The total frequency of
avoidance processes is proportional to the probablility $P_b(\vec{x},t)$
of behavior $\vec{x}$ within subpopulation $b$.
\end{itemize}
A more detailled discussion of the different kinds of interaction processes
and of {\em ansatz} (\ref{concrates}) is given in publications of
{\sc Helbing} (1992b,c,e).
\par
For $R^a(\vec{x}'|\vec{x};t)$ we take the quite general form\alpheqn{Util}
\begin{equation}
 R^a(\vec{x}'|\vec{x};t) = \frac{\mbox{e}^{U^a(\weg{x}',t) - U^a(\weg{x},t)}}
 {D_a(\vec{x}',\vec{x};t)} 
 \label{util}
\end{equation}
with
\begin{displaymath}
 D_a(\vec{x}',\vec{x};t) = D_a(\vec{x},\vec{x}';t) > 0 
\end{displaymath}
(cf. {\sc Weidlich} and {\sc Haag}, 1988, {\sc Helbing}, 1992c). 
Then, the readiness $R^a(\vec{x}'|\vec{x};t)$
for an individual of subpopulation $a$ to change behavior from $\vec{x}$
to $\vec{x}'$ will be greater, 
\begin{itemize}
\item the greater the {\em difference} in the
{\em utilities} $U^a(.,t)$ of behaviors $\vec{x}'$ and $\vec{x}$, 
\item the smaller the {\em incompatibility 
(``distance'')} $D_a(\vec{x}',\vec{x};t)$
between the behaviors $\vec{x}$ and $\vec{x}'$. 
\end{itemize}
Similar to (\ref{util}) we use 
\begin{equation}
 R_a(\vec{x}'|\vec{x};t) = \frac{\mbox{e}^{U_a(\weg{x}',t) - U_a(\weg{x},t)}}
 {D_a(\vec{x}',\vec{x};t)} \, ,
 \label{util2}
\end{equation}\reseteqn
and, therefore, allow the utility function $U_a(\vec{x},t)$ for spontaneous
(or externally induced)
behavioral changes to differ from the utility function $U^a(\vec{x},t)$
for behavioral changes in pair interactions.
{\em Ansatz} (\ref{Util}) is related to the {\em multinomial 
logit model} ({\sc Domencich} and {\sc McFadden}, 1975,
{\sc Ort\'{u}zar} and {\sc Willumsen}, 1990).
It assumes {\em utility maximization} with 
incomplete information about the exact
utility of a behavioral change from $\vec{x}$ to $\vec{x}'$, which
is, therefore, estimated and stochastically varying 
(cf. {\sc Helbing}, 1992c). 
\par
Computer simulations of the {\sc Boltzmann}-like
equations (\ref{Boltz}), (\ref{concrates}), (\ref{Util}) 
are discussed and illustrated in 
{\sc Helbing} (1992b,c,e) (cf. also sect. \ref{s4}).

\subsection{Special fields of application in the social sciences} \label{s2.2}

The {\sc Boltzmann}-like equations (\ref{Boltz}), (\ref{concrates}) 
include a variety of special cases, which have become
very important in the social sciences: 
\begin{itemize}
\item The {\em logistic equation} ({\sc Pearl}, 1924, {\sc Verhulst}, 1845)
describes limited growth
processes. Let us consider the situation of two behaviors $\vec{x} \equiv x
\in \{ 1, 2 \}$ (i.e., $P_a(1,t) = 1 - P_a(2,t)$) 
and one subpopulation ($A=1$).
$x=2$ may, for example, have the meaning to apply a certain
strategy, and $x = 1$ not to do so. If only imitative
processes
\begin{equation}
 2, 2 \longleftarrow 1, 2
\end{equation}
and processes of spontaneous replacement
\begin{equation}
 1 \longleftarrow 2
\end{equation}
are considered, one arrives at the {\em logistic equation}
\begin{eqnarray}
 \frac{d}{dt} P_1(2,t) &=& - \nu_1(t) R_1(1|2;t)P_1(2,t)
 + \nu_{11}(t)f_{11}^1(t)R^1(2|1;t)\Big( 1 - P_1(2,t) \Big) P_1(2,t) \nonumber
 \\
 &\equiv & A(t) P_1(2,t) \Big( B(t) - P_1(2,t) \Big) \, .
\end{eqnarray}
\item The {\em gravity model} ({\sc Zipf}, 1946,
{\sc Ravenstein}, 1876) describes processes of 
exchange between different places $\vec{x}$. Its dynamical version results for
$R_a(\vec{x}'|\vec{x};t) \equiv 0$, $f_{ab}^1(t) \equiv 1$,
$f_{ab}^2(t) \equiv 0$, and $A=1$:
\begin{equation}
 \frac{d}{dt} P(\vec{x},t) = \nu(t) \sum_{\weg{x}'\in \Omega} \left[ 
 \frac{\mbox{e}^{U(\weg{x},t) - U(\weg{x}',t)}}{D(\vec{x},\vec{x}')} 
 - \frac{\mbox{e}^{U(\weg{x}',t) - U(\weg{x},t)}}{D(\vec{x}',\vec{x})}
 \right] P(\vec{x},t)P(\vec{x}',t) \, .
\end{equation}
Here, we have dropped the index $a$ because of $a=1$. $P(\vec{x},t)$ is the
probability of being at place $\vec{x}$. The absolute rate
of exchange from $\vec{x}$ to $\vec{x}'$ is proportional to the
probabilities $P(\vec{x},t)$ and $P(\vec{x}',t)$ at the places $\vec{x}$
and $\vec{x}'$. $D(\vec{x},\vec{x}')$ is often chosen as a function 
of the metric distance $\| \vec{x} - \vec{x}' \|$ between
$\vec{x}$ and $\vec{x}'$: $D(\vec{x},\vec{x}') \equiv
D(\| \vec{x} - \vec{x}' \|)$.
\item The {\em behavioral model} of {\sc Weidlich} and {\sc Haag}
(1983, 1988, {\sc Weidlich}, 1991, 1994) is based on
spontaneous transitions. We obtain
this model for $f_{ab}^1(t) \equiv 0 \equiv f_{ab}^2(t)$ and
\begin{equation}
 U_a(\vec{x},t) := \delta_a(\vec{x},t) 
 + \sum_{b=1}^A \kappa_{ab} \, P_b(\vec{x},t)
 \, . \label{kappa}
\end{equation}
Because of the dependence of the utilities $U_a(\vec{x},t)$ on the
behavioral distributions $P_b(\vec{x},t)$ the model assumes 
{\em indirect interactions}, which are, for example, mediated by
the newspapers, TV or radio.
$\delta_a(\vec{x},t)$ is the {\em preference} of subpopulation $a$ for 
behavior $\vec{x}$. $\kappa_{ab}$ are {\em coupling parameters} describing the
influence of the behaviorial distribution 
within subpopulation~$b$ on the behavior of
subpoplation $a$. For $\kappa_{ab} > 0$, $\kappa_{ab}$ reflects the
{\em social pressure} of behavioral majorities.
\item The {\em game dynamical equations} ({\sc Hofbauer} and {\sc Sigmund},
1988, {\sc Schuster} {\em et. al.}, 1981, {\sc Helbing}, 1992c,e, 1993) 
result for $f_{ab}^1(t) \equiv \delta_{ab}$, $f_{ab}^2(t) \equiv 0$, and
\begin{equation}
 R^a(\vec{x}'|\vec{x};t) := \max \Big( E_a(\vec{x}',t) - E_a(\vec{x},t) , 0
 \Big) \, ,
\end{equation}
where
\begin{equation}
 \delta_{ab} := \left\{
\begin{array}{ll}
 1 & \mbox{if } a=b \\
 0 & \mbox{if } a\ne b
\end{array}\right.
\qquad \mbox{and} \qquad
\max (x,y) := \left\{
\begin{array}{ll}
 x & \mbox{if } x\ge y \\
 y & \mbox{if } y > x \, .
\end{array}\right.
\end{equation}
For a detailled interpretation of these relations see {\sc Helbing} (1992c,e).
\par
The explicit form of the game dynamical equations is\alpheqn{game}
\begin{eqnarray}
 \frac{d}{dt} P_a(\vec{x},t) &=& \sum_{\weg{x}'\in \Omega} \Big[
 w_a(\vec{x}|\vec{x}';t)P_a(\vec{x}',t) - w_a(\vec{x}'|\vec{x};t)P_a(\vec{x},t)
 \Big] \\
 &+& \nu_{aa}(t) P_a(\vec{x},t) \Big[ E_a(\vec{x},t) - \langle E_a \rangle
 \Big] \, .
\end{eqnarray}\reseteqn
Whereas (\ref{game}a) again describes spontaneous behavioral changes
({\em ``mutations''}, innovations), 
(\ref{game}b) reflects competition processes
leading to a {\em ``selection''} of behaviors 
with a {\em success} $E_a(\vec{x},t)$
that exceeds the {\em average success}
\begin{equation}
 \langle E_a \rangle := \sum_{\weg{x}'\in \Omega} 
 E_a(\vec{x}',t) P_a(\vec{x}',t) \, .
\end{equation}
The success $E_a(\vec{x},t)$ is connected with the socalled {\em payoff
matrices} $\underline{A}_{ab} \equiv \Big( A_{ab}(\vec{x},\vec{y}) \Big)$ by
\begin{equation}
 E_a(\vec{x},t) := A_a(\vec{x}) 
 + \sum_{b=1}^A \sum_{\weg{y}\in \Omega} A_{ab}(\vec{x},\vec{y})
 P_b(\vec{y},t) 
\end{equation}
({\sc Helbing}, 1992c,e). $A_a(\vec{x})$ means the success of 
behavior $\vec{x}$ with respect to the environment.
\par
Since the game dynamical equations (\ref{game}) 
agree with the {\em selection mutation equations} ({\sc Hofbauer} and
{\sc Sigmund}, 1988) they are not only a 
powerful tool in social sciences and economy ({\sc Axelrod}, 1984, {\sc von
Neumann} and {\sc Morgenstern}, 1944, {\sc Schuster} {\em et. al.},
1981, {\sc Helbing}, 1992c,e, 1993), but also in
evolutionary biology ({\sc Fisher}, 1930, {\sc Eigen}, 1971, {\sc Eigen}
and {\sc Schuster}, 1979, {\sc Fei\-stel} and {\sc Ebeling}, 1989).
\end{itemize}

\section{The {\sc Boltzmann-Fokker-Planck} equations} \label{s3}

We shall now assume the set $\Omega$ of possible behaviors forms a
{\em continuous} space. The $n$ dimensions
of this space correspond to different characteristic
{\em aspects} of the considered behaviors.
In the continuous formulation, the sums in (\ref{Boltz}), (\ref{effrate})
have to be replaced by integrals:\alpheqn{Cont}
\begin{equation}
 \frac{d}{dt} P_a(\vec{x},t) = \int\limits_\Omega d^n x' \, \Big[
 w^a(\vec{x}|\vec{x}';t) P_a(\vec{x}',t) 
 - w^a(\vec{x}'|\vec{x};t) P_a(\vec{x},t) \Big] \, ,
\end{equation}
\begin{equation}
 w^a(\vec{x}'|\vec{x};t) := w_a(\vec{x}'|\vec{x};t)
+ \sum_{b=1}^A \int\limits_\Omega d^n y 
\int\limits_\Omega d^n y' \, N_b \, \widetilde{w}_{ab}
(\vec{x}',\vec{y}'|\vec{x},\vec{y};t) P_b(\vec{y},t) \, .
\end{equation}\reseteqn
A reformulation of the {\sc Boltzmann}-like equations (\ref{Cont}) via  
a second order {\sc Taylor} {\em approximation}
({\sc Kramers-Moyal} expansion ({\sc Kramers}, 1940, {\sc Moyal},
1949)) leads to {\em diffusion equations}
({\sc Helbing}, 1992a,c):\alpheqn{bfp}
\begin{equation}
 \frac{\partial}{\partial t}P_a(\vec{x},t) =
- \sum_{i=1}^n \frac{\partial}{\partial x_{i}} \Big[K_{a i}(\vec{x},t)
P_a(\vec{x},t)\Big]
+ \frac{1}{2} \sum_{i, j=1}^n 
\frac{\partial}{\partial x_{i}}\frac{\partial}
{\partial x_{j}} \Big[Q_{a i j}(\vec{x},t)
P_a(\vec{x},t)\Big] 
 \label{BFP}
\end{equation}
with the effective {\em drift coefficients}
\begin{equation}
K_{a i}(\vec{x},t) 
:= \int\limits_\Omega \!  d^n x' \, (x'_i - x_i) w^a(\vec{x}'|\vec{x};t) 
\label{effdr}
\end{equation}
and the effective {\em diffusion 
coefficients}\footnote{In a paper of {\sc Helbing} (1992a)
the expression for 
$Q_{aij}(\vec{x},t)$ contains additional terms
due to another derivation of (\ref{bfp}).
However, they make no contributions, since they
result in vanishing surface integrals 
({\sc Helbing}, 1992c).}
\begin{equation}
Q_{a i j}(\vec{x},t) 
:= \displaystyle \int\limits_\Omega \!  d^n x' \, (x'_i - x_i) (x'_j - x_j)
 w^a(\vec{x}'|\vec{x};t) \, .
\label{effdiff}
\end{equation}\reseteqn
Because of their relation with the {\sc Boltzmann} equation ({\sc Boltzmann},
1964) and the {\sc Fokker-Planck} equation ({\sc Fokker}, 1914, {\sc Planck},
1917) equations (\ref{bfp}) will be called the
{\sc Boltzmann-Fokker-Planck} {\em equations} in the following
(cf. {\sc Helbing} 1992a,c). In the {\sc Boltzmann-Fokker-Planck} equations
the drift coefficients $K_{ai}(\vec{x},t)$ govern the systematic
change (``drift'', motion) of the distribution $P_a(\vec{x},t)$, whereas
the diffusion coefficients $Q_{aij}(\vec{x},t)$ describe the
spread of the distribution $P_a(\vec{x},t)$ due to fluctuations
resulting from the individual variation of behavioral changes.
\par
For {\em ansatz} (\ref{concrates}), 
the effective drift and diffusion coefficients
can be split into contributions due to spontaneous 
(or externally induced) transitions
($k=0$), imitative processes ($k=1$), and avoidance processes 
($k=2$):\alpheqn{SUM}
\begin{equation}
 K_{ai}(\vec{x},t) = \sum_{k=0}^2 K_{ai}^k(\vec{x},t)\, , \qquad
 Q_{aij}(\vec{x},t) = \sum_{k=0}^2 Q_{aij}^k(\vec{x},t)\, ,
\label{split}
\end{equation}
where
\begin{eqnarray}
K_{ai}^0(\vec{x},t) &:=& \nu_a(t) \int \!  d^n x' \, (x'_i - x_i)
 R_a(\vec{x}'|\vec{x};t)\, , \nonumber \\
K_{ai}^1(\vec{x},t) &:=& \sum_{b=1}^A \nu_{ab}(t) f_{ab}^1(t) 
\int \!  d^n x' \, (x'_i - x_i)
 R^a(\vec{x}'|\vec{x};t)P_b(\vec{x}',t) \, , \nonumber \\
K_{ai}^2(\vec{x},t) &:=& \sum_{b=1}^A \nu_{ab}(t) f_{ab}^2(t) 
\int \!  d^n x' \, (x'_i - x_i)
 R^a(\vec{x}'|\vec{x};t)P_b(\vec{x},t) 
\label{splitforce}
\end{eqnarray}
and
\begin{eqnarray}
Q_{aij}^0(\vec{x},t) &:=& \nu_a(t) \int \!  d^n x' \, (x'_i-x_i) (x'_j-x_j)
 R_a(\vec{x}'|\vec{x};t)\, , \nonumber \\
Q_{aij}^1(\vec{x},t) &:=& \sum_{b=1}^A \nu_{ab}(t) f_{ab}^1(t) 
\int \!  d^n x' \, (x'_i - x_i)
(x'_j - x_j) R^a(\vec{x}'|\vec{x};t)P_b(\vec{x}',t) \, , \nonumber \\
Q_{aij}^2(\vec{x},t) &:=& \sum_{b=1}^A \nu_{ab}(t) f_{ab}^2(t) 
\int \!  d^n x' \, (x'_i - x_i)
(x'_j - x_j) R^a(\vec{x}'|\vec{x};t)P_b(\vec{x},t) \, .
\end{eqnarray}\reseteqn
The behavioral changes induced by the {\em environment} are included in
$K_{ai}^0(\vec{x},t)$ and $Q_{aij}^0(\vec{x},t)$.

\subsection{Social force and social field} \label{s3.1}

The {\sc Boltzmann-Fokker-Planck} equations (\ref{bfp}) are equivalent to 
the stochastic equations ({\sc Langevin} equations,
1908)\alpheqn{langevin}
\begin{equation}
 \frac{d x_i}{dt} = F_{ai}(\vec{x},t) 
 + \sum_{j=1}^n G_{aij}(\vec{x},t) \xi_j(t)
 \label{Langevin}
\end{equation}
with 
\begin{equation}
 K_{ai}(\vec{x},t) = F_{ai}(\vec{x},t) + \frac{1}{2} \sum_{j,k=1}^n
 \left[ \frac{\partial}{\partial x_k} G_{aij}(\vec{x},t) \right]
 G_{ajk}(\vec{x},t) 
 \label{Drift}
\end{equation}
and
\begin{equation}
 Q_{aij}(\vec{x},t) = \sum_{k=1}^n G_{aik}(\vec{x},t) G_{akj}(\vec{x},t)
 \label{Diff}
\end{equation}\reseteqn
(cf. {\sc Stratonovich}, 1963, 
{\sc Helbing}, 1992c). For an individual of subpopulation $a$
the vector $\vec{\zeta}_a(\vec{x},t)$ with the components
\begin{equation}
 \zeta_{ai}(\vec{x},t) = \sum_{j=1}^n G_{aij}(\vec{x},t) \xi_j(t)
\end{equation}
describes the contribution to the 
change of behavior $\vec{x}$ that is caused
by behavioral fluctuations $\vec{\xi}(t)$ (which are assumed
to be delta-correlated and {\sc Gauss}ian ({\sc Helbing}, 1992c)). 
Since the diffusion coefficients
$Q_{aij}(\vec{x},t)$ and the coefficients
$G_{aij}(\vec{x},t)$ are usually
small quantities, we have $F_{ai}(\vec{x},t) \approx
K_{ai}(\vec{x},t)$ (cf. (\ref{Drift})), 
and (\ref{Langevin}) can be put into the form
\begin{equation}
 \frac{d\vec{x}}{dt}  \approx \vec{K}_a(\vec{x},t) + \mbox{\em fluctuations.}
 \label{result}
\end{equation}
Whereas the fluctuation term describes individual behavioral variations, the
vectorial quantity
\begin{equation}
 \vec{K}_a(\vec{x},t) := \left(
\begin{array}{c}
K_{a1}(\vec{x},t) \\
\vdots \\
K_{an}(\vec{x},t)
\end{array}\right)
\end{equation}
drives the systematic change of the behavior $\vec{x}(t)$ of individuals of
subpopulation $a$. Therefore, it is justified to denote $\vec{K}_a(\vec{x},t)$
as {\em social force} acting on individuals of subpopulation $a$. 
With that
we have attained a very intuitive formulation of social processes, 
according to which behavioral changes are caused by social forces.
\par
On the one hand, social forces influence the behavior of the individuals, 
but on the other hand, 
due to interactions, the behavior of the individuals
also influences the social forces via the behavioral distributions
$P_a(\vec{x},t)$ (cf. (\ref{Cont}b), (\ref{effdr})). 
That means, $K_a(\vec{x},t)$ is a
function of the social processes within the given population.
\par
Under the integrability conditions
\begin{equation}
 \frac{\partial}{\partial x_j} K_{ai}(\vec{x},t) =
 \frac{\partial}{\partial x_i} K_{aj}(\vec{x},t) \qquad
 \mbox{for all } i,j
 \label{intcond}
\end{equation}
there exists a time-dependent {\em potential}
\begin{equation}
 V_a(\vec{x},t) := - \int\limits^{\weg{x}} d\vec{x}' \cdot
 \vec{K}_a(\vec{x}',t) \equiv - \sum_{i=1}^n \int\limits^{\weg{x}}
 dx'_i \, K_{ai}(\vec{x}',t) \, ,
 \label{potential}
\end{equation}
so that the social force is given by its derivative (by its gradient $\nabla$):
\begin{equation}
 \vec{K}_a(\vec{x},t) = - \nabla V_a(\vec{x},t)\, , \qquad \mbox{i.e.,} \qquad
 K_{ai}(\vec{x},t) = - \frac{\partial}{\partial x_i} V_a(\vec{x},t) \, .
 \label{gradient}
\end{equation}
The potential $V_a(\vec{x},t)$ can be understood as {\em social field}.
It reflects the social influences and interactions relevant for 
behavioral changes: the public opinion, trends, social norms, etc.

\subsection{Discussion of the concept of force} \label{s3.2}

Clearly, the social force is not a force obeying the {\sc Newton}ian 
(1687) laws of
mechanics (cf. {\sc Greenwood}, 1988). 
Instead, the social force $\vec{K}_a(\vec{x},t)$ is a vectorial
quantity with the following properties:
\begin{itemize}
\item $\vec{K}_a(\vec{x},t)$ drives the temporal change $d\vec{x}/dt$ of
another vectorial quantity: the behavior $\vec{x}(t)$ of an individual
of subpopulation $a$.
\item The component
\begin{equation}
 \vec{K}_{ab}(\vec{x},t) := \nu_{ab}(t) \int\limits_\Omega d^n x' \, (\vec{x}'
 - \vec{x}) \Big[ f_{ab}^1(t) P_b(\vec{x}',t)
 + f_{ab}^2(t) P_b(\vec{x},t) \Big] R^a(\vec{x}'|\vec{x};t) 
\end{equation}
of the social force $\vec{K}_a(\vec{x},t)$
describes the reaction of subpopulation $a$ on the behavioral distribution
within subpopulation $b$ and usually differs from 
$\vec{K}_{ba}(\vec{x},t)$, which describes the influence of subpopulation $a$
on subpopulation $b$.
\item Neglecting fluctuations, the behavior $\vec{x}(t)$ does not change
if $\vec{K}_a(\vec{x},t)$ vanishes. 
$\vec{K}_a(\vec{x},t) = \mbox{\bf 0}$ corresponds
to an {\em extremum} of the social field $V_a(\vec{x},t)$, because it means
\begin{equation}
 \nabla V_a(\vec{x},t) = \mbox{\bf 0} \, , \qquad \mbox{i.e.,} \qquad
 \frac{\partial}{\partial x_i} V_a(\vec{x},t) = 0 \quad \mbox{for all }
 i \in \{ 1,\dots,n\} \, .
\end{equation}
\end{itemize}
We will now compare our results with
{\sc Lewin}'s (1951) {\em social field theory}. From social psychology
it is well-known that the behavior of an individual is determined
by the totality of {\em environmental influences} 
and his or her {\em personality}.
Inspired by electro-magnetic field theory, social field theory claims
that environmental influences can be considered as a {\em dynamical
force field}, which should be mathematically representable. A temporal
change of this field will evoke a {\em psychical tension} which, then,
induces a {\em (behavioral) compensation}.
\par
In the following it will turn out that our model allows 
a fully mathematical
specification of the field theoretical ideas, which was still to be found:
\begin{itemize}
\item Let us assume that an individual's objective is to behave in an
optimal way with respect to the social field $V_a(\vec{x},t)$, that
means, he or she tends to a behavior corresponding to a {\em minimum} of the
social field.
\item If the behavior $\vec{x}$ does not agree with a minimum of the social
field $V_a(\vec{x},t)$ this evokes a force
\begin{equation}
 \vec{K}_a(\vec{x},t) = - \nabla V_a(\vec{x},t) 
\end{equation}
that is given by the gradient of the social field $V_a(\vec{x},t)$
(pulling into the direction of steepest descent of $V_a(\vec{x},t)$).
The force $\vec{K}_a(\vec{x},t)$ plays the role of the {\em psychical
tension}.
It induces a behavioral change according to
\begin{equation}
 \frac{d\vec{x}}{dt} \approx \vec{K}_a(\vec{x},t) \, .
\label{accor}
\end{equation}
\item The behavioral change $d\vec{x}/dt$ drives the behavior $\vec{x}(t)$
towards a minimum $\vec{x}_a^*$ of the social field $V_a(\vec{x},t)$.
When the minimum $\vec{x}_a^*$ is reached, then
\begin{equation}
 \nabla V_a(\vec{x},t) = \mbox{\bf 0}
\end{equation}
holds and, therefore, $\vec{K}_a(\vec{x},t) = \mbox{\bf 0}$. As a consequence,
the psychical tension vanishes, that means, it is {\em compensated} by
the previous behavioral changes.
\par
When the psychical tension $\vec{K}_a(\vec{x},t)$ vanishes then, 
except for fluctuations,
no behavioral changes take place---in accordance with (\ref{accor}). 
The individual has reached an {\em equilibrium} within the social field, then.
\item Note, that the social fields $V_a(\vec{x},t)$ of 
different subpopulations $a$ usually have different minima $\vec{x}_a^*$.
This means that individuals of different types $a$ of behavior
will normally feel different psychical tensions
$\vec{K}_a(\vec{x},t)$. In other words, index $a$ distinguishes different 
{\em personalities}.
\end{itemize}

\section{Computer simulations} \label{s4}

The {\sc Boltzmann-Fokker-Planck} equations are able to describe a broad
spectrum of social phenomena. In the following, some of the results shall be
illustrated by computer
simulations. We shall examine the case of $A=2$ subpopulations,
and a situation for which the interesting aspect of the individual behavior 
can be described
by a certain {\em position} $x$ on a one-dimensional continuous
scale (i.e., $n=1$, $\vec{x} \equiv x$). 
A concrete example for this situation would be the case of {\em
opinion formation}. Here, conservative and progressive
thinking individuals could be distinguished by different
subpopulations. The position $x$ would describe
the grade of approval or disapproval with respect to a certain
political option (for example, SDI, power stations, Golf war, etc.).
\par
\begin{figure}[htbp]
\parbox[b]{7.4cm}{
\epsfxsize=7.3cm 
\centerline{\rotate[r]{\hbox{\epsffile[28 28 570
556]{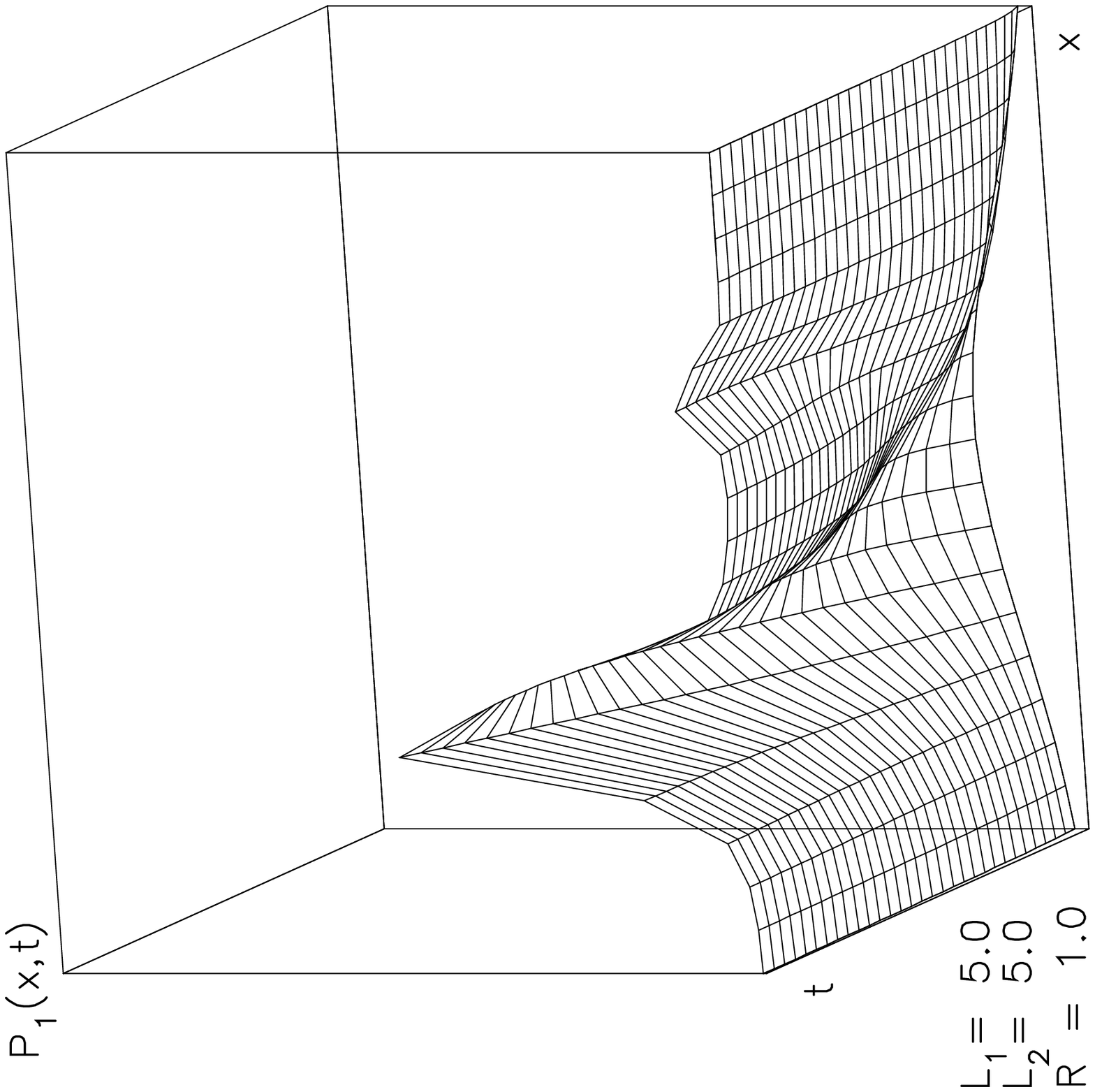}}}}
}\hfill
\parbox[b]{7.4cm}{
\epsfxsize=7.3cm 
\centerline{\rotate[r]{\hbox{\epsffile[28 28 570
556]{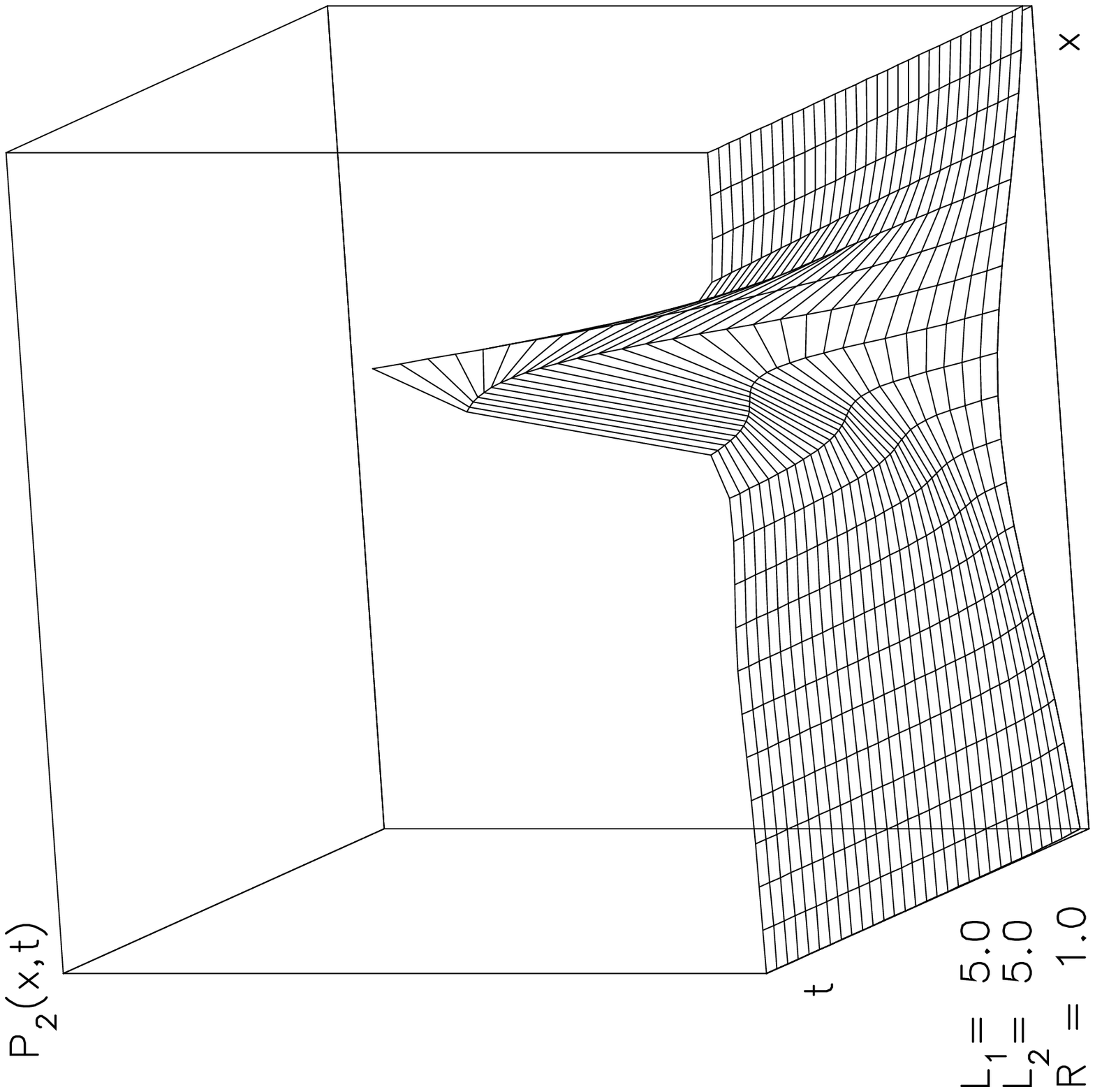}}}}
}  
\parbox[b]{7.4cm}{
\epsfxsize=7.3cm 
\centerline{\rotate[r]{\hbox{\epsffile[28 28 570
556]{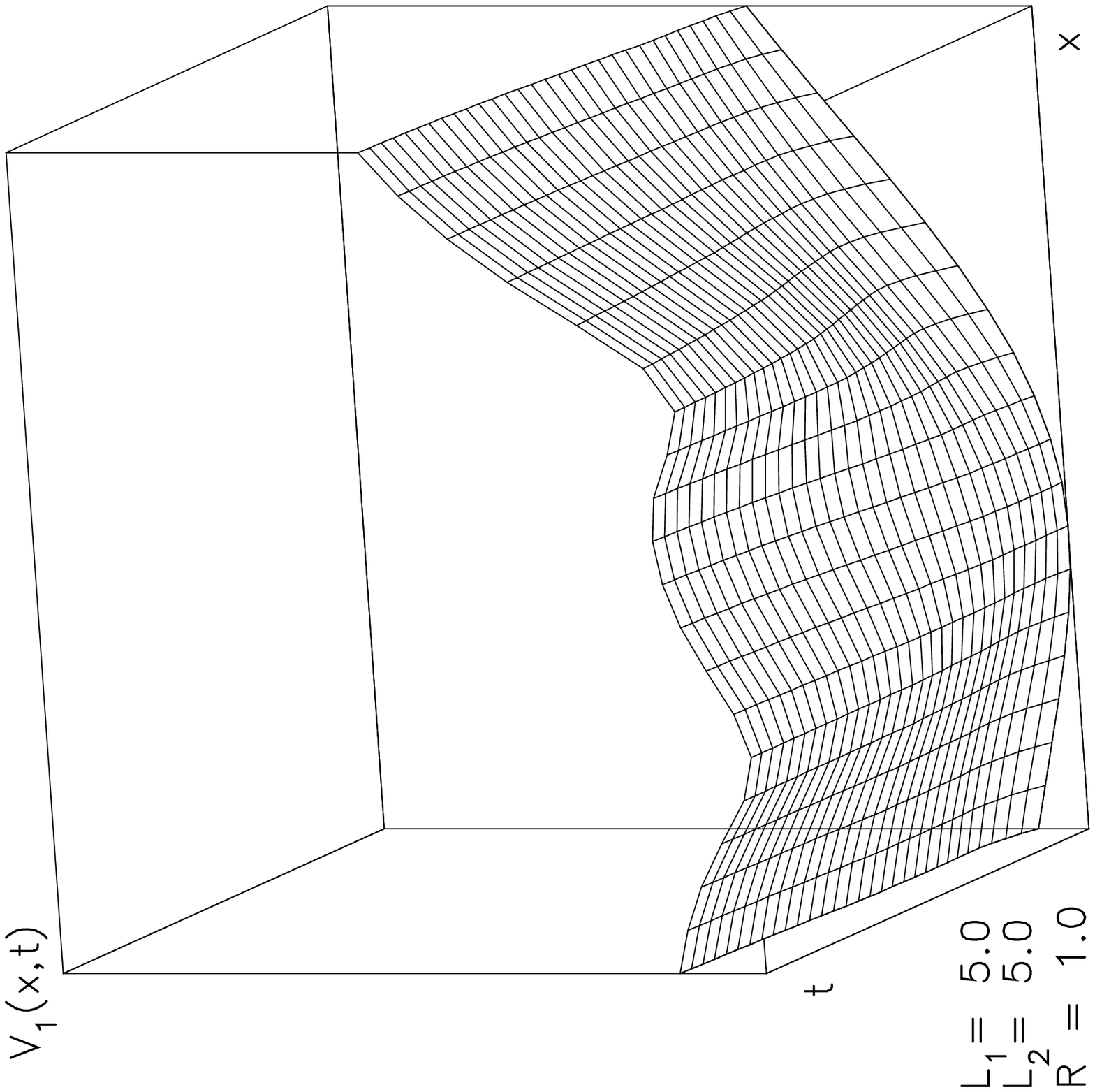}}}}
}\hfill
\parbox[b]{7.4cm}{
\epsfxsize=7.3cm 
\centerline{\rotate[r]{\hbox{\epsffile[28 28 570
556]{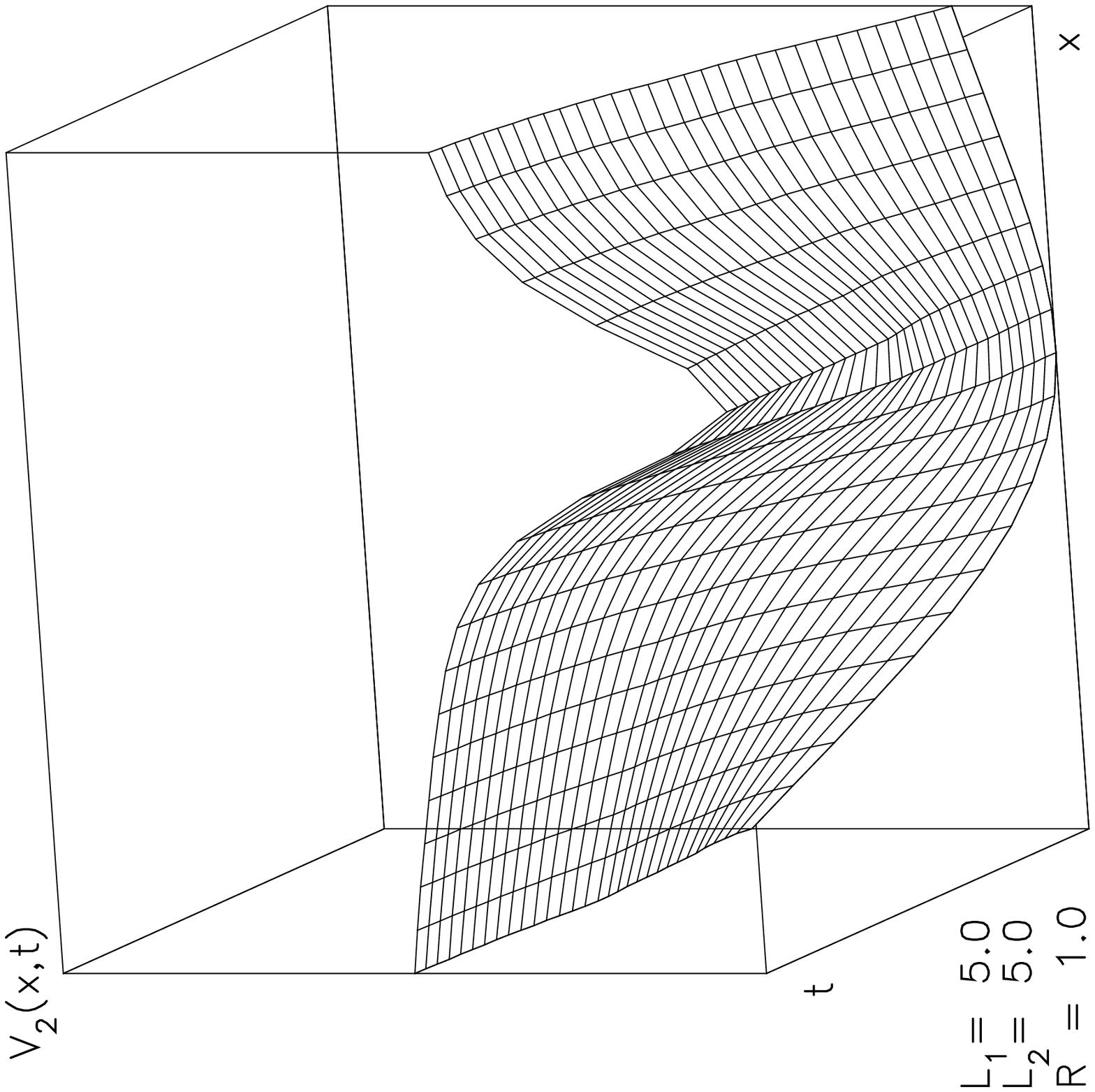}}}}
}
\parbox[b]{7.4cm}{
\epsfxsize=7.3cm 
\centerline{\rotate[r]{\hbox{\epsffile[28 28 570
556]{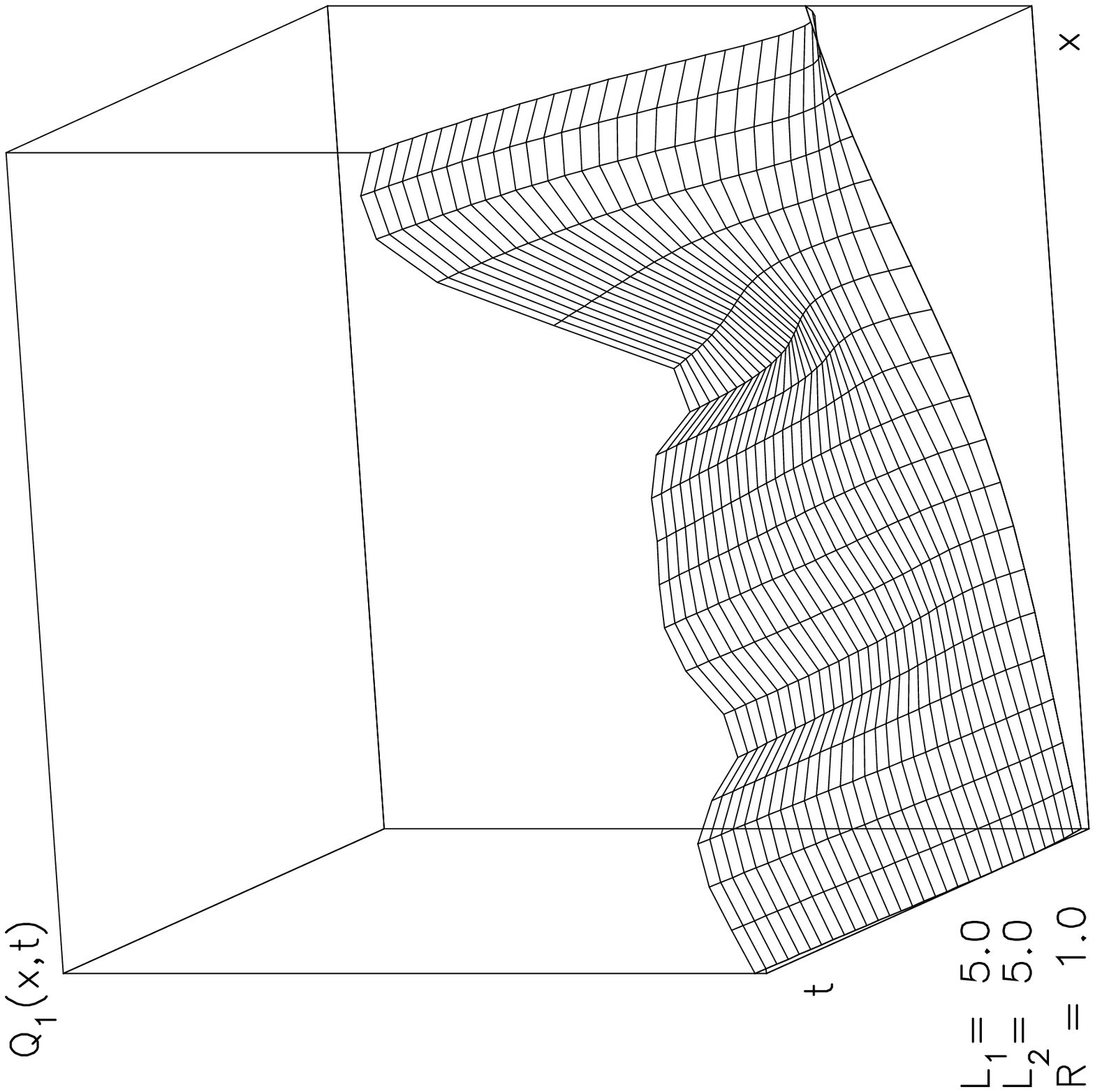}}}}
}\hfill
\parbox[b]{7.4cm}{
\epsfxsize=7.3cm 
\centerline{\rotate[r]{\hbox{\epsffile[28 28 570
556]{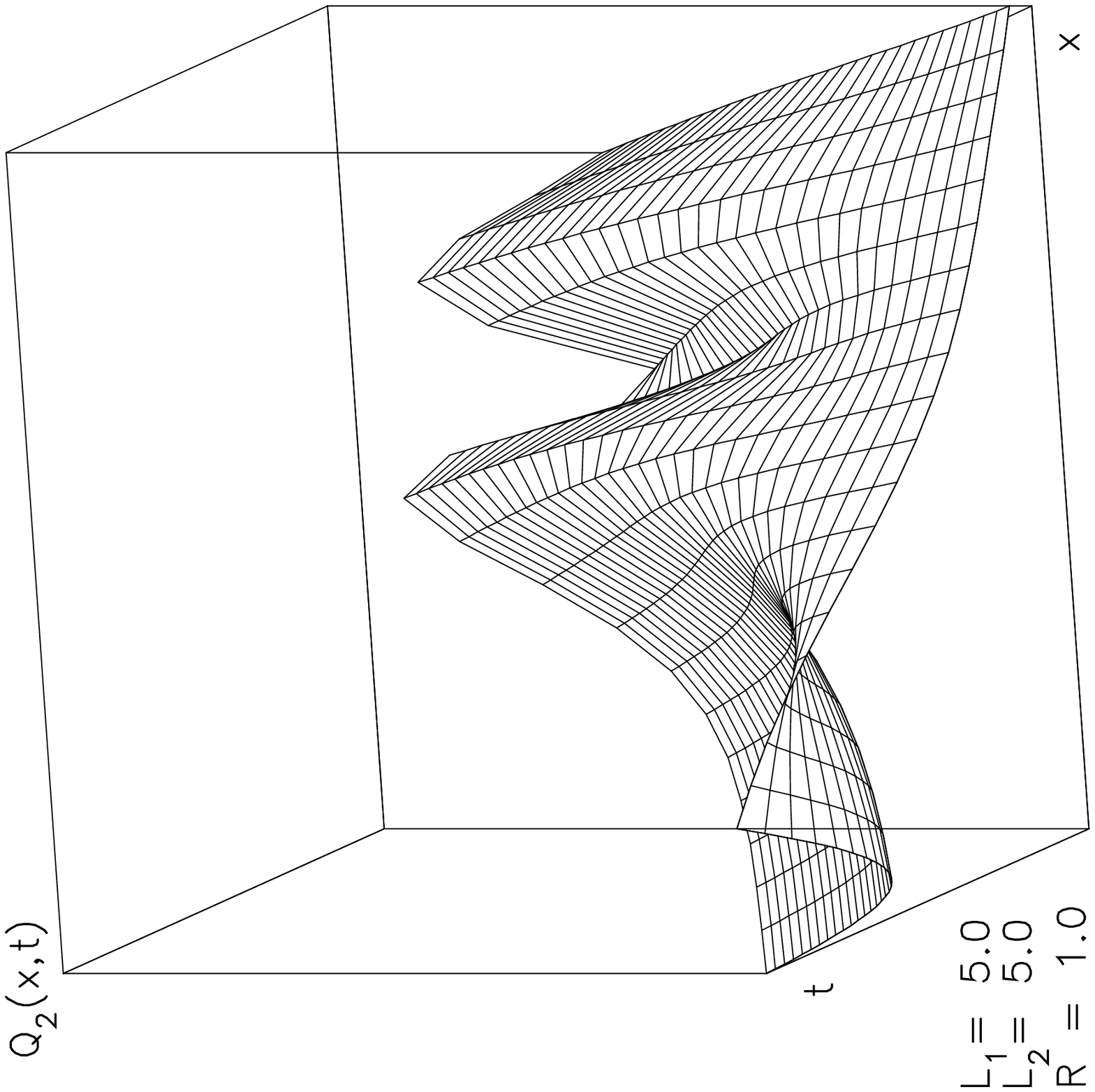}}}}
}
\parbox{15cm}{
\caption{Imitative processes in the case of one-sided sympathy and
low indifference $L_a$ with respect to behavioral changes.\label{fi1}}
}
\end{figure}
\begin{figure}[htbp]
\parbox[b]{7.4cm}{
\epsfxsize=7.3cm 
\centerline{\rotate[r]{\hbox{\epsffile[28 28 570
556]{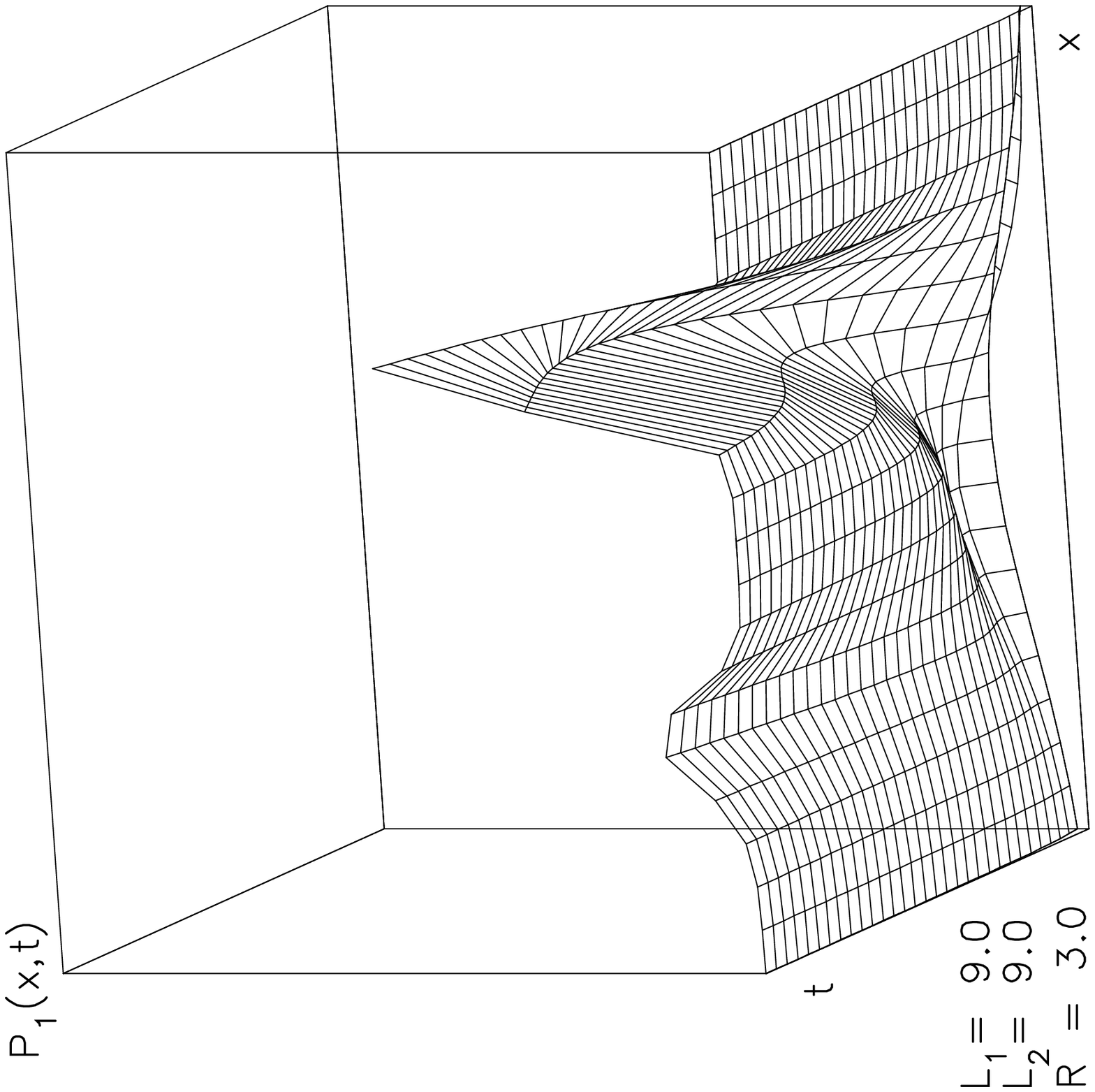}}}}
}\hfill
\parbox[b]{7.4cm}{
\epsfxsize=7.3cm 
\centerline{\rotate[r]{\hbox{\epsffile[28 28 570
556]{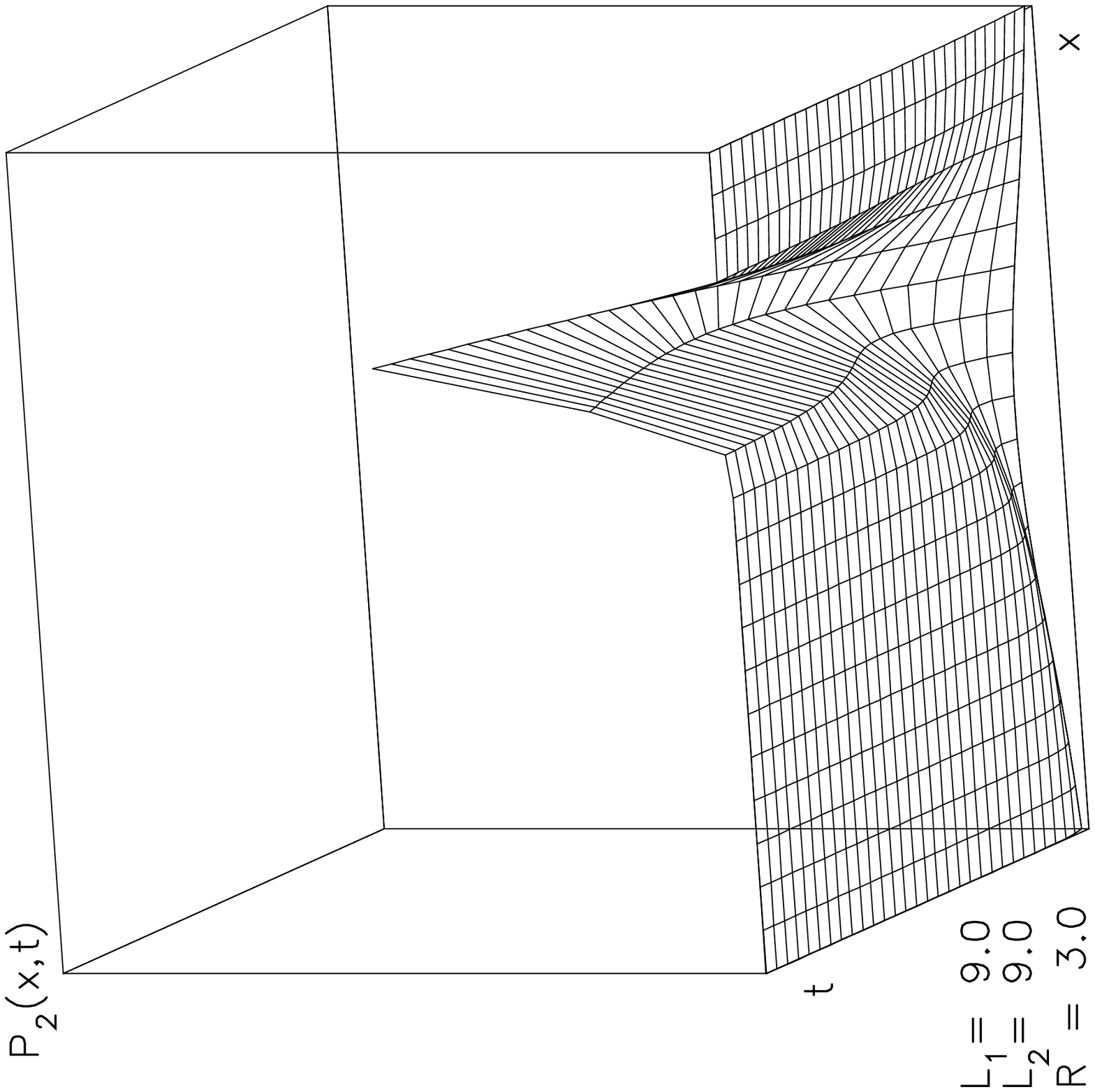}}}}
}
\parbox[b]{7.4cm}{
\epsfxsize=7.3cm 
\centerline{\rotate[r]{\hbox{\epsffile[28 28 570
556]{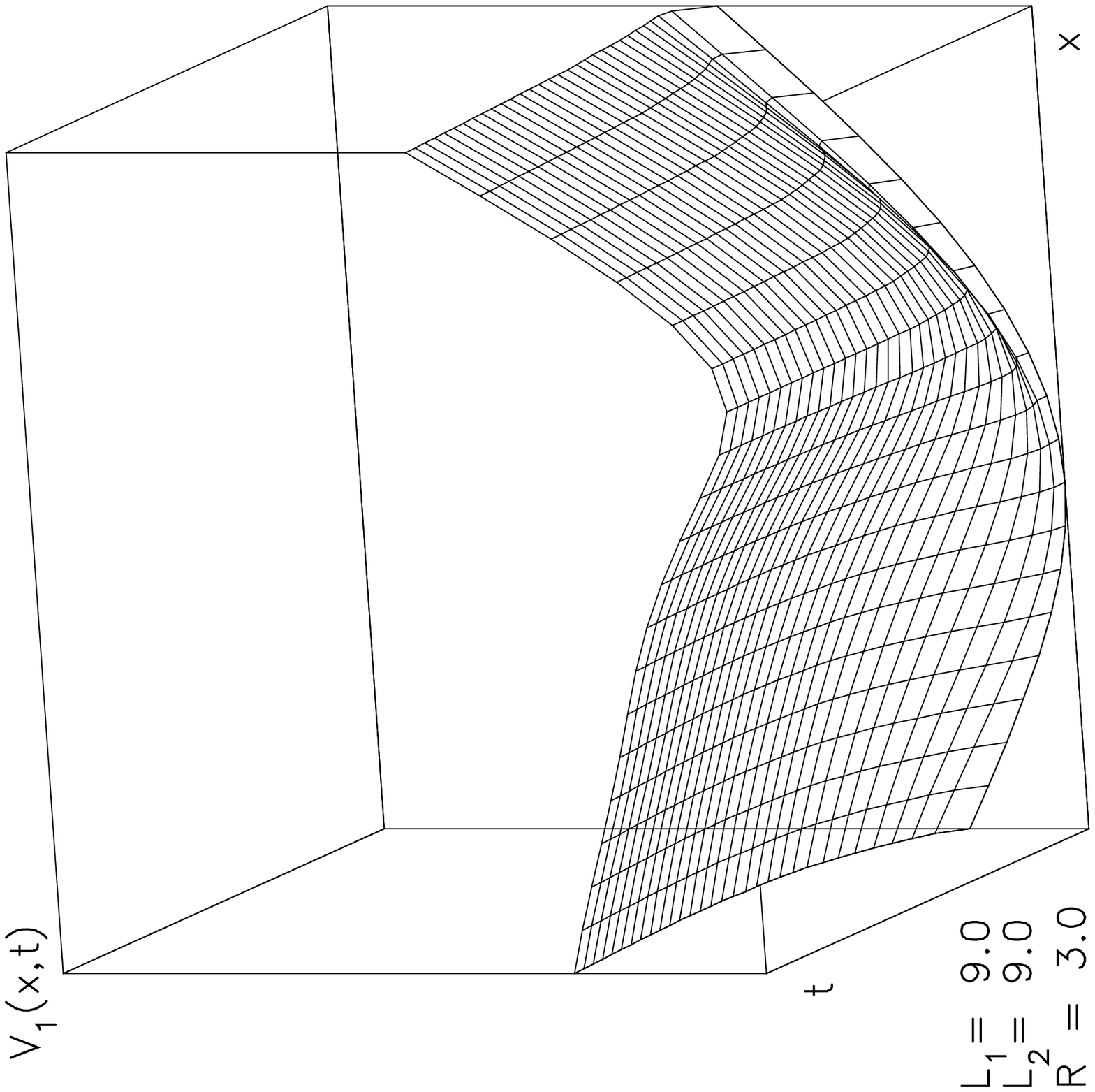}}}}
}\hfill
\parbox[b]{7.4cm}{
\epsfxsize=7.3cm 
\centerline{\rotate[r]{\hbox{\epsffile[28 28 570
556]{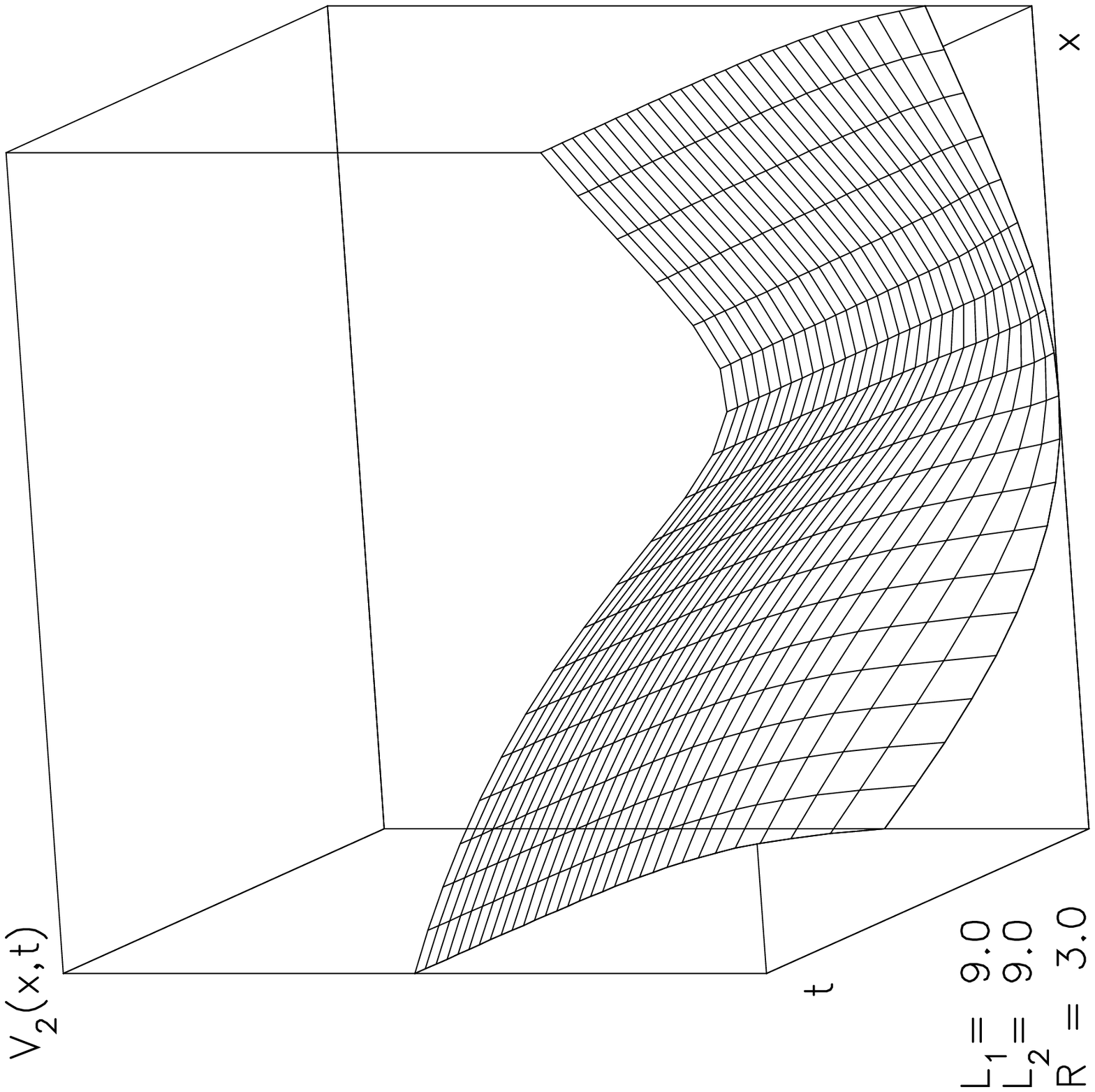}}}}
}
\parbox[b]{7.4cm}{
\epsfxsize=7.3cm 
\centerline{\rotate[r]{\hbox{\epsffile[28 28 570
556]{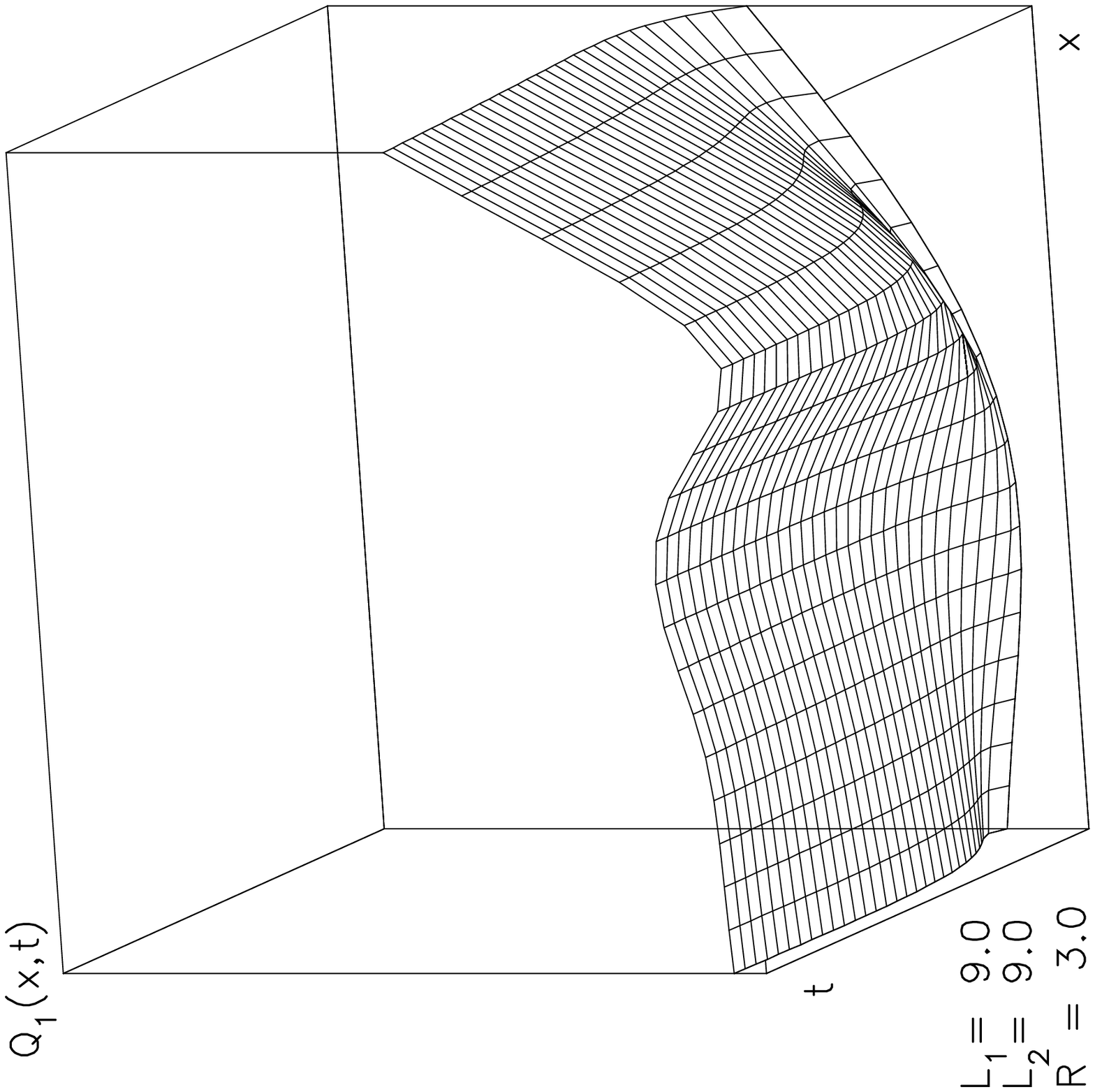}}}}
}\hfill
\parbox[b]{7.4cm}{
\epsfxsize=7.3cm 
\centerline{\rotate[r]{\hbox{\epsffile[28 28 570
556]{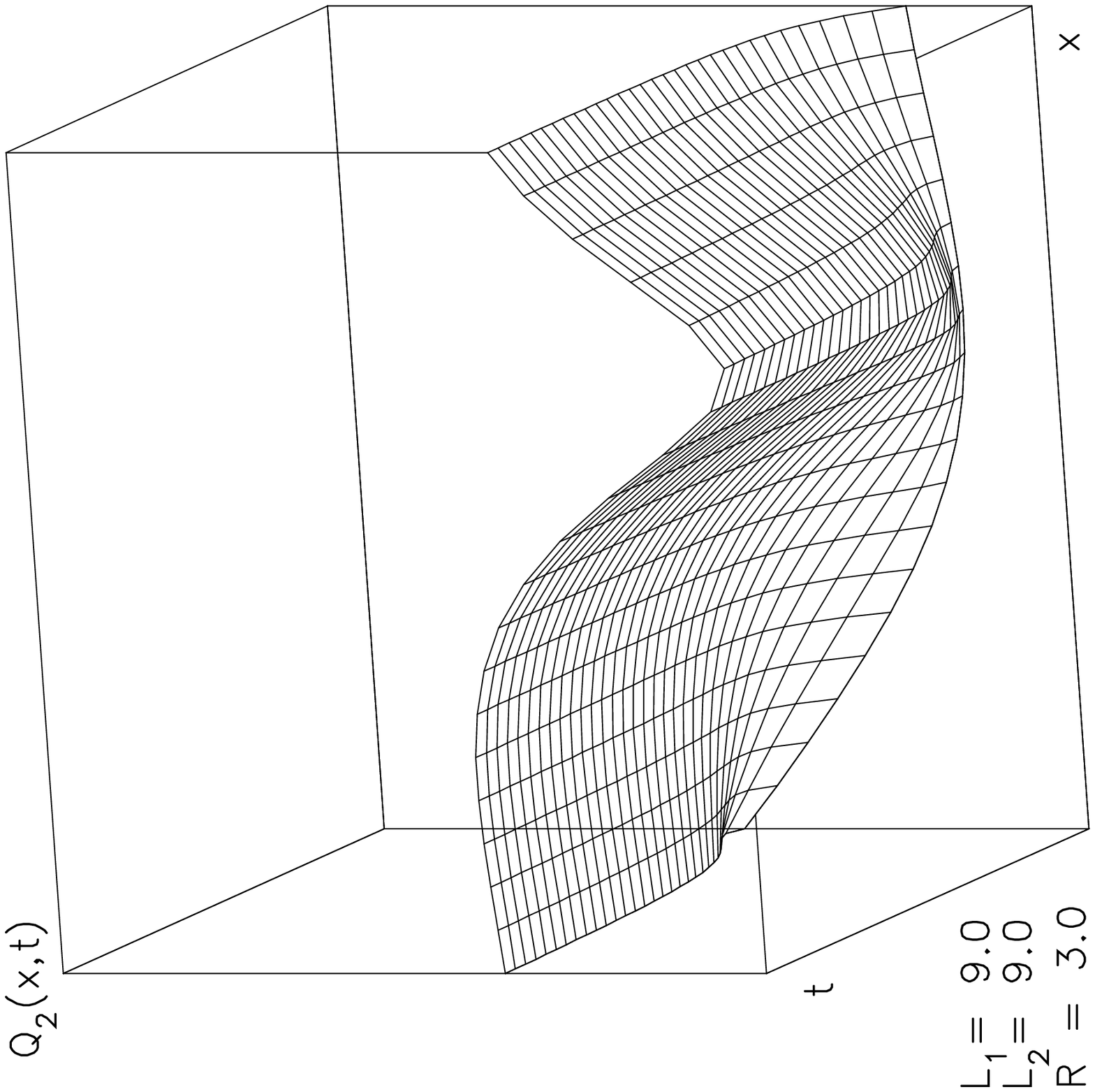}}}}
}
\parbox{15cm}{
\caption{As figure 1, but for high indifference $L_a$ with respect
to behavioral changes.\label{fi2}}
}
\end{figure}
\begin{figure}[htbp]
\parbox[b]{7.4cm}{
\epsfxsize=7.3cm 
\centerline{\rotate[r]{\hbox{\epsffile[28 28 570
556]{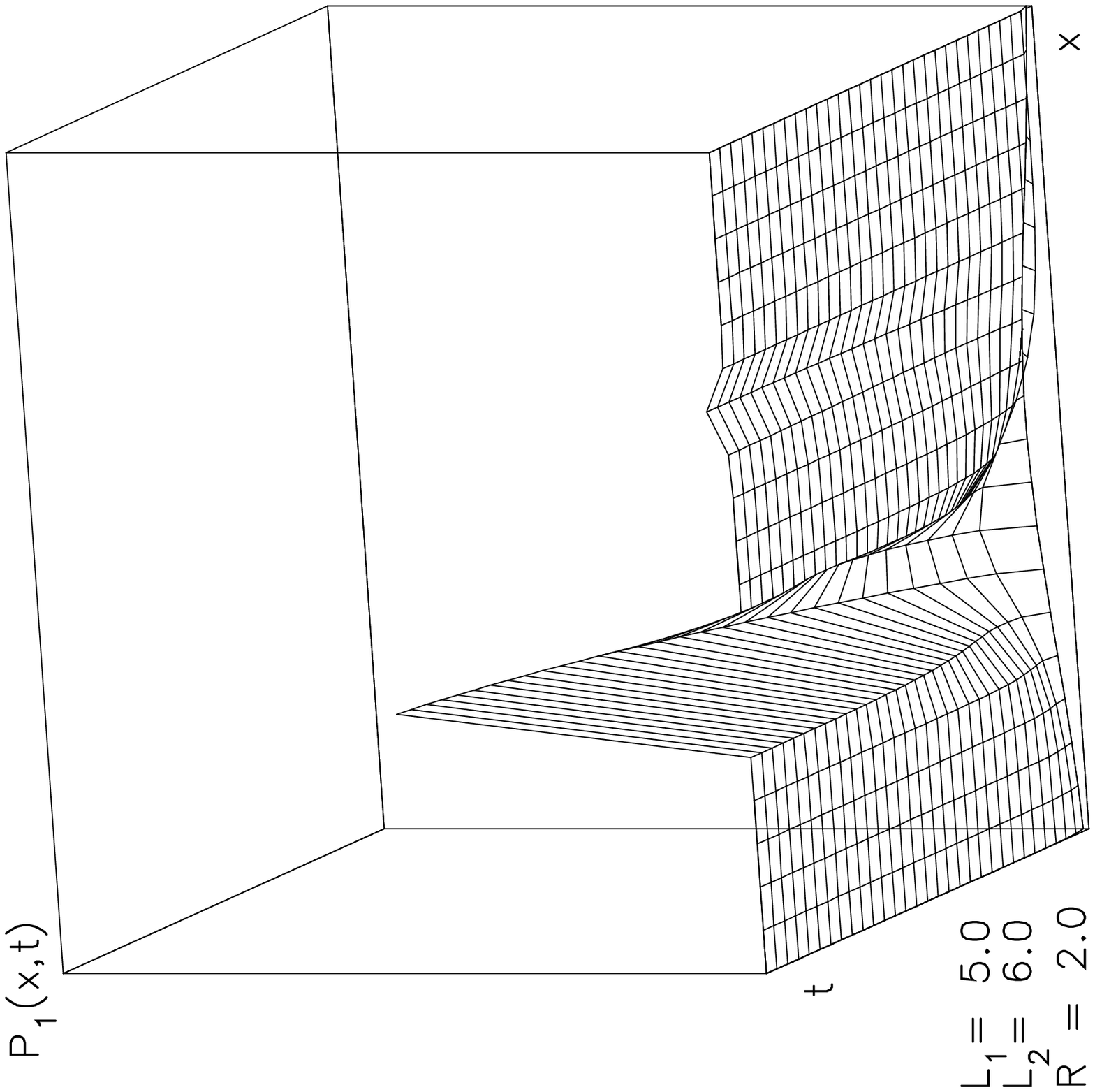}}}}
}\hfill
\parbox[b]{7.4cm}{
\epsfxsize=7.3cm 
\centerline{\rotate[r]{\hbox{\epsffile[28 28 570
556]{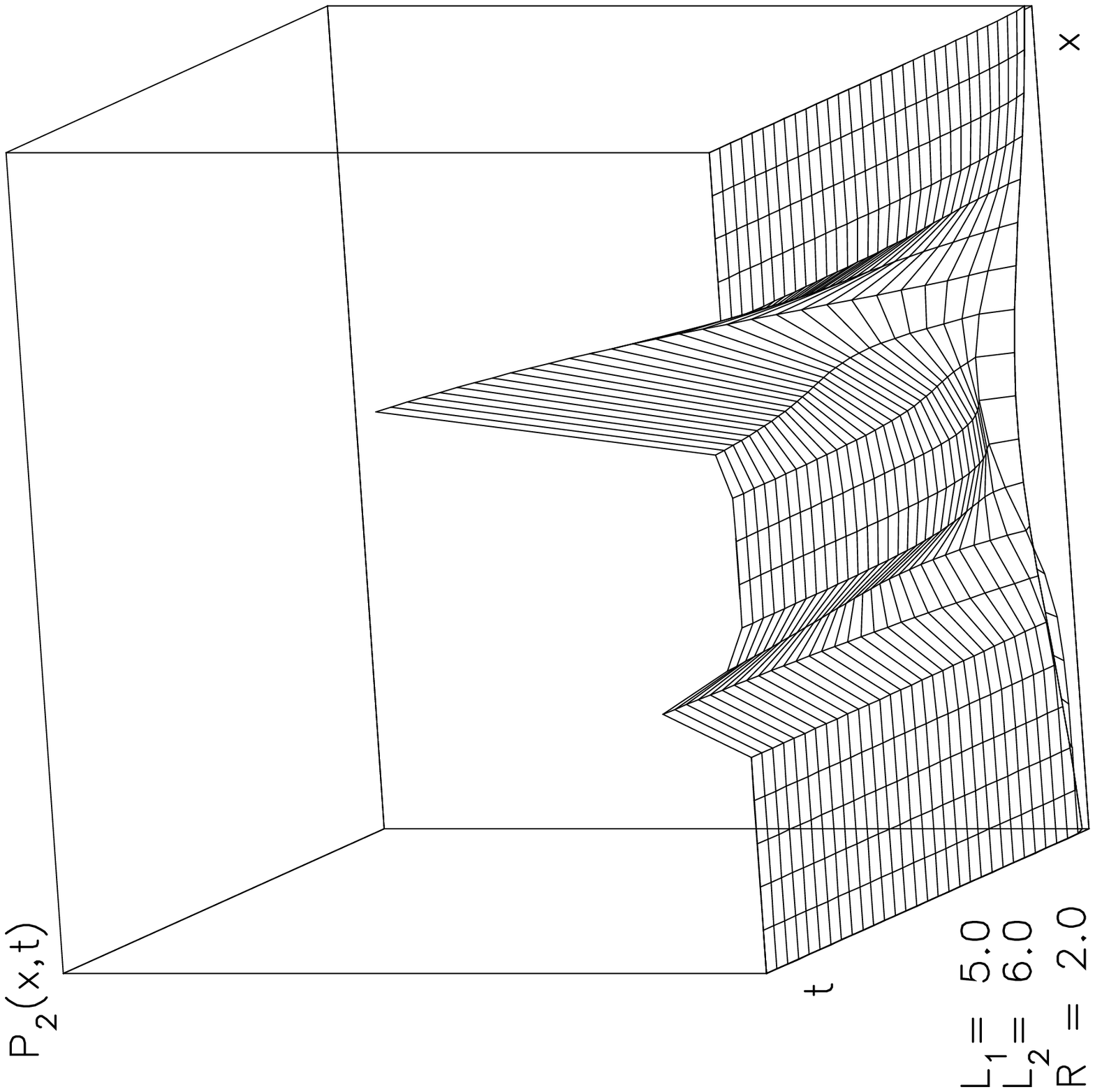}}}}
}
\parbox[b]{7.4cm}{
\epsfxsize=7.3cm 
\centerline{\rotate[r]{\hbox{\epsffile[28 28 570
556]{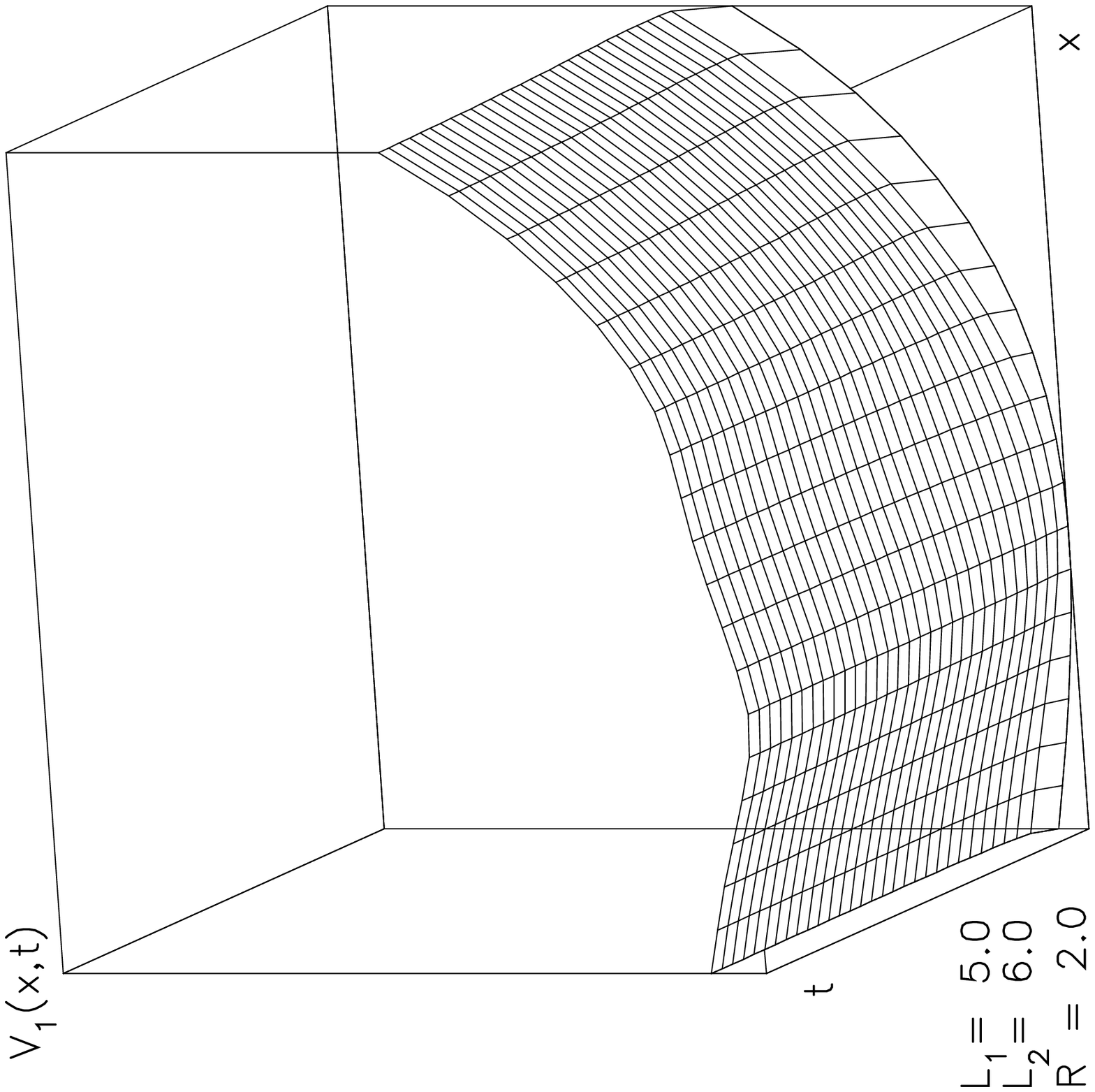}}}}
}\hfill
\parbox[b]{7.4cm}{
\epsfxsize=7.3cm 
\centerline{\rotate[r]{\hbox{\epsffile[28 28 570
556]{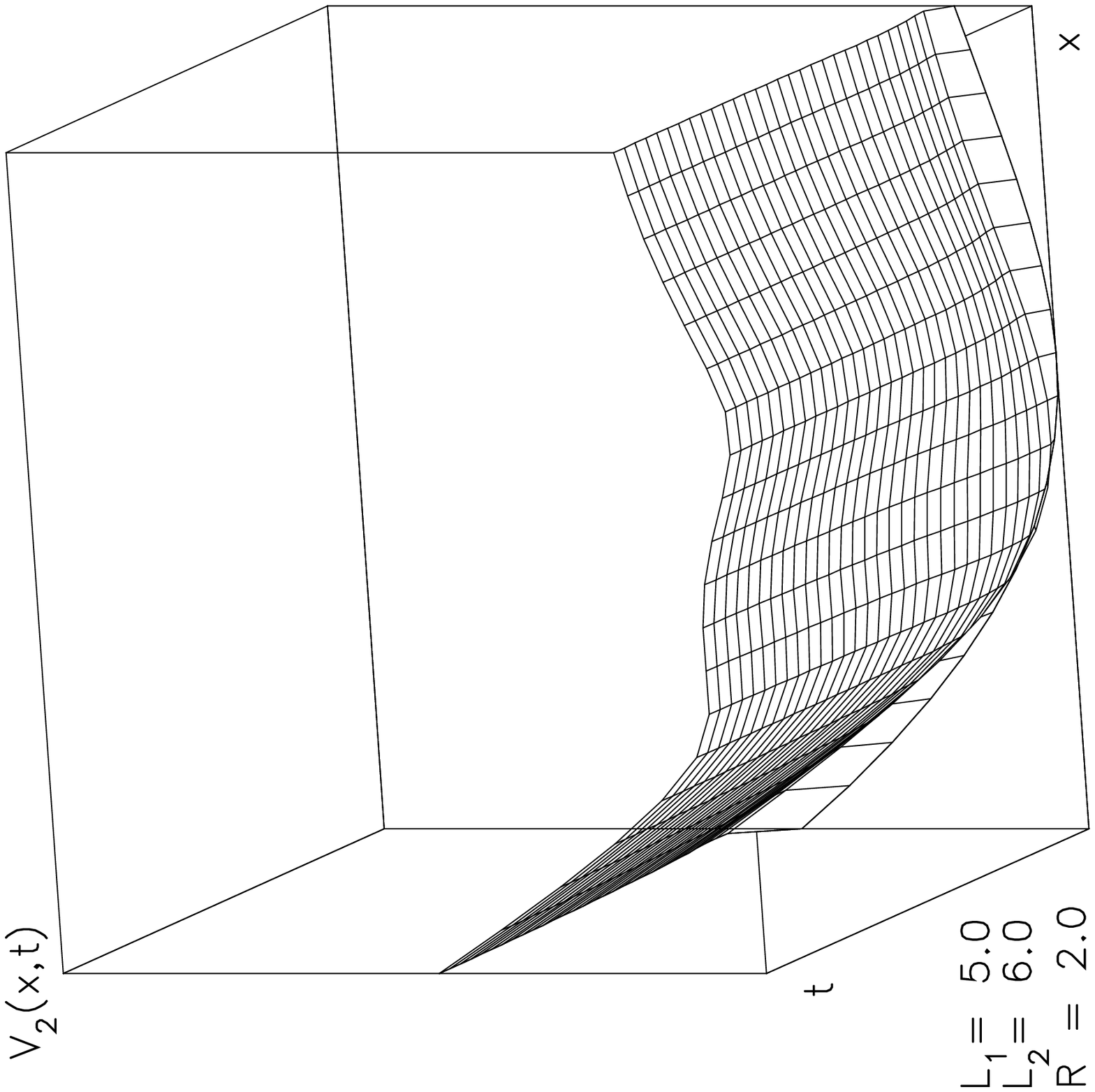}}}}
}
\parbox[b]{7.4cm}{
\epsfxsize=7.3cm 
\centerline{\rotate[r]{\hbox{\epsffile[28 28 570
556]{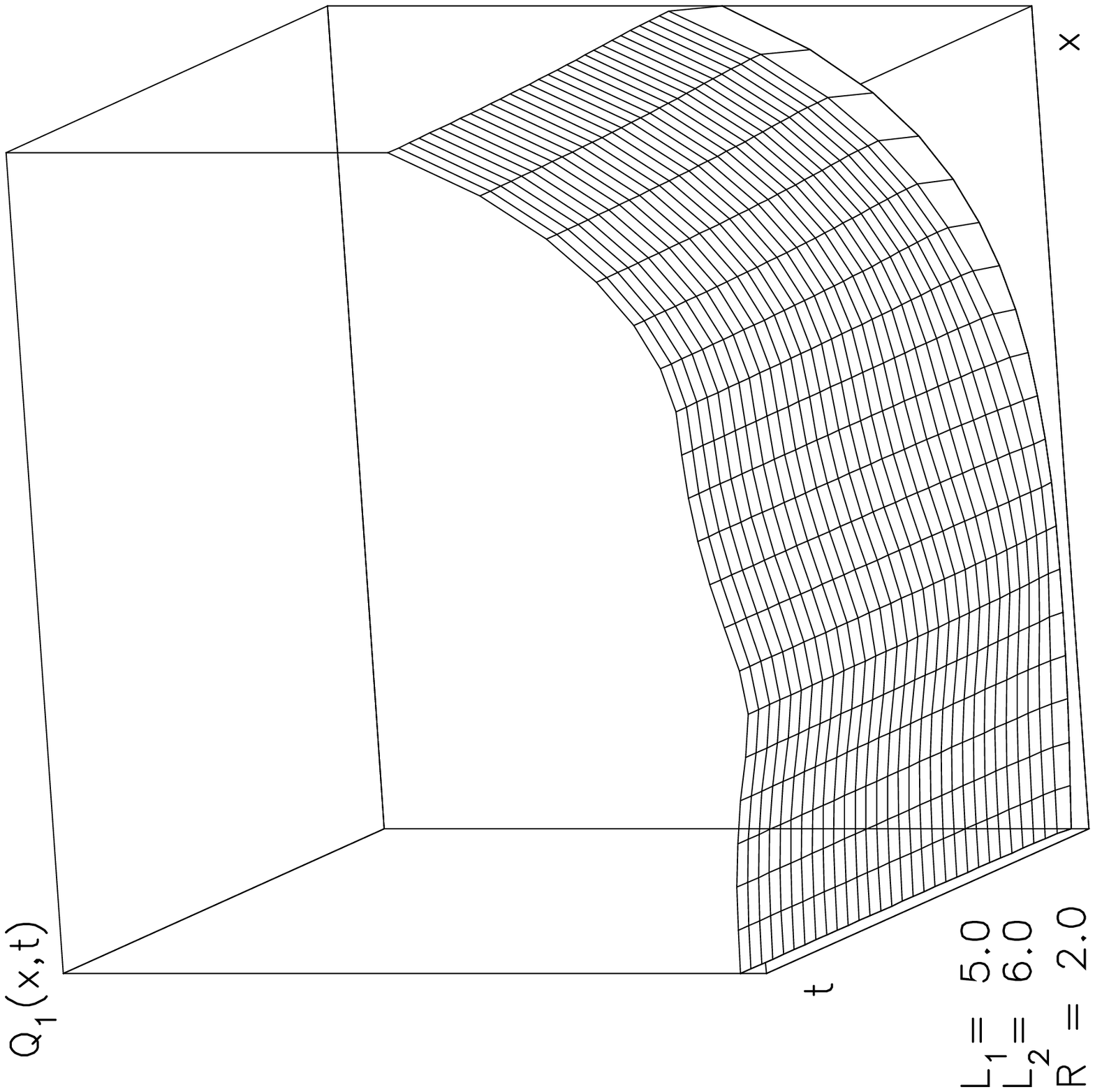}}}}
}\hfill
\parbox[b]{7.4cm}{
\epsfxsize=7.3cm 
\centerline{\rotate[r]{\hbox{\epsffile[28 28 570
556]{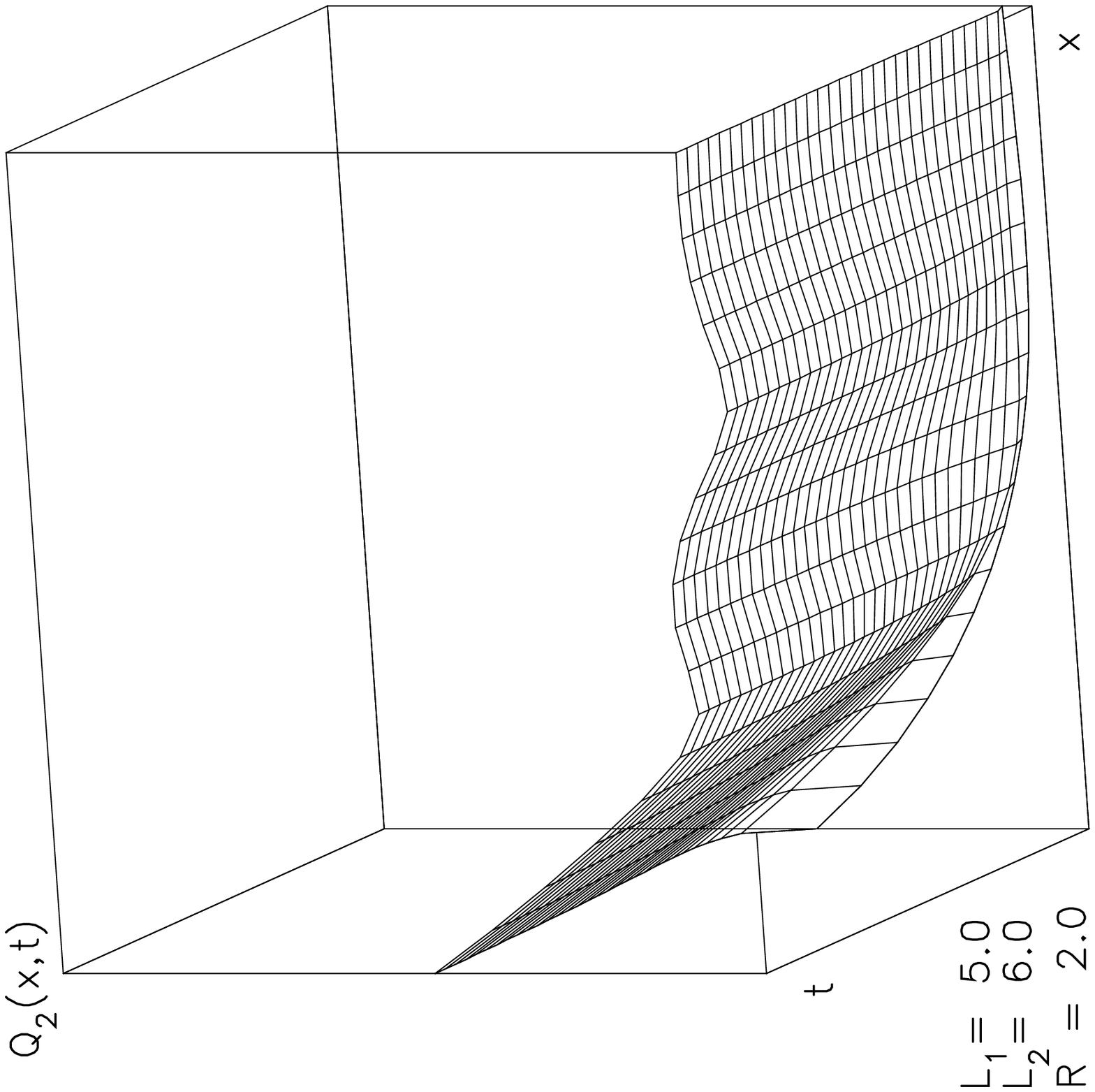}}}}
}
\parbox{15cm}{
\caption{Imitative processes for mutual sympathy and low indifference $L_a$
in both subpopulations.\label{fi3}}
}
\end{figure}
\begin{figure}[htbp]
\parbox[b]{7.4cm}{
\epsfxsize=7.3cm 
\centerline{\rotate[r]{\hbox{\epsffile[28 28 570
556]{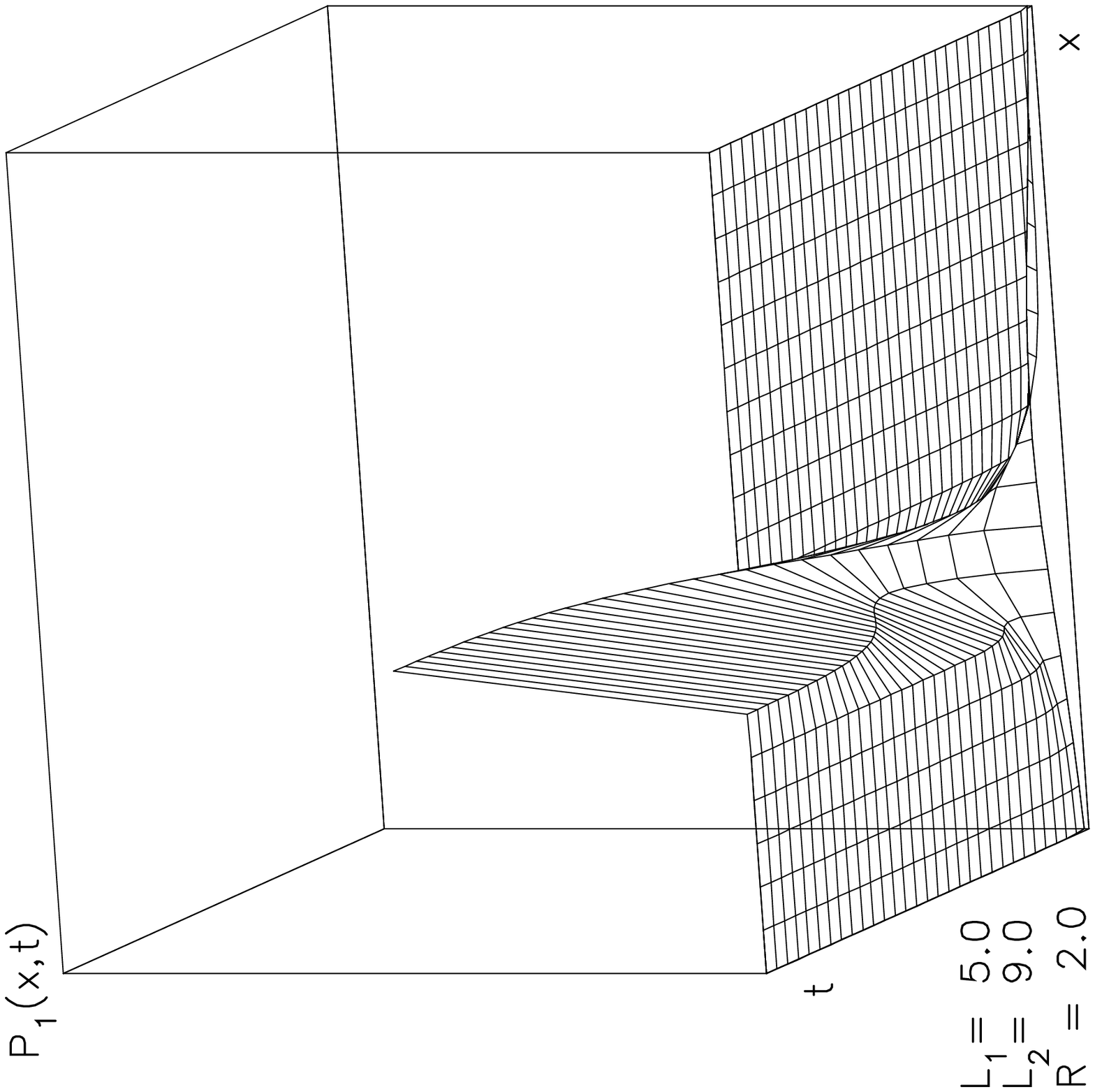}}}}
}\hfill
\parbox[b]{7.4cm}{
\epsfxsize=7.3cm 
\centerline{\rotate[r]{\hbox{\epsffile[28 28 570
556]{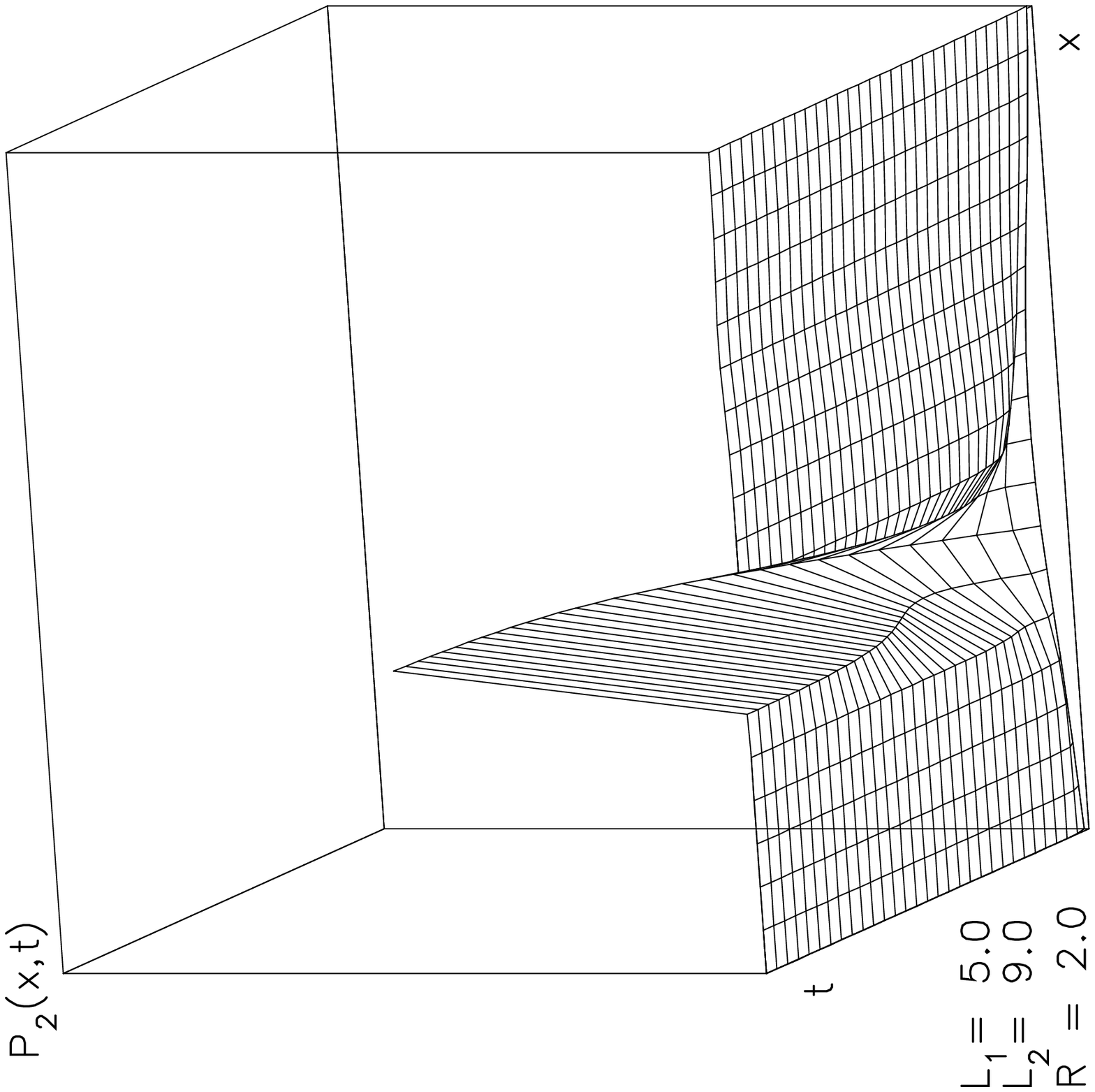}}}}
}
\parbox[b]{7.4cm}{
\epsfxsize=7.3cm 
\centerline{\rotate[r]{\hbox{\epsffile[28 28 570
556]{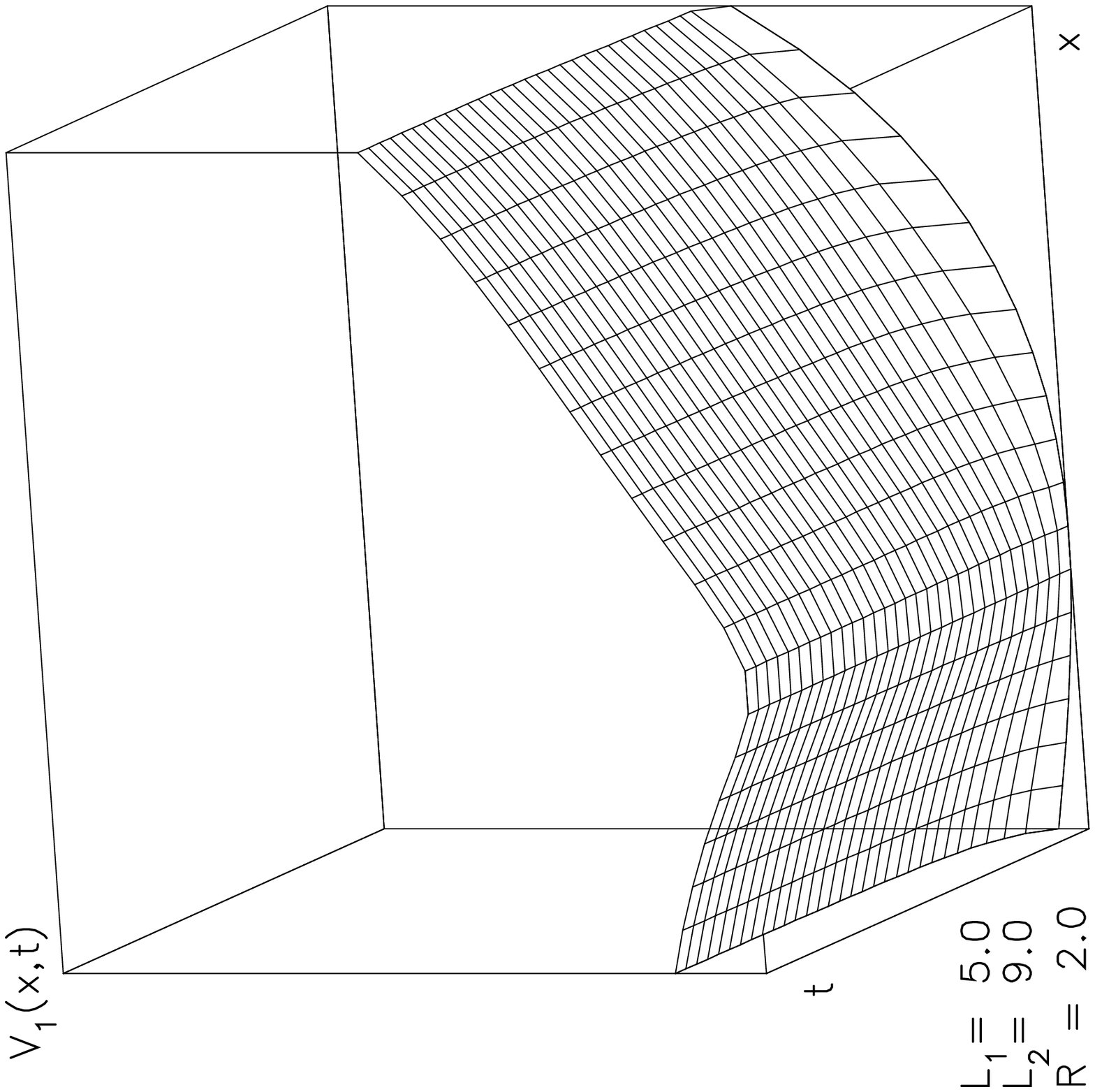}}}}
}\hfill
\parbox[b]{7.4cm}{
\epsfxsize=7.3cm 
\centerline{\rotate[r]{\hbox{\epsffile[28 28 570
556]{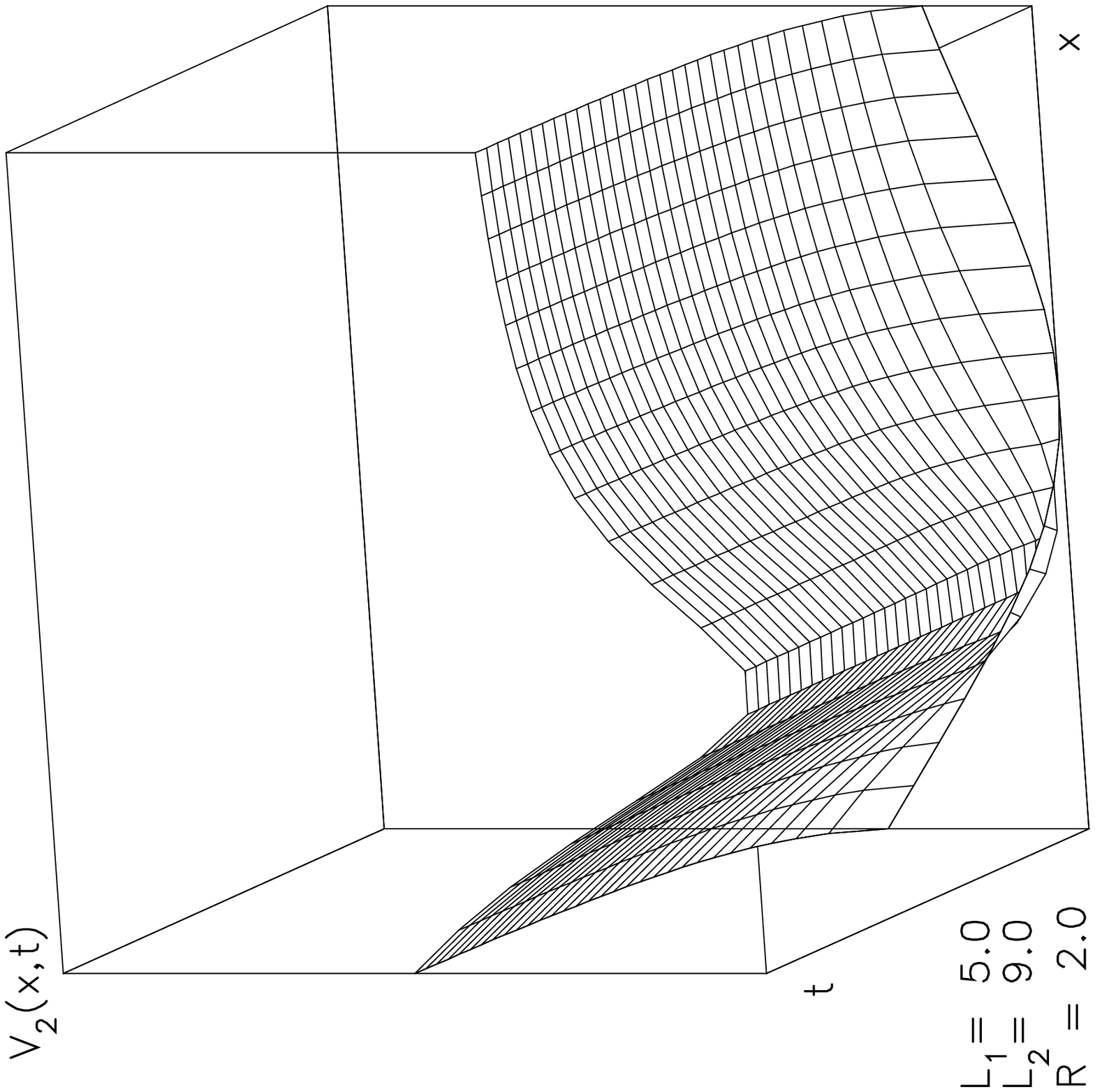}}}}
}
\parbox[b]{7.4cm}{
\epsfxsize=7.3cm 
\centerline{\rotate[r]{\hbox{\epsffile[28 28 570
556]{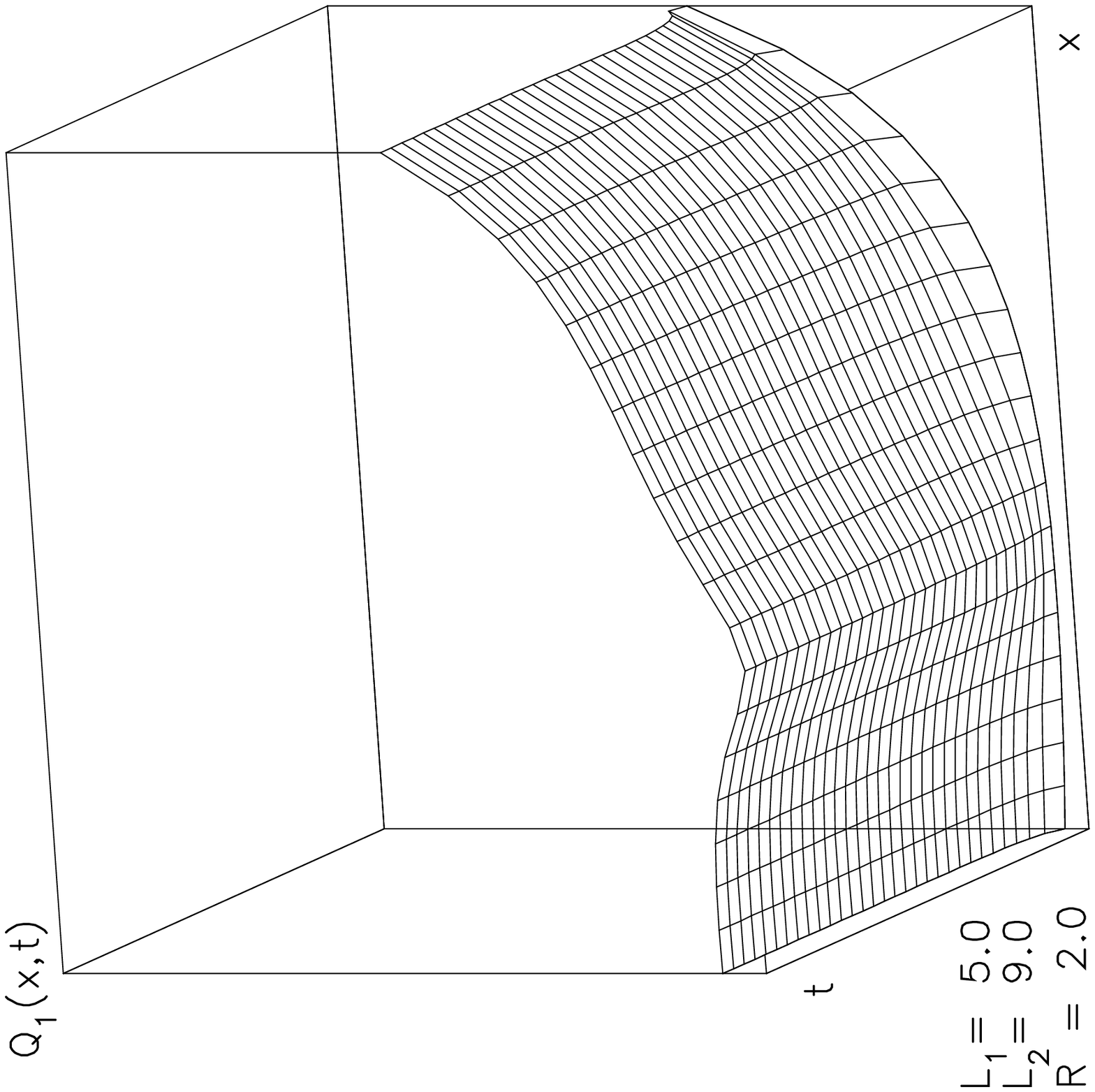}}}}
}\hfill
\parbox[b]{7.4cm}{
\epsfxsize=7.3cm 
\centerline{\rotate[r]{\hbox{\epsffile[28 28 570
556]{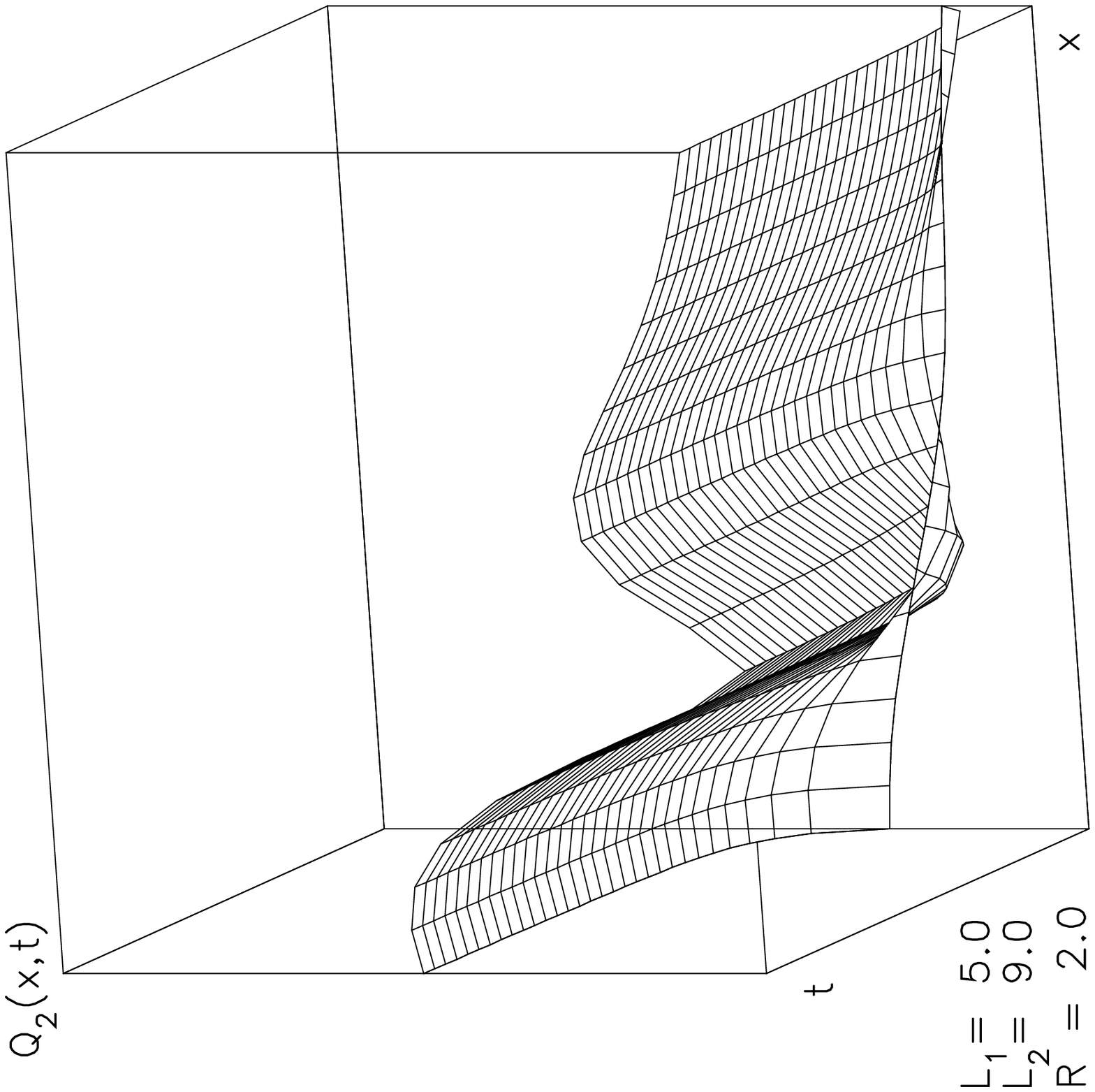}}}}
}
\parbox{15cm}{
\caption{As figure 3, but for high indifference $L_2$ in
subpopulation 2.\label{fi4}}
}
\end{figure}

In the one-dimensional case,
the integrability conditions
(\ref{intcond}) are automatically fulfilled, and the social field
\begin{equation}
 V_a(x,t) = - \int\limits_{x_0}^x dx' \, K_a(x',t) - c_a(t)
\end{equation}
is well-defined. The parameter $c_a(t)$ can be chosen arbitrarily. 
We will take for $c_a(t)$ the value that shifts the absolute
minimum of $V_a(x,t)$ to zero, that means,
\begin{equation}
 c_a(t) := \min_x \left( - \int\limits_{x_0}^x dx' \, K_a(x',t) \right) \, .
\end{equation}
\begin{itemize}
\item Since we will restrict the simulations to the case of
imitative or avoidance processes, the shape of the social field
$V_a(x,t)$ changes with time only due to changes of the probability
distributions $P_a(x,t)$ (cf. (\ref{SUM})), 
that means, due to behavioral changes
of the individuals (see figures \ref{fi1} to \ref{fi6}).
\end{itemize}
In the one-dimensional case one can find the formal {\em stationary solution}
\begin{equation}
 P_a(x) = P_a(x_0) \frac{Q_{a}(x_0)}{Q_{a}(x)}
\exp\left( 2 \int\limits_{x_0}^x dx' \, \frac{K_{a}(x')}{Q_{a}(x')}
\right) \, ,
\label{ABh}
\end{equation}
which we expect to be approached in the limit of large times
$t \rightarrow \infty$. Due to the dependence of $K_a(x)$ and $Q_a(x)$
on $P_a(x)$, equations (\ref{ABh}) are only {\em implicit} equations.
However, from (\ref{ABh}) we can derive the following conclusions:
\begin{itemize}
\item If the diffusion coefficients are constant ($Q_a(x) \equiv Q_a$),
(\ref{ABh}) simplifies to
\begin{equation}
 P_a(x) = P_a(x_0) \exp \left( - \frac{2}{Q_a} \Big[ V_a(x) + c_a \Big]
 \right) \, ,
\end{equation}
that means, the stationary solution $P_a(x)$ is completely determined
by the social field $V_a(x)$. Especially, $P_a(x)$ has its maxima at the
positions $x_a^*$, where the social field $V_a(x)$ has its minima
(see fig. \ref{fi1}).
The diffusion constant $Q_a$ regulates the width of the behavioral
distribution $P_a(x)$. For $Q_a = 0$ there were no individual behavioral
variations, and the behavioral distribution $P_a(x)$ were
sharply peaked at the deepest minimum $x_a^*$ of $V_a(x)$.
\item If the diffusion coefficients $Q_a(x)$ are varying functions of the
position $x$, the behavioral distribution $P_a(x)$
is also influenced by the
concrete form of $Q_a(x)$. From (\ref{ABh}) one expects
high behavioral probabilities $P_a(x)$ where the diffusion coefficients
$Q_a(x)$ are small (see fig. \ref{fi2}, where the probability distribution
$P_1(x)$ cannot be explained solely by the social field $V_1(x)$).
\item Since the stationary solution $P_a(x)$ depends on both, $K_a(x)$ and
$Q_a(x)$, different combinations of $K_a(x)$ and $Q_a(x)$ can lead to the
same probability distribution $P_a(x)$ (see fig. \ref{fi4}
in the limit of large times).
\end{itemize}
For the following simulations, we shall assume $x \in [1/20,1]$
and use the {\em ansatz}\alpheqn{Ans} 
\begin{equation}
 R^a(x'|x;t) = \frac{\mbox{e}^{U^a(x',t) - U^a(x,t)}}
 {D_a(x',x;t)} 
\end{equation}
for the readiness $R^a(x'|x;t)$ to change from $x$ to $x'$
(cf. (\ref{util})).
With the utility function
\begin{equation}
 U^a(x,t) := - \frac{1}{2} \left( \frac{x - x_a}{l_a} \right)^2
 \, , \qquad l_a := \frac{L_a}{20}
\end{equation}
subpopulation $a$ prefers behavior $x_a$. $L_a$ means the
{\em indifference} of subpopulation $a$
with respect to variations of the position $x$.
Moreover, we take
\begin{equation}
 \frac{\nu_{ab}(t)}{D_a(x',x;t)} := \mbox{e}^{-|x' - x|/r} \, ,
 \qquad r = \frac{R}{20} \, ,
 \label{Dist}
\end{equation}\reseteqn
where $R$ can be interpreted as measure for the {\em range of interaction}.
According to (\ref{Dist}), the rate of behavioral changes is the smaller
the greater they are. Only small changes of the position (i.e., between
neighboring positions) contribute with an appreciable rate. 
Figure \ref{fi1} to \ref{fi6} show the respective values of 
$R$, $L_1$, and $L_2$ used in the simulations. 
\par
Note, that $R$ and the readiness $R^a$ are different quantities. For
very small values of the range $R$ of interaction the diffusion coefficients
$Q_a(x)$ can be neglected and the fluctuations play a neglible role,
that means, behavioral changes are mainly given by the
social field (see fig. 1). 
For greater but still small values of $R$, the diffusion
coefficients have to be taken into account in order to fully understand
the temporal development of the behavioral distribution $P_a(x,t)$
(see fig. 2). If $R$ exceeds a certain value, the {\sc Taylor}
approximation is invalid, and the {\sc Boltzmann}-like
equations should be applied instead of the {\sc Boltzmann-Fokker-Planck}
equations.

\subsection{Sympathy and interaction frequency} \label{s4.1}

Let $s_{ab}(t)$ be the degree of {\em sympathy} 
which individuals of subpopulation
$a$ feel towards individuals of subpopulation $b$. 
Then, one expects the following:
Whereas the frequency $f_{ab}^1(t)$ of imitative processes will be
increasing with $s_{ab}(t)$, the frequency $f_{ab}^2(t)$ of avoidance processes
will be decreasing with $s_{ab}(t)$.
This functional relationship can, for example, be described by
\begin{eqnarray}
 f_{ab}^1(t) &:=& f_a^1(t) \, s_{ab}(t) \, , \nonumber \\
 f_{ab}^2(t) &:=& f_a^2(t) \Big( 1 - s_{ab}(t) \Big)  
 \label{symp}
\end{eqnarray}
with
\begin{equation}
 0 \le s_{ab}(t) \le 1 \, .
\end{equation}
$f_a^1(t)$ is a measure for the frequency of imitative processes within
subpopulation $a$, $f_a^2(t)$ a measure for the frequency of avoidance
processes. If we assume the sympathy between individuals of the same
subpopulation to be be maximal, we have $s_{11}(t) \equiv 1 \equiv s_{22}(t)$.

\subsection{Imitative processes ($f_a^1(t) \equiv 1$, $f_a^2(t) \equiv 0$)}
\label{s4.2}
In the following simulations of imitative processes we assume the preferred
positions to be $x_1 = 6/20$ and $x_2 = 15/20$. With 
\begin{equation}
 \Big( s_{ab}(t) \Big) \equiv \Big( f_{ab}^1(t) \Big) := \left(
\begin{array}{cc}
1 & 1 \\
0 & 1 
\end{array} \right) \, ,
\end{equation}
the individuals of subpopulation $a=1$ like the individuals of subpopulation
$a=2$, but not the other way round. That means,
subpopulation 2 influences subpopulation 1, but not vice versa.
\par
As expected, in both behavioral distributions 
$P_a(x,t)$ there appears a maximum
around the preferred behavior $x_a$. In addition, due to imitative processes of
subpopulation 1, a second maximum of $P_1(x,t)$ develops around the
preferred behavior $x_2$ of subpopulation~2. 
This second maximum is small, if the indifference $L_1$ of subpopulation 1
with respect to variations of the position $x$ is low (see fig. \ref{fi1}).
For high values of the indifference 
$L_1$ even the {\em majority} of individuals
of subpopulation 1 imitates the behavior of subpopulation 2 (see fig. 
\ref{fi2})! One could say, 
the individuals of subpopulation 2 act as {\em trendsetters}.
This phenomenon is typical for {\em fashion}.
\par
We shall now consider the case
\begin{equation}
 \Big( s_{ab}(t) \Big) \equiv \Big( f_{ab}^1(t) \Big) := \left(
\begin{array}{cc}
1 & 1 \\
1 & 1 
\end{array} \right) \, ,
\end{equation}
for which the subpopulations influence 
each other mutually with equal strengths.
If the indifference $L_a$ with respect to changes of the position $x$
is small in both subpopulations~$a$, {\em each} probability distribution 
$P_a(x,t)$ has {\em two} maxima. The higher maximum is located around the 
preferred position $x_a$. A second maximum can be found around the position
preferred in the {\em other} subpopulation. It is the higher, the greater
the indifference $L_a$ is (see fig. \ref{fi3}).
\par
However, if $L_a$ exceeds a certain value in at least one subpopulation,
a socalled {\em phase transition} (that means, a qualitative different
situation) occurs, since
only {\em one} maximum develops in each behavioral distribution $P_a(x,t)$!
Despite the fact
that the social fields $V_a(x,t)$
and diffusion coefficients $Q_a(x,t)$ of the subpopulations $a$
are different because of their different preferred positions $x_a$ 
(and different utility functions $U^a(x,t)$), 
the behavioral distributions $P_a(x,t)$ agree after some time!
Especially, the maxima $x_a^*$ of the distributions $P_a(x,t)$
are located at the {\em same} position $x^*$ in both subpopulations. 
One could say, the two subpopulations made a {\em compromise}.
The compromise $x^*$ is nearer to the position $x_a$ of the
subpopulation $a$ with the lower indifference $L_a$ (see fig.~\ref{fi4}). 

\subsection{Avoidance processes ($f_a^1(t) \equiv 0$, $f_a^2(t) \equiv 1$)} 
\label{s4.3}
\begin{figure}[htbp]
\parbox[b]{7.4cm}{
\epsfxsize=7.3cm 
\centerline{\rotate[r]{\hbox{\epsffile[28 28 570
556]{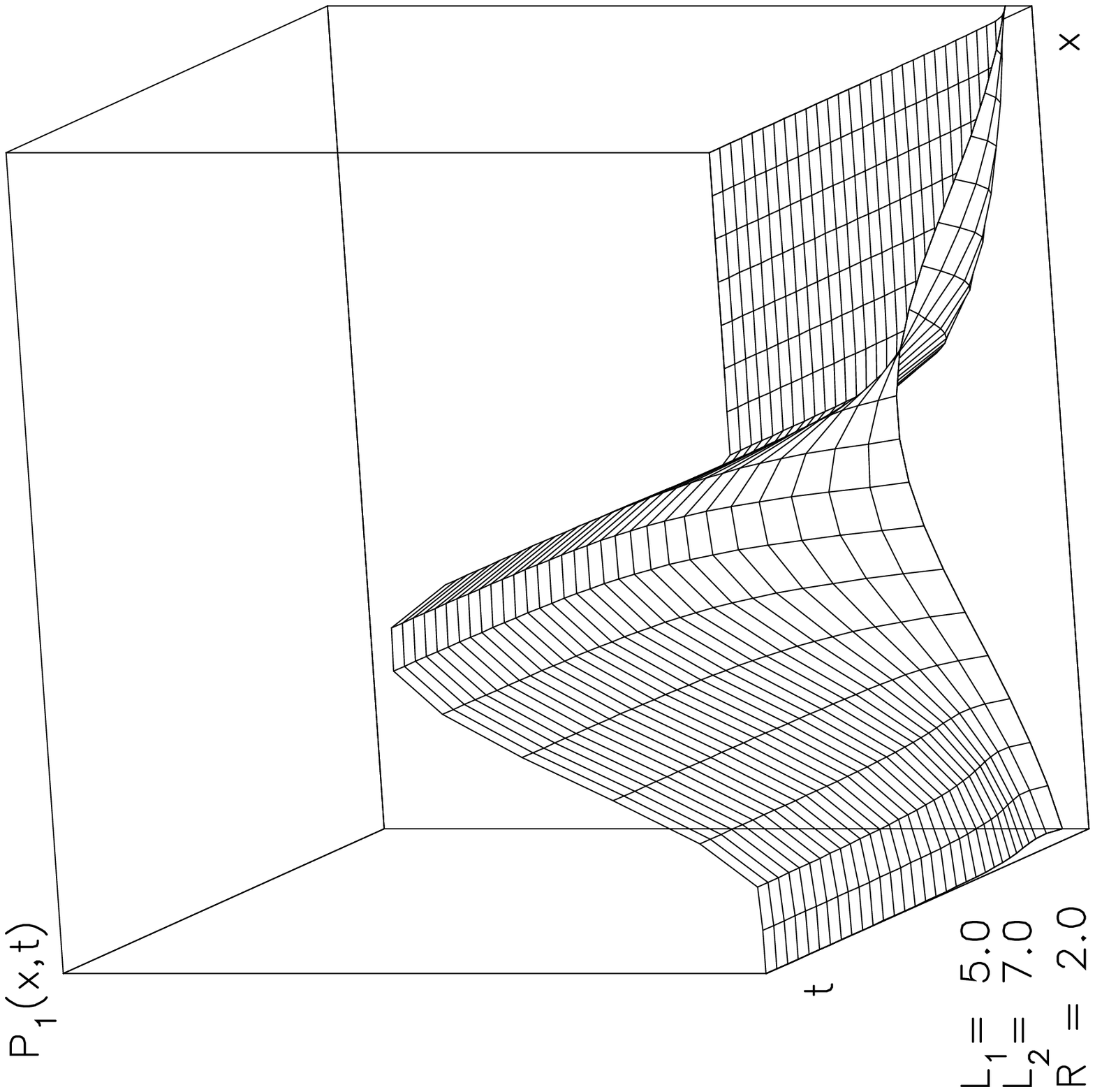}}}}
}\hfill
\parbox[b]{7.4cm}{
\epsfxsize=7.3cm 
\centerline{\rotate[r]{\hbox{\epsffile[28 28 570
556]{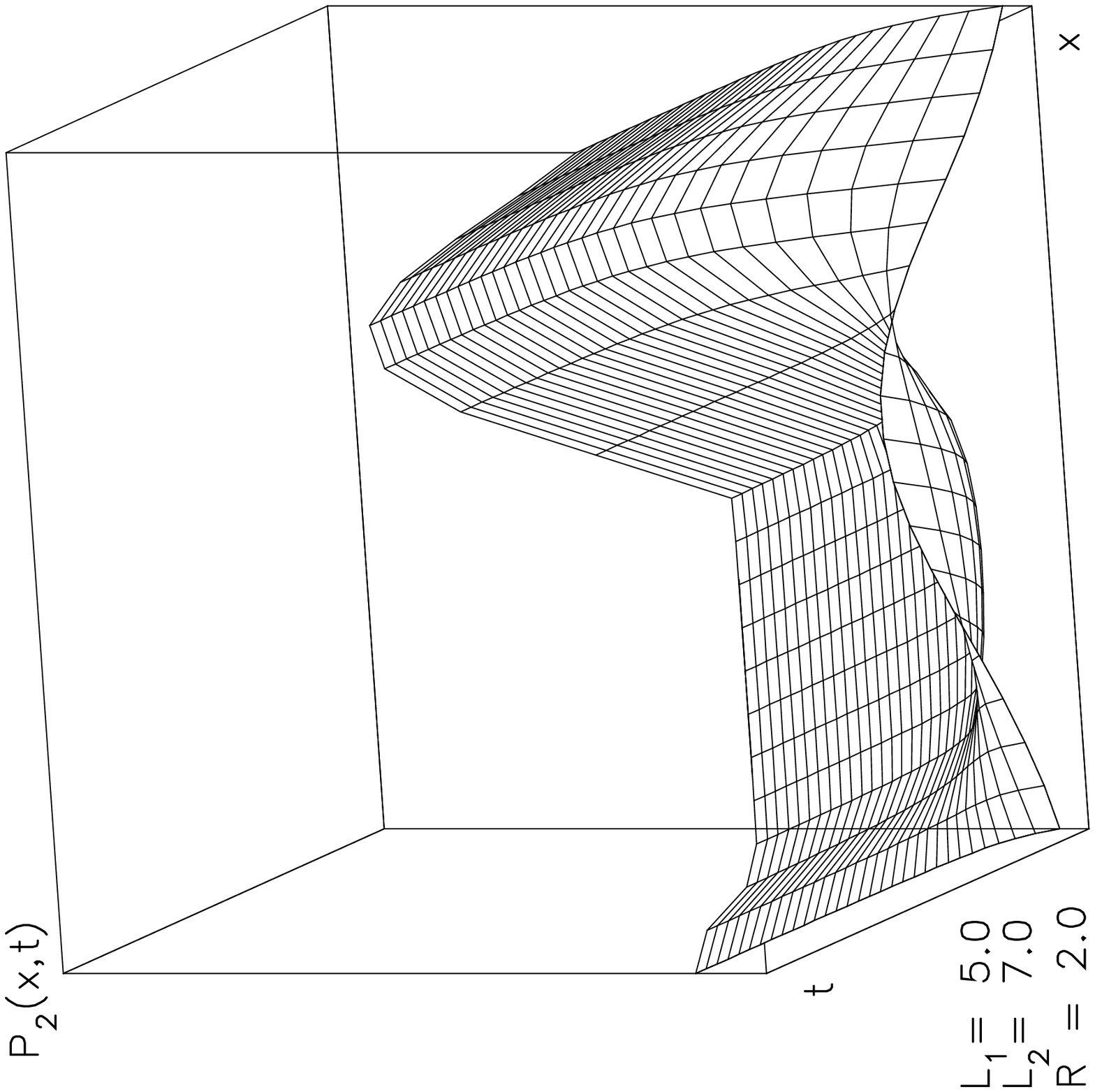}}}}
}
\parbox[b]{7.4cm}{
\epsfxsize=7.3cm 
\centerline{\rotate[r]{\hbox{\epsffile[28 28 570
556]{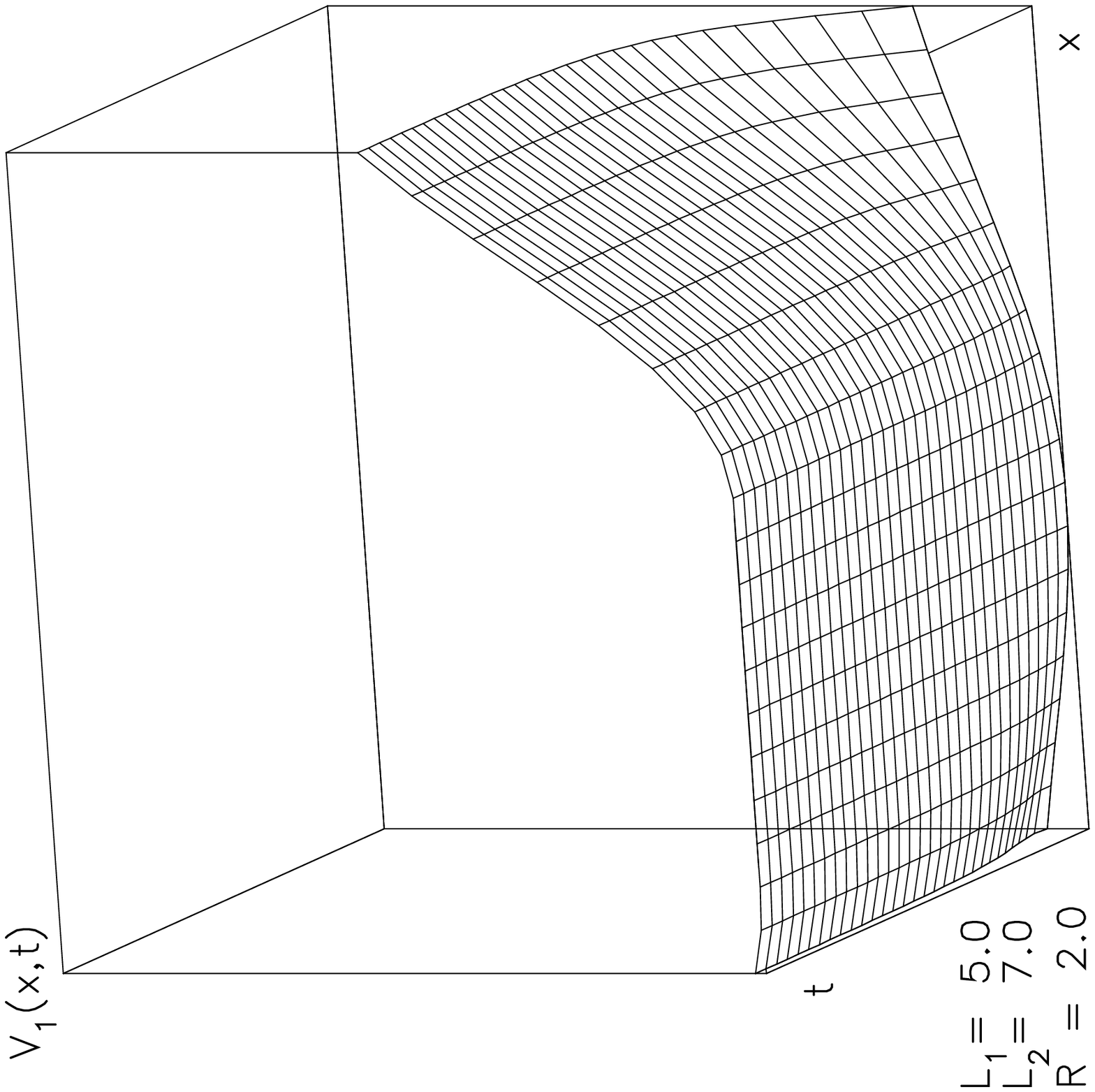}}}}
}\hfill
\parbox[b]{7.4cm}{
\epsfxsize=7.3cm 
\centerline{\rotate[r]{\hbox{\epsffile[28 28 570
556]{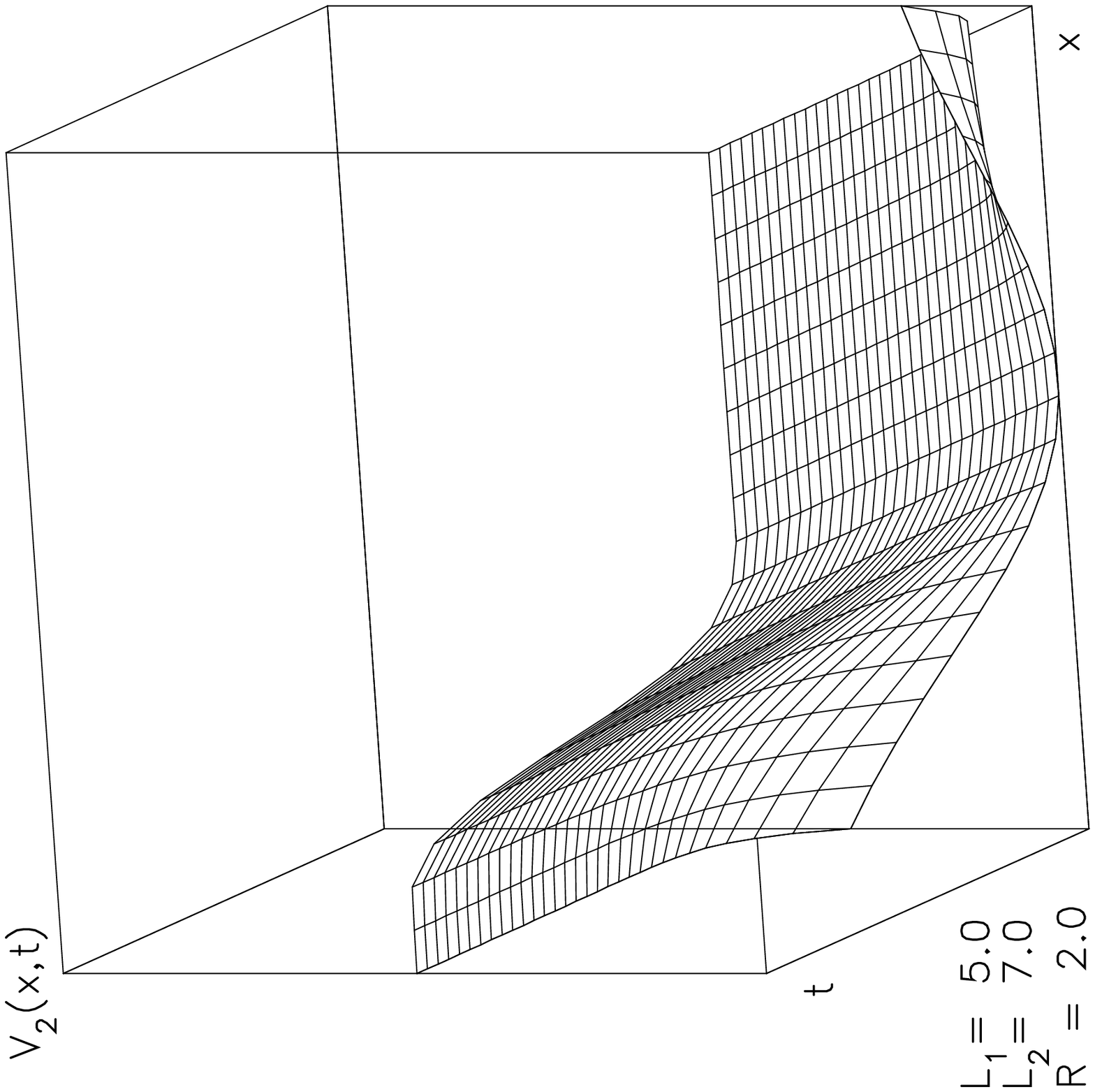}}}}
}
\parbox[b]{7.4cm}{
\epsfxsize=7.3cm 
\centerline{\rotate[r]{\hbox{\epsffile[28 28 570
556]{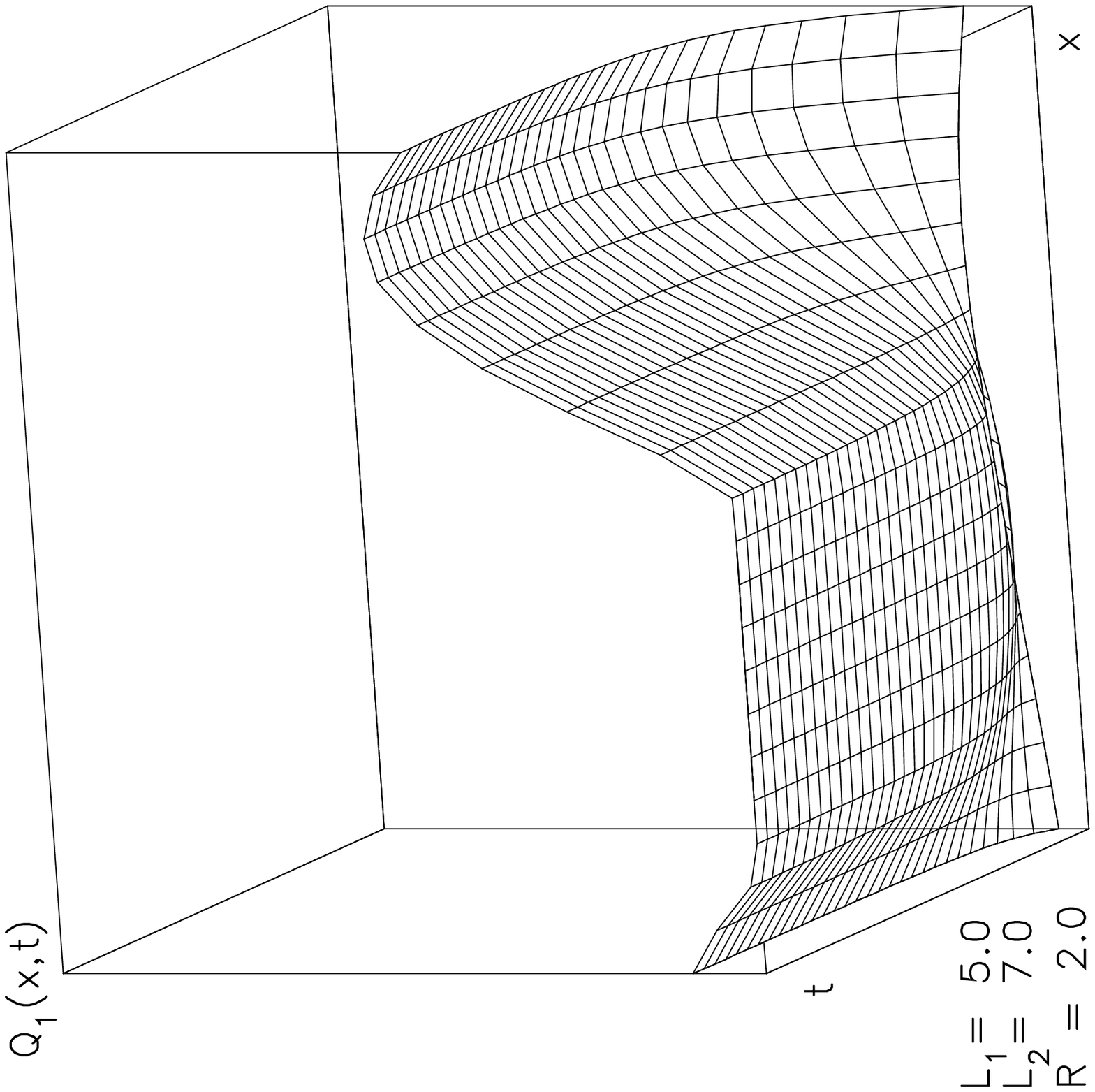}}}}
}\hfill
\parbox[b]{7.4cm}{
\epsfxsize=7.3cm 
\centerline{\rotate[r]{\hbox{\epsffile[28 28 570
556]{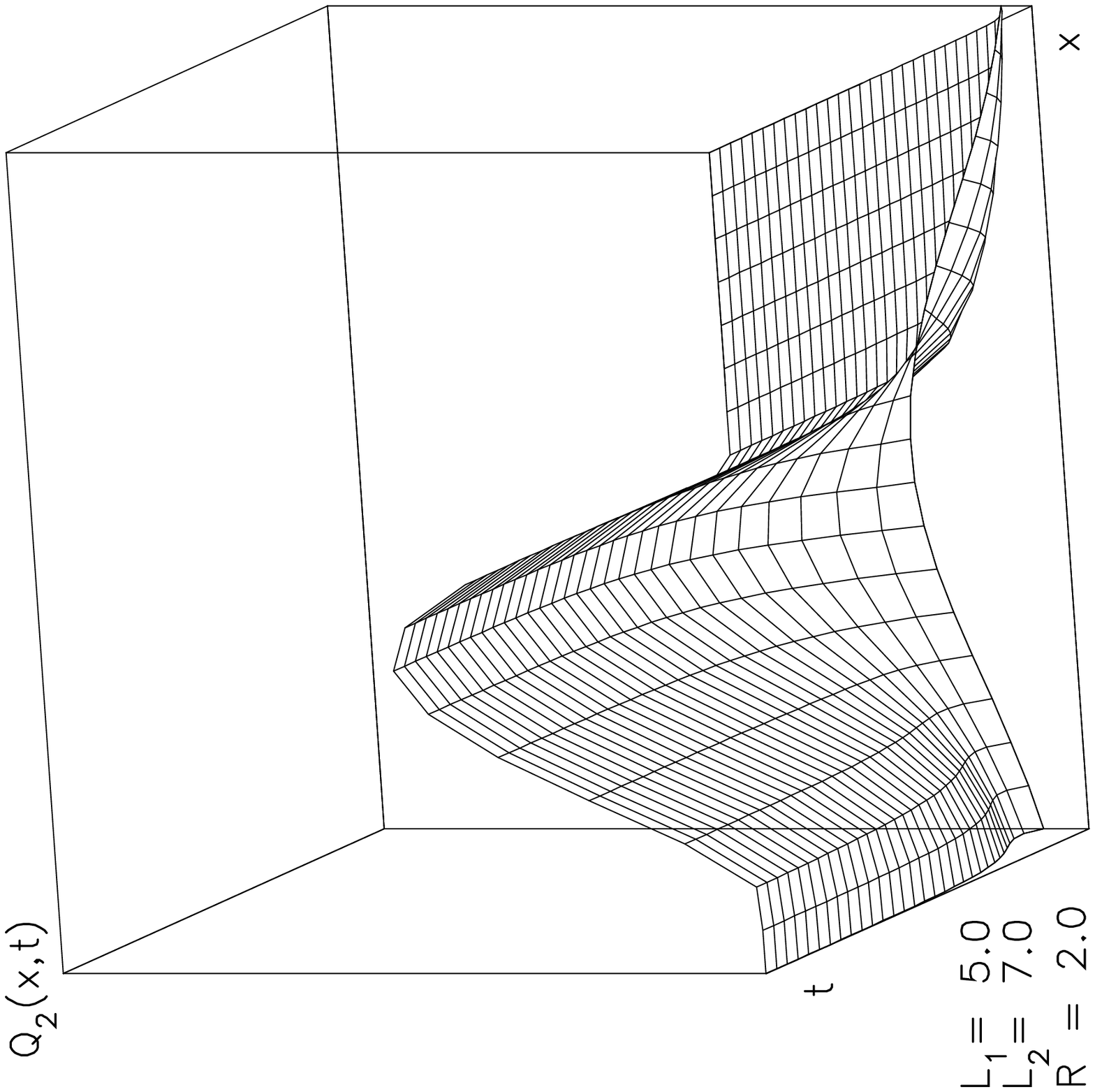}}}}
}
\parbox{15cm}{
\caption{Avoidance processes for mutual dislike of both
subpopulations.\label{fi5}}
}
\end{figure}

\begin{figure}[htbp]
\parbox[b]{7.4cm}{
\epsfxsize=7.3cm 
\centerline{\rotate[r]{\hbox{\epsffile[28 28 570
556]{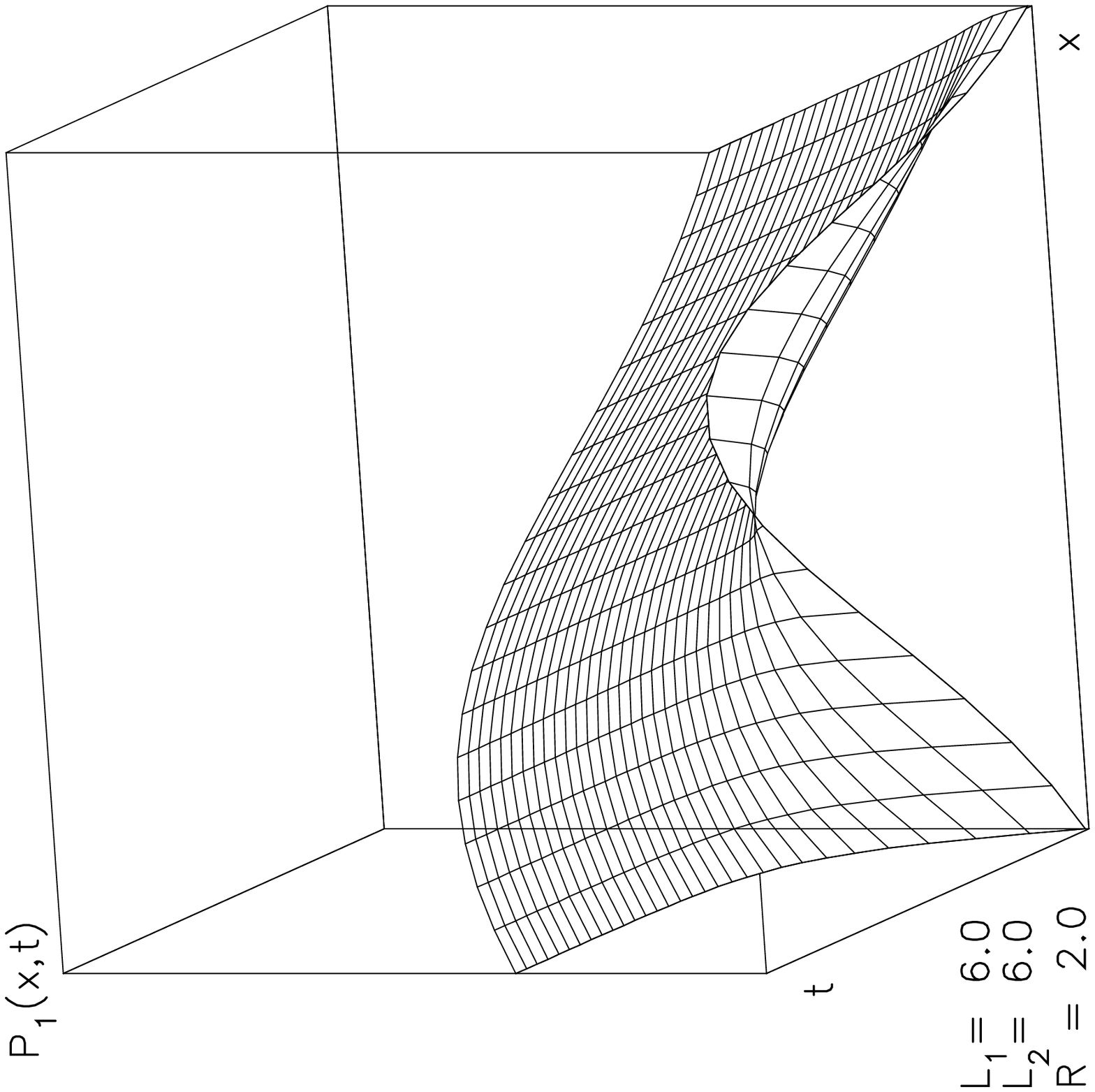}}}}
}\hfill
\parbox[b]{7.4cm}{
\epsfxsize=7.3cm 
\centerline{\rotate[r]{\hbox{\epsffile[28 28 570
556]{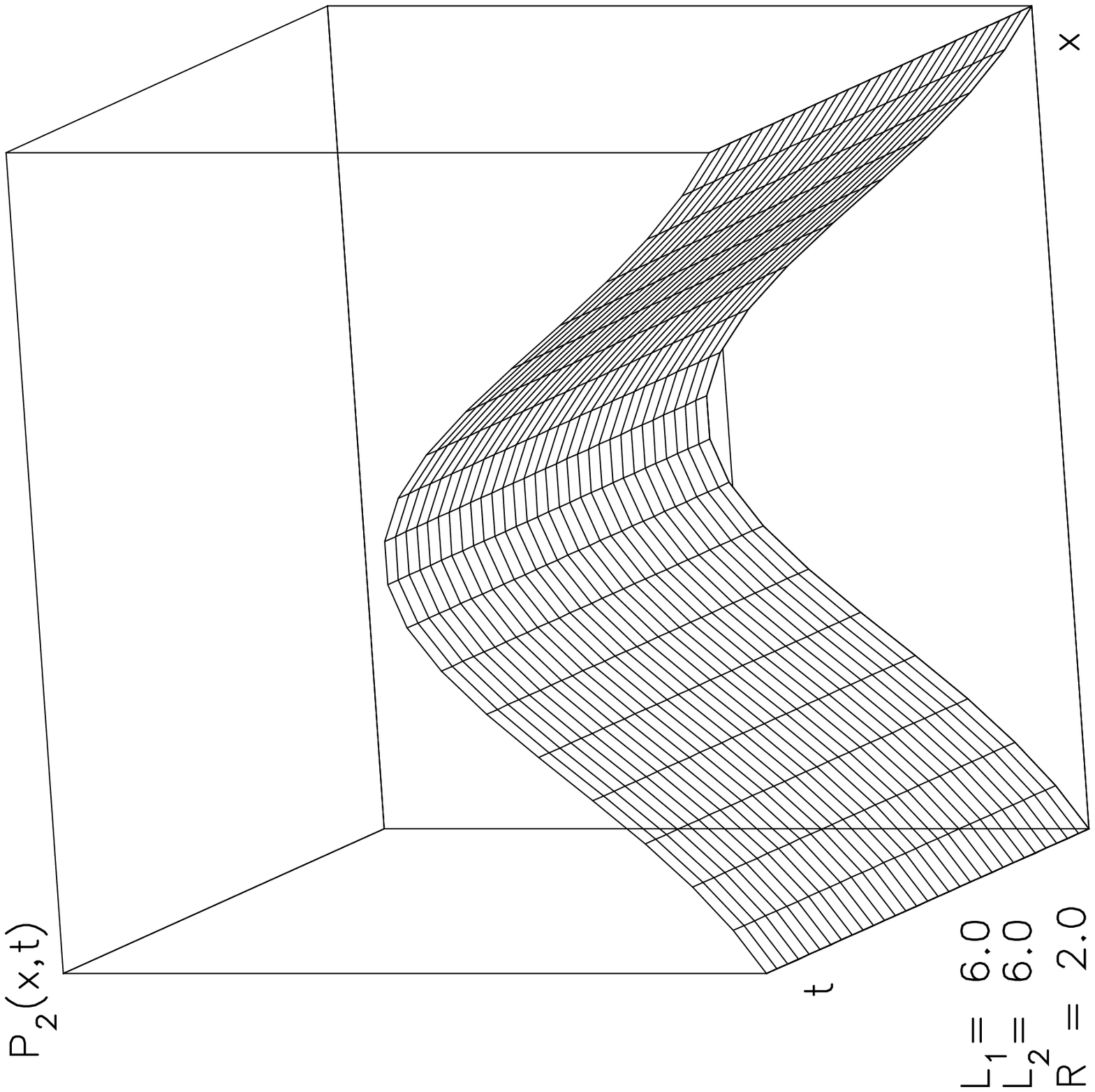}}}}
}
\parbox{15cm}{
\caption{Avoidance processes for one-sided dislike.\label{fi6}}
}
\end{figure}

For the simulation of avoidance processes we assume with
$x_1 = 9/20$ and $x_2 = 12/20$ that both subpopulations nearly prefer the same
behavior. Figure \ref{fi5} shows the case, where the individuals of
different subpopulations dislike each other:
\begin{equation}
 \Big( s_{ab}(t) \Big) := \left(
\begin{array}{cc}
1 & 0 \\
0 & 1 
\end{array} \right) \, , \qquad \mbox{i.e.,} \qquad
 \Big( f_{ab}^2(t) \Big) \equiv \left(
\begin{array}{cc}
0 & 1 \\
1 & 0 
\end{array} \right) \, .
\end{equation}
This corresponds to a mutual influence of each 
subpopulation on the other.
The computational results indicate that
\begin{itemize}
\item individuals avoid behaviors which are found in the 
other subpopulation. 
\item The subpopulation $a=1$ with the lower indifference $L_1 < L_2$
is distributed around the preferred behavior $x_1$ and
pushes away the other subpopulation!
\end{itemize}
Despite the fact that the initial behavioral 
distribution $P_a(x,0)$ agrees in both subpopulations, there is nearly
no overlapping of $P_1(x,t)$ and $P_2(x,t)$ after some time. This
is typical of {\em polarization phenomena} in the society. 
Well-known examples are
the development of {\em ghettos} or the formation of {\em extremist
groups}. 
\par
In figure \ref{fi6}, we assume that the 
individuals of subpopulation 2 like the
individuals of subpopulation 1 and, therefore, do not react to the behaviors
in subpopulation 1 with avoidance processes:
\begin{equation}
 \Big( s_{ab}(t) \Big) := \left(
\begin{array}{cc}
1 & 0 \\
1 & 1 
\end{array} \right) \, , \qquad \mbox{i.e.,} \qquad
 \Big( f_{ab}^2(t) \Big) \equiv \left(
\begin{array}{cc}
0 & 1 \\
0 & 0 
\end{array} \right) \, .
\end{equation}
As a consequence, 
$P_2(x,t)$ remains unchanged with time, whereas $P_1(x,t)$
drifts away from the preferred behavior $x_1$ due to avoidance processes.
Surprisingly, the polarization effect 
is much smaller than in figure \ref{fi5}!
The distributions $P_1(x,t)$ and $P_2(x,t)$ overlap considerably. This is,
because the slope of $P_2(x,t)$ is smaller than in figure \ref{fi5}
(and remains constant). As a consequence, the probability for an individual
of subpopulation~1 to meet a disliked individual of subpopulation 2 with
the same behavior $x$ can hardly be decreased by a small behavioral change.
One may conclude, that polarization effects (which often lead to an
escalation) can be reduced, if individuals do not return dislike
with dislike. 

\section{Empirical determination of the model parameters}\label{s5}

For practical purposes one has, of course, to determine the model parameters
from empirical data. Therefore, let us assume we know empirically the
distribution functions $P_a^{\rm e}(\vec{x},t_l)$, 
[the interaction rates $\nu_{ab}^{\rm e}(t_l)$,] and the effective
transition rates $w_{\rm e}^a(\vec{x}'|\vec{x};t_l)$ ($\vec{x}' \ne \vec{x}$)
for a couple of times $t_l \in \{ t_0,\dots, t_L\}$. 
The corresponding effective
social fields $V_a^{\rm e}(\vec{x},t_l)$ and diffusion
coefficients $Q_{aij}^{\rm e}(\vec{x},t_l)$ are, then, 
easily obtained as\alpheqn{empi}
\begin{equation}
 V_a^{\rm e}(\vec{x},t_l) 
 := - \int\limits^{\weg{x}} d\vec{x}' \cdot
 \vec{K}_a^{\rm e}(\vec{x}',t_l) 
 \equiv - \sum_{i=1}^n \int\limits^{\weg{x}}
 dx'_i \, K_{ai}^{\rm e}(\vec{x}',t_l) 
\end{equation}
with
\begin{equation}
K_{a i}^{\rm e}(\vec{x},t_l) 
:= \int\limits_\Omega \!  d^n x' \, 
(x'_i - x_i) w_{\rm e}^a(\vec{x}'|\vec{x};t_l) \, ,
\end{equation}\reseteqn
and 
\begin{equation}
Q_{a i j}^{\rm e}(\vec{x},t_l) 
:= \displaystyle \int\limits_\Omega \!  d^n x' \, (x'_i - x_i) (x'_j - x_j)
 w_{\rm e}^a(\vec{x}'|\vec{x};t_l) \, .
\end{equation}
Much more difficult is the determination of the utility functions
$U_a^{\rm e}(\vec{x},t_l)$, $U^a_{\rm e}(\vec{x},t_l)$, the distance functions
$D_a^{\rm e}(\vec{x}',\vec{x};t_l)$, and the rates $\nu_a^{\rm e}(t_l)$,
$\nu_{ab}^{\rm 1e}(t_l) := \nu_{ab}^{\rm e}(t_l) f_{ab}^{\rm 1e}(t_l)$, 
$\nu_{ab}^{\rm 2e}(t_l) := \nu_{ab}^{\rm e}(t_l) f_{ab}^{\rm 2e}(t_l)$. 
This can be done by numerical minimization of the {\em error function} 
\begin{equation}
 F := \sum_{a=1}^A \sum_{l=0}^L 
 \sum_{\weg{x}, \weg{x}'\in \Omega \atop (\weg{x}'\ne \weg{x})} 
 \frac{1}{2} \left\{ \left[
 w_{\rm e}^a(\vec{x}'|\vec{x};t_l) - \frac{1}{D_a(\vec{x}',\vec{x};t_l)}
 g_a(\vec{x}',\vec{x};t_l) \right] P_a^{\rm e}(\vec{x},t_l) \right\}^2 \, ,
 \label{error}
\end{equation}
for example with the method of {\em steepest descent} 
(cf. {\sc Forsythe} {\em et. al.}, 1977). 
In (\ref{error}), we have introduced the abbreviation
\begin{equation}
 g_a(\vec{x}',\vec{x};t_l) := \nu_a(t_l) \mbox{e}^{U_a(\weg{x}',t_l)
 - U_a(\weg{x},t_l)} 
 + \sum_{b=1}^A \Big[ \nu_{ab}^1(t_l) P_b^{\rm e}(\vec{x}',t_l)
 + \nu_{ab}^2(t_l) P_b^{\rm e} (\vec{x},t_l) \Big]
 \mbox{e}^{U^a(\weg{x}',t_l) - U^a(\weg{x},t_l)} \, .
\label{assum}
\end{equation}
It turns out (cf. {\sc Helbing}, 1992c), 
that the rates $\nu_a(t_l)$ have to be taken
constant during the minimization process (e.g., $\nu_a(t_l) \equiv 1$),
whereas the parameters $U_a(\vec{x},t_l)$, $U^a(\vec{x},t_l)$, 
$\nu_{ab}^1(t_l):= \nu_{ab}^{\rm e}(t_l) f_{ab}^{\rm 1}(t_l)$ 
and $\nu_{ab}^2(t_l):= \nu_{ab}^{\rm e}(t_l) f_{ab}^{\rm 2}(t_l)$ 
are to be varied.
For $1/D_a(\vec{x}',\vec{x};t_l)$ one inserts\alpheqn{dista}
\begin{equation}
 \frac{1}{D_a(\vec{x}',\vec{x};t_l)} 
= \frac{n_a(\vec{x}',\vec{x};t_l)}{d_a(\vec{x}',\vec{x};t_l)}
\end{equation}
with
\begin{equation}
 n_a(\vec{x}',\vec{x};t_l) :=
w_{\rm e}^a(\vec{x}'|\vec{x};t_l) g_a(\vec{x}',\vec{x};t_l)
\Big[ P_a^{\rm e}(\vec{x},t_l) \Big]^2 \! + w_{\rm e}^a(\vec{x}|\vec{x}';t_l)
g_a(\vec{x},\vec{x}';t_l) \Big[ P_a^{\rm e}(\vec{x}',t_l) \Big]^2 
\end{equation}
and
\begin{equation}
 d_a(\vec{x}',\vec{x};t_l)
:= \Big[ g_a(\vec{x}',\vec{x};t_l) P_a^{\rm e}(\vec{x},t_l) \Big]^2 \!
+ \Big[ g_a(\vec{x},\vec{x}';t_l) P_a^{\rm e}(\vec{x}',t_l) \Big]^2 \, .
\end{equation}\reseteqn
(\ref{dista}) follows from the minimum 
condition for $D_a(\vec{x}',\vec{x};t_l)$
(cf. {\sc Helbing}, 1992c). 
\par
Since $F$ may have a couple of minima due to its nonlinearity, suitable
start parameters have to be taken. Especially, the numerically determined
rates $\nu_{ab}^1(t_l)$ and $\nu_{ab}^2(t_l)$ have to be
non-negative.
\par
If $F$ is minimal for the parameters 
$U_a(\vec{x},t_l)$, $U^a(\vec{x},t_l)$, 
$D_a(\vec{x}',\vec{x};t_l)$, $\nu_a(t_l)$,
$\nu_{ab}^1(t_l)$ and $\nu_{ab}^2(t_l)$, this is (as can easily be checked)
also true for the scaled parameters
\begin{eqnarray}
 U_a^{\rm e}(\vec{x},t_l) &:=& U_a(\vec{x},t_l) - C_a(t_l) \, , \nonumber \\
 U^a_{\rm e}(\vec{x},t_l) &:=& U^a(\vec{x},t_l) - C^a(t_l) \, , \nonumber \\
 D_a^{\rm e}(\vec{x}',\vec{x};t_l) &:=& \frac{D_a(\vec{x}',\vec{x};t_l)}
 {D_a(t_l)} \, , \nonumber \\
 \nu_a^{\rm e}(t_l) &:=& \frac{\nu_a(t_l)}{D_a(t_l)} \, , \nonumber \\
 \nu_{ab}^{1{\rm e}}(t_l) &:=& \frac{\nu_{ab}^1(t_l)}{D_a(t_l)} \, ,
 \nonumber \\
 \nu_{ab}^{2{\rm e}}(t_l) &:=& \frac{\nu_{ab}^2(t_l)}{D_a(t_l)} \, .
\label{scal}
\end{eqnarray}
In order to obtain unique results we put
\begin{equation}
 \sum_{\weg{x}\in \Omega} U_a^{\rm e}(\vec{x},t_l) \stackrel{!}{\equiv} 0 \, ,
 \qquad \sum_{\weg{x}\in \Omega} U^a_{\rm e}(\vec{x},t_l) 
 \stackrel{!}{\equiv} 0 \, ,
\end{equation}
and
\begin{equation}
 \sum_{\weg{x}, \weg{x}'\in \Omega \atop (\weg{x}'\ne \weg{x})} 
 \frac{1}{D_a^{\rm e}(\vec{x}',\vec{x};t_l)}
 \stackrel{!}{\equiv} 
 \sum_{\weg{x}, \weg{x}'\in \Omega \atop (\weg{x}'\ne \weg{x})} 1 \, ,
\end{equation}
which leads to
\begin{equation}
 C_a(t_l) := \frac{\displaystyle \sum_{\weg{x}\in \Omega} U_a(\vec{x},t_l)}
 {\displaystyle \sum_{\weg{x}\in \Omega} 1} \, , \qquad
 C^a(t_l) := \frac{\displaystyle \sum_{\weg{x}\in \Omega} U^a(\vec{x},t_l)}
 {\displaystyle \sum_{\weg{x}\in \Omega} 1} \, ,
\end{equation}
and
\begin{equation}
 \frac{1}{D_a(t_l)} := \frac{\displaystyle 
 \sum_{\weg{x}, \weg{x}'\in \Omega \atop (\weg{x}'\ne\weg{x})} 
 \frac{1}{D_a(\vec{x}',\vec{x};t_l)}}
 {\displaystyle 
 \sum_{\weg{x}, \weg{x}'\in \Omega \atop (\weg{x}'\ne \weg{x})} 1} \, .
\label{undi}
\end{equation}
$C_a(t_l)$ and $C^a(t_l)$ are {\em mean utilities}, whereas $D_a(t_l)$
is a kind of {\em unit of distance}.
\par
The distances $D_a^{\rm e}(\vec{x}',\vec{x};t)$ are suitable quantities for
{\em multidimensional scaling} ({\sc Kruskal} and {\sc Wish}, 1978,
{\sc Young} and {\sc Hamer}, 1987). 
They reflect the ``psychical structure''
(psychical topology) of individuals of subpopulation $a$, 
since they determine which behaviors are
more or less related (compatible) (comp. to {\sc Osgood} {\em et. al.}, 1957). 
By the dependence on 
$a$, $D_a^{\rm e}(\vec{x}',\vec{x};t)$ distinguishes 
different psychical structures 
resulting in different types $a$ of behavior and, therefore, different
``characters'' (personalities).

\subsection{Evaluation of the German migration data}

It is not easy to find suitable data for the determination of the 
model parameters, since
usually there only exist data for the temporal development of a certain
behavioral distribution $P_a(\vec{x},t_l)$, but not for the corresponding 
effective transition rates $w^a_{\rm e}(\vec{x}'|\vec{x};t_l)$.
However, for a few countries all necessary data 
are known about {\em migration} between different
regions (see {\sc Weidlich} and {\sc Haag}, 1988). In this case,
the behavior $\vec{x} \equiv x$ means {\em to live in region}
$x \in \{1,2,\dots,S\}$, where $S$ is the number of distinguished regions.
\par
\begin{figure}[htbp]
\epsfxsize=6cm 
\centerline{\hbox{\epsffile[0 0 168
224]{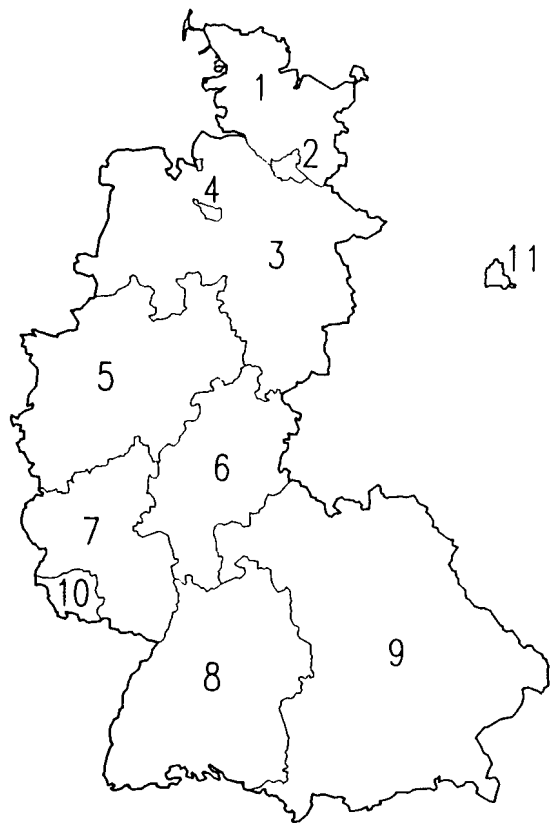}}}
\caption{The subdivision of West Germany into eleven federal states
(from {\sc Weidlich} and {\sc Haag}, 1988).\label{fi7}}
\end{figure}
\begin{figure}[htbp]
\epsfxsize=8cm 
\centerline{\rotate[r]{\hbox{\epsffile[85 40 525 756]
{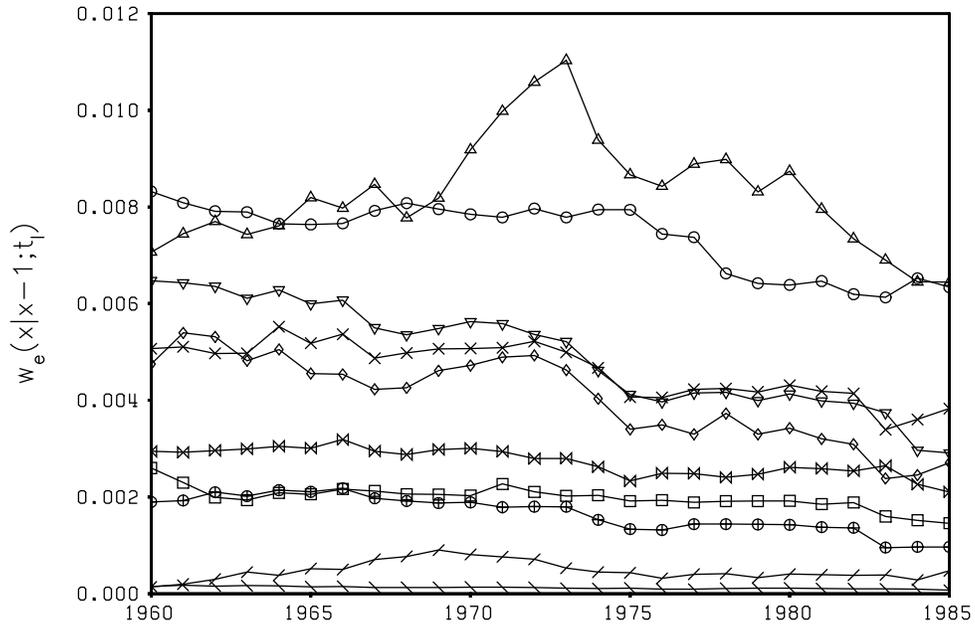}}}}
\caption{Time dependence of some effective transition rates
of migration in West Germany.\label{fi8}}
\end{figure}

In the following, the results for migration in West Germany 
shall be presented. West Germany is divided into 10 federal states
and the region of West Berlin (see figure \ref{fi7} and table \ref{ta1}).
The data for these $S=11$ regions can be found in
the {\em Statistische Jahrb\"ucher} 
of the years 1960 to 1985 on an annual basis ($t_0 = 1960$, 
$t_1 = 1961$, $\dots$, $t_L = 1985$). 
Our data analysis will assume---with respect to migration---one more or less
homogeneous population,
that means, we have $A=1$, and the index $a$ can be dropped. 
Figure \ref{fi8} shows some examples for the temporal
variation of the effective transition 
rates $w_{\rm e}(x'|x;t_l)$.
\par
\begin{figure}[htbp]
\epsfxsize=8cm 
\centerline{\rotate[r]{\hbox{\epsffile[85 40 525 756]
{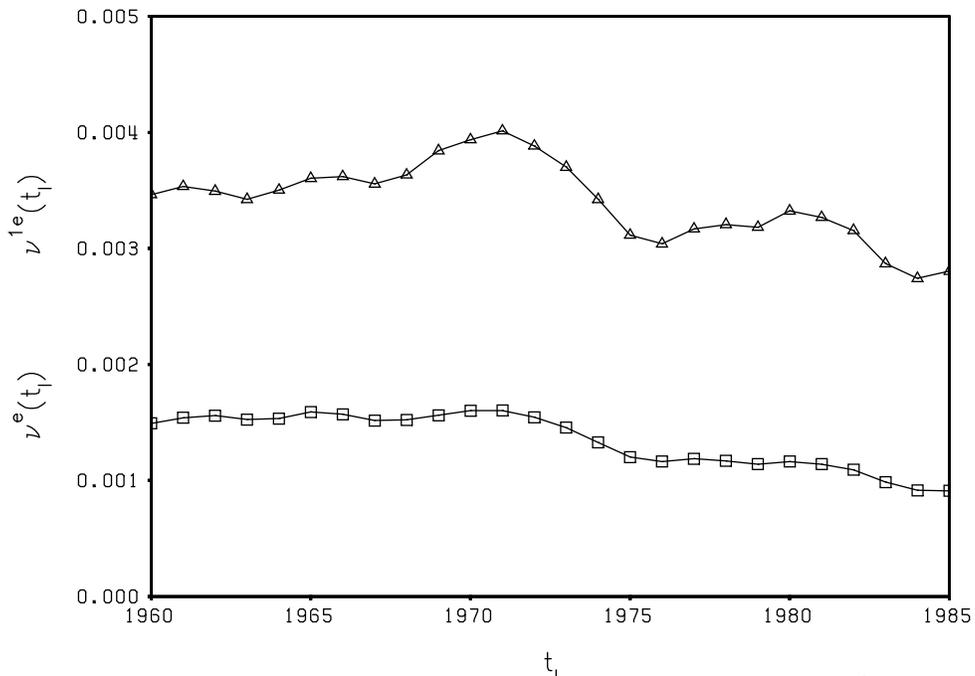}}}}
\caption{Rate $\nu^{\rm e}(t_l)$ of spontaneous behavioral changes
($\Box$) and rate $\nu^{\rm 1e}(t_l)$ of imitative processes
($\bigtriangleup$) calculated from the migration data of
West Germany.\label{fi9}}                                      
\end{figure}
\begin{figure}[htbp]
\epsfxsize=8cm                                                  
\centerline{\rotate[r]{\hbox{\epsffile[85 40 525 756]
{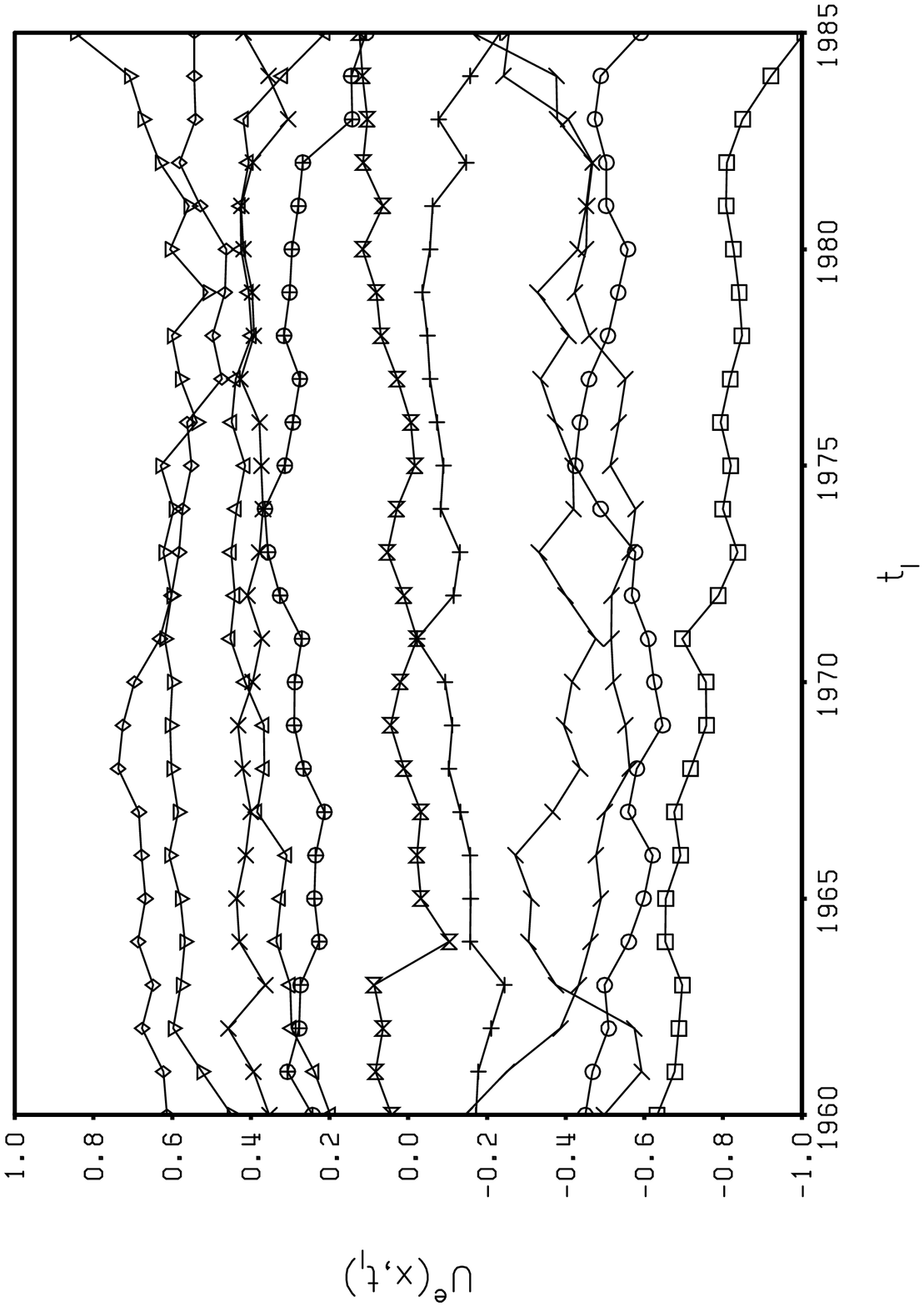}}}}
\caption{Utility functions for spontaneous removals from one
federal state to another in West Germany.\label{fi10}}
\end{figure}
\begin{figure}[htbp]
\epsfxsize=8cm 
\centerline{\rotate[r]{\hbox{\epsffile[85 40 525 756]
{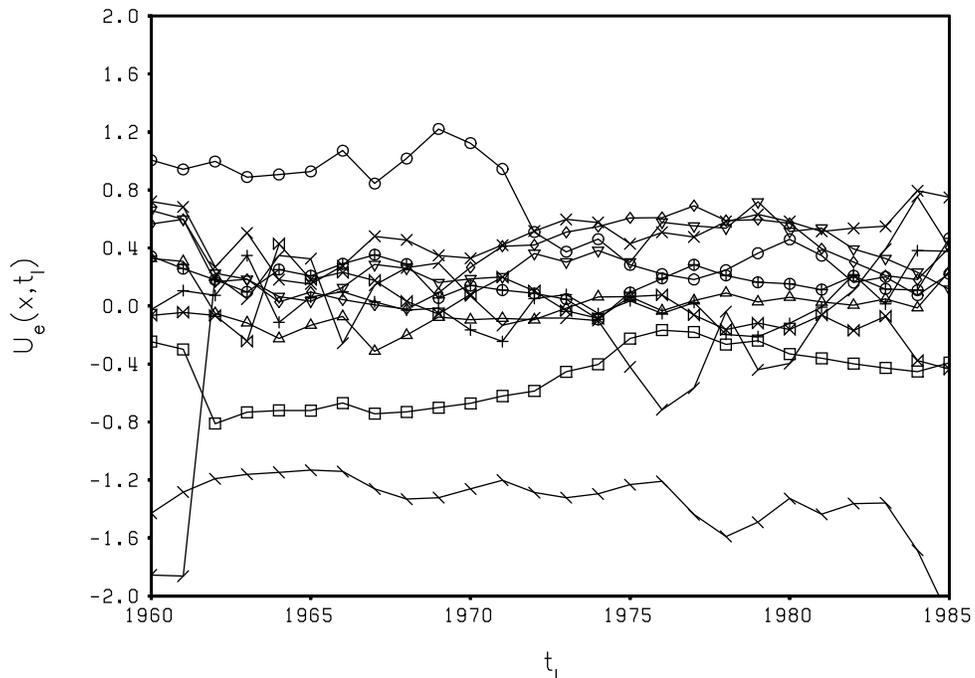}}}}
\caption{Utility functions for removals from one federal state
to another due to imitative processes.\label{fi11}}
\end{figure}

Using the method of steepest descent, the minimization of the error function
(\ref{error}), (\ref{assum}) gives the following results:
\begin{equation}
 \nu^{\rm 2}(t_l) \approx 0 \, ,
\end{equation}
that means, avoidance processes are negligible. The rate
$\nu^{\rm e}(t_l)$ of spontaneous behavioral changes and the rate 
$\nu^{1e}(t_l)$ of imitative processes are depicted in figure \ref{fi9}.
The utility functions $U^{\rm e}(x,t_l)$ for spontaneous changes
and the utility functions $U_{\rm e}(x,t_l)$ for
imitative processes are illustrated in figures 
\ref{fi10} and \ref{fi11}. 
\par
\begin{figure}[htbp]
\epsfxsize=8cm 
\centerline{\rotate[r]{\hbox{\epsffile[85 40 525 756]
{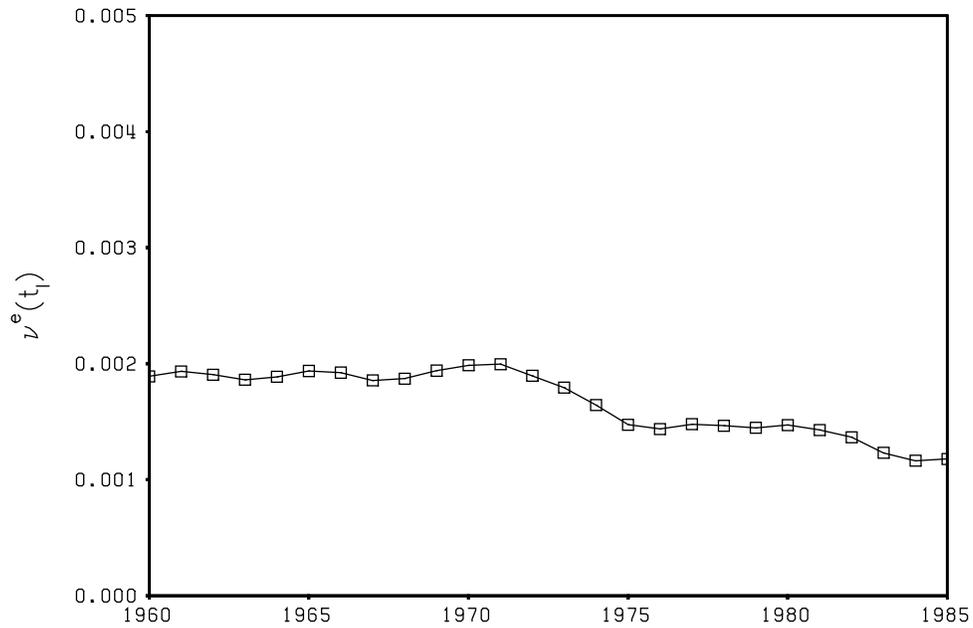}}}}
\caption{Rate of spontaneous migration in West Germany for the
model of {\sc Weidlich} and {\sc Haag}.\label{fi12}}
\end{figure}
\begin{figure}[htbp]
\epsfxsize=8cm 
\centerline{\rotate[r]{\hbox{\epsffile[85 40 525 756]
{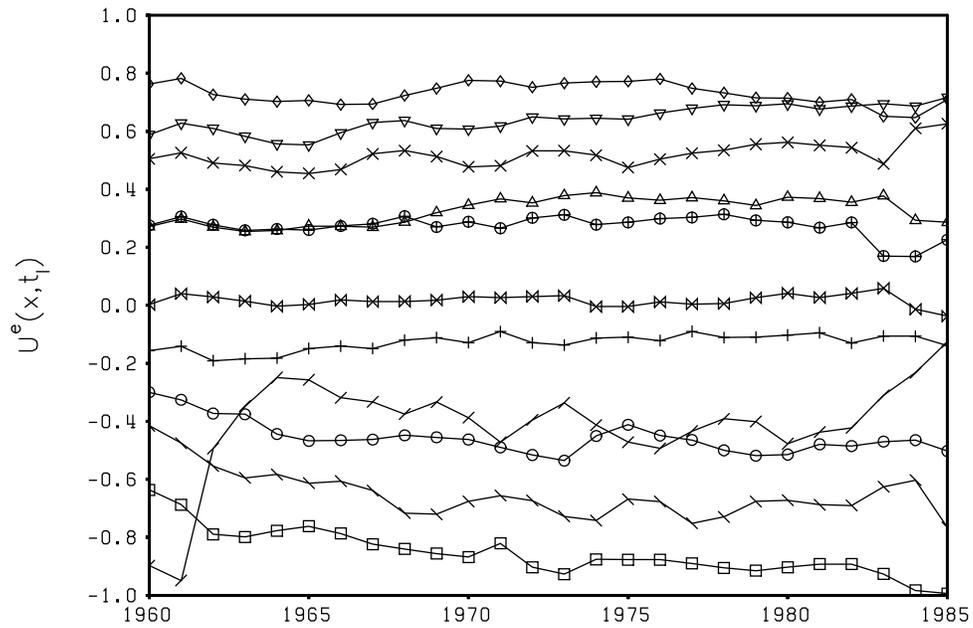}}}}
\caption{Utility functions of the federal states of West Germany
for the model of {\sc Weidlich} and {\sc Haag}. Note, that the
utility of West Berlin (/) is extremely dependent on changes of
the political situation. For example, there is an remarkable increase
of the utility after the erection of the Berlin wall in the year
1961.\label{fi13}}
\end{figure}

The irregulatity of the utility 
functions $U_{\rm e}(x,t_l)$ indicates that they probably fit random
fluctuations of the migration data. Indeed, a mathematical analysis proves
that the term
\begin{equation}
 \nu^1(t_l)P^{\rm e}(x,t_l) \frac{\mbox{e}^{U(x',t_l) - U(x,t_l)}}
{D(x',x;t_l)} = \nu^{\rm 1e}(t_l)P^{\rm e}(x,t_l) 
\frac{\mbox{e}^{U_{\rm e}(x',t_l) - U_{\rm e}(x,t_l)}}{D^{\rm e}(x',x;t_l)}
\end{equation}
only explains 5.2 percent of the variance of the effective transition
rates $w_{\rm e}(x'|x;t)$. Therefore, it makes {\em no significant 
contribution} to their mathematical description, and the migration rates
$w_{\rm e}(x'|x;t)$ of West Germany can already be represented
by the model
\begin{equation}
 w(x'|x;t_l) := \nu(t_l) \frac{\mbox{e}^{U(x',t_l) - U(x,t_l)}}{D(x',x;t_l)}
= \nu^{\rm e}(t_l) \frac{\mbox{e}^{U^{\rm e}(x',t_l) - U^{\rm e}(x,t_l)}}
{D^{\rm e}(x',x;t_l)} \, .
\end{equation}
This result agrees with the model of {\sc Weidlich} and {\sc Haag}
(1988)! Figure \ref{fi12} shows the corresponding rate $\nu^{\rm e}(t_l)$
of spontaneous changes, and figure \ref{fi13} depicts the utility functions
$U^{\rm e}(x,t_l)$. The distances $D^{\rm e}(x',x;t_l)$
can be calculated from the formulas (\ref{dista}), (\ref{scal}), 
and (\ref{undi}).
They are not only a measure for the mean {\em geographical distances}, but
also for {\em transaction costs} (e.g., removal costs) 
and {\em psychical differences}
(of the language, mentality, etc.).
\par
The replacement of the time dependent distances 
$D^{\rm e}(x',x;t_l)$ with the time independent values
$D_*^{\rm e}(x',x)$ defined by
\begin{equation}
 \frac{1}{D_*^{\rm e}(x',x)} := \frac{1}{L+1} \sum_{l=0}^L 
 \frac{1}{D^{\rm e}(x',x;t_l)}
\end{equation}
(see table \ref{ta2})
allows for a further {\em model reduction}. The optimal value of the
rate of spontaneous behavioral changes is, then, given by
\begin{equation}
 \nu_*^{\rm e}(t_l) := \nu^{\rm e}(t_l) \cdot
\sum_{x,x'=1 \atop (x' \ne x)}^{{ }\atop S} \frac{1}{D^{\rm e}(x',x;t_l)} \, .
\end{equation}
Although the reduced model 
\begin{equation}
 w(x'|x;t_l) := \nu_*^{\rm e}(t_l) \frac{\mbox{e}^{U^{\rm e}(x',t_l) 
 - U^{\rm e}(x,t_l)}}{D_*^{\rm e}(x',x)}
\end{equation}
only needs 
$(S+1)\cdot (L+1) + S \cdot (S-1)/2 = 367$ variables for the description of
$S \cdot (S-1) \cdot (L+1) = 2860$ effective transition
rates $w^{\rm e}(x'|x;t_l)$, it attains a very good 
correlation of 0.984 with the empirical data.
For a more detailled discussion of this model, see
{\sc Weidlich} and {\sc Haag} (1988).

\section{Summary and outlook}\label{s6}

In this article, a behavioral model has been proposed that incorporates in a
consistent way many models of social theory:
the diffusion models, the multinomial logit model, {\sc Lewin}'s field
theory, the logistic equation, the gravity model, the
{\sc Weidlich-Haag} model, and the game dynamical equations.
This very general model opens new 
perspectives concerning a theoretical description and
understanding of behavioral changes, since it is formulated  fully 
mathematically. It takes into account spontaneous (or externally induced)
behavioral changes and behavioral changes due to pair interactions.
Two important kinds of pair interactions have been distinguished: imitative
processes and avoidance processes. 
The model turns out to be suitable for computational simulations,
but it can also be applied to concrete empirical data.

\subsection{Memory effects}\label{s6.1}

The formulation of the model in the previous sections has neglected {\em memory
effects} that may also influence behavioral changes. However, memory
effects can be easily included by generalizing the {\sc Boltzmann}-like
equations to\alpheqn{memb}
\begin{equation}
\frac{d}{dt}P_a(\vec{x},t) = 
\int\limits_{t_0}^t dt' \sum_{\weg{x}'\in \Omega} 
\Big[ w_{t-t'}^a(\vec{x}|\vec{x}';t')P_a(\vec{x}',t') 
- w_{t-t'}^a(\vec{x}'|\vec{x};t')P_a(\vec{x},t') \Big] 
\label{memboltz}
\end{equation}
with the effective transition rates
\begin{equation}
w_{t-t'}^a(\vec{x}'|\vec{x};t') := w^{t-t'}_a(\vec{x}'|\vec{x};t') 
+ \sum_{b=1}^A \sum_{\weg{y}\in \Omega} \sum_{\weg{y}'\in \Omega} 
w^{t-t'}_{ab}(\vec{x}',\vec{y}'|\vec{x},\vec{y};t')P_b(\vec{y},t') \, ,
\label{memrates}
\end{equation}\reseteqn
and generalizing the {\sc Boltzmann-Fokker-Planck} equations to\alpheqn{membfp}
\begin{equation}
\begin{array}{rcl}
\displaystyle \frac{\partial}{\partial t}P_a(\vec{x},t) 
&=& \displaystyle \int\limits_{t_0}^t dt' \, \bigg\{
- \sum_{i=1}^n \frac{\partial}{\partial x_{i}} \Big[K_{a i}^{t - t'}
(\vec{x},t') P_a(\vec{x},t')\Big] \\
& & \qquad \quad \displaystyle +  \frac{1}{2} \sum_{i, j=1}^n 
\frac{\partial}{\partial x_{i}}\frac{\partial}
{\partial x_{j}} \Big[Q_{a i j}^{t - t'}(\vec{x},t')
P_a(\vec{x},t')\Big] \bigg\}
\end{array} 
\label{memBFP}
\end{equation}
with the effective drift coefficients
\begin{equation}
K_{a i}^{t - t'}(\vec{x},t') 
:= \int\limits_\Omega \!  d^n x' \, (x'_i - x_i) 
w^a_{t - t'}(\vec{x}'|\vec{x};t') \, ,
\label{memdrift}
\end{equation}
the effective diffusion coefficients
\begin{equation}
Q_{a i j}^{t - t'}(\vec{x},t') 
:= \int\limits_\Omega \!  d^n x' \, (x'_i - x_i) (x'_j - x_j)
 w^a_{t - t'}(\vec{x}'|\vec{x};t') \, ,
\label{memdiff}
\end{equation}
and
\begin{equation}
w_{t-t'}^a(\vec{x}'|\vec{x};t') := w^{t-t'}_a(\vec{x}'|\vec{x};t') 
+ \sum_{b=1}^A \int\limits_\Omega d^n y \int\limits_\Omega d^n y' \, 
w^{t-t'}_{ab}(\vec{x}',\vec{y}'|\vec{x},\vec{y};t')P_b(\vec{y},t') \, .
\end{equation}\reseteqn
Obviously, in these formulas there only appears
an additional integration over past
times $t'$ ({\sc Helbing}, 1992c). 
The influence of the past results in a dependence
of $w_{t-t'}^a(\vec{x}'|\vec{x};t')$, $K_{a i}^{t - t'}(\vec{x},t')$, 
and $Q_{a i j}^{t - t'}(\vec{x},t')$ on $(t-t')$. The 
{\sc Boltzmann}-like equations (\ref{Boltzlike}) resp. the 
{\sc Boltzmann-Fokker-Planck} equations (\ref{bfp}) used in the
previous sections result from (\ref{memb}) resp.
(\ref{membfp}) in the 
limit 
of short memory. 

\subsection{Group dynamics}\label{s6.2}

The force model described in section \ref{s3} can serve as a new mathematical
modelling concept. For example, it has successfully been applied to the
simulation of {\em pedestrian behavior} (cf. {\sc Helbing}, 1991). 
\par
Another interesting field is an application of the force model to {\em group 
dynamics} and {\em group formation}.
In this case we take $A=N$, that means, each subpopulation consists of
one individual only ($N_a = 1$). Since $N_a \gg 1$ is violated, then,
the temporal change $dP_a(\vec{x},t)/dt$ of $P_a(\vec{x},t)$, which
describes the probability of individual $a$ to show the behavior $\vec{x}$
at time $t$, is additionally subject to fluctuations 
(cf. {\sc Helbing}, 1992a,c).
The interaction rates $\nu_{ab} \equiv \widetilde{\nu}_{ab}$ are
related to the {\em adjacency matrix} and describe the {\em social
(interpersonal) network} ({\sc Burt}, 1982).
Moreover, the sympathy matrix $\Big( s_{ab}(t)\Big)$
is affected by the social processes. The crucial task for simulating group 
dynamics is, therefore, to set up equations for the
temporal change of $s_{ab}(t)$. 
The topic of group dynamics 
will be treated in a forthcoming
paper. 
\clearpage
\paragraph{Acknowledgements}\mbox{ }\\[7mm]
This work has been financially supported by the {\em Volkswagen Stiftung} 
and the {\em Deutsche
Forschungsgemeinschaft} (SFB 230).
The author is grateful to Prof. Dr. W. Weidlich and Dr. R. Reiner for
valuable discussions and commenting on the manuscript.

%
%
\clearpage
\thispagestyle{empty}
\clearpage
\thispagestyle{empty}
\clearpage
\thispagestyle{empty}
\clearpage
\thispagestyle{empty}
\clearpage
\thispagestyle{empty}
\clearpage
\thispagestyle{empty}
\vfill
\clearpage
\thispagestyle{empty}
\begin{table}[hbtp]
\begin{center}
\begin{tabular}{|c|c|l|}
\hline
symbol & region $x$ & name \\
\hline
\hline
$+$ & 1 & Schleswig-Holstein \\
\hline
$\odot $ & 2 & Hamburg \\
\hline
$\bigtriangleup $ & 3 & Niedersachsen \\
\hline 
$\Box $ & 4 & Bremen \\
\hline
$\Diamond $ & 5 & Nordrhein-Westfalen \\
\hline
$\oplus $ & 6 & Hessen\\
\hline
$\Join $ & 7 & Rheinland-Pfalz \\
\hline
$\times $ & 8 & Baden-W\"urttemberg \\
\hline
$\bigtriangledown $ & 9 & Bayern \\
\hline
$\setminus $ & 10 & Saarland \\
\hline
/ & 11 & Berlin \\                                       
\hline
\end{tabular}
\end{center}
\caption{The eleven federal states of West Germany, their names, and
the symbols used in the following figures.\label{ta1}}
\end{table}
\clearpage
\thispagestyle{empty}
\vfill
\clearpage       
\thispagestyle{empty}
\vfill
\clearpage
\thispagestyle{empty}                                  
\vfill
\clearpage
\thispagestyle{empty}
\begin{table}[hbtp]
\begin{center}
\begin{tabular}{|c||c|c|c|c|c|c|c|c|c|c|c|}
\hline
$x$ & 1 & 2 & 3 & 4 & 5 & 6 & 7 & 8 & 9 & 10 & 11 \\
\hline
\hline
1 & -- & 0.16 & 0.56 & 1.19 & 0.99 & 1.93 & 3.01 & 1.79 & 2.13 & 8.35 & 1.26 \\
\hline
2 & 0.16 & -- & 0.44 & 1.65 & 1.58 & 2.18 & 5.03 & 2.34 & 2.61 & 12.89 & 1.55 \\
\hline
3 & 0.56 & 0.44 & -- & 0.25 & 0.44 & 0.90 & 2.13 & 1.26 & 1.55 & 6.38 & 0.79 \\
\hline
4 & 1.19 & 1.65 & 0.25 & -- & 1.91 & 3.10 & 6.16 & 3.31 & 4.30 & 17.37 & 2.58 \\
\hline
5 & 0.99 & 1.58 & 0.44 & 1.91 & -- & 0.65 & 0.53 & 0.79 & 0.93 & 2.56 & 0.92 \\
\hline
6 & 1.93 & 2.18 & 0.90 & 3.10 & 0.65 & -- & 0.47 & 0.61 & 0.75 & 2.23 & 1.15 \\
\hline
7 & 3.01 & 5.03 & 2.13 & 6.16 & 0.53 & 0.47 & -- & 0.59 & 1.39 & 0.54 & 2.33 \\
\hline
8 & 1.79 & 2.34 & 1.26 & 3.31 & 0.79 & 0.61 & 0.59 & -- & 0.38 & 1.56 & 1.14 \\
\hline
9 & 2.13 & 2.61 & 1.55 & 4.30 & 0.93 & 0.75 & 1.39 & 0.38 & -- & 3.49 & 1.09 \\
\hline
10 & 8.35 & 12.89 & 6.38 & 17.37 & 2.56 & 2.23 & 0.54 & 1.56 & 3.49 & -- & 5.07 
\\ \hline
11 & 1.26 & 1.55 & 0.79 & 2.58 & 0.92 & 1.15 & 2.33 & 1.14 & 1.09 & 5.07 & -- \\
\hline
\end{tabular}
\end{center}
\caption{Time independent effective distances $D^{\rm e}_*(x',x)$
between the eleven federal states of West Germany.\label{ta2}}
\end{table}
\clearpage
\end{document}